\documentclass[11pt]{article}
\pdfoutput=1
\usepackage{jcappub,natbib}
\input{colordvi.tex}

\def\ga{\mathrel{\raise.3ex\hbox{$>$\kern-.75em\lower1ex\hbox{$\sim$}}}}
\def\la{\mathrel{\raise.3ex\hbox{$<$\kern-.75em\lower1ex\hbox{$\sim$}}}}

\title{The phenomenology of trapped inflation}

\author[a,b,e]{Lauren Pearce,}
\author[a,c]{Marco Peloso,}
\author[d]{Lorenzo Sorbo,}

\affiliation[a]{School of Physics and Astronomy, University of Minnesota, Minneapolis, 55455 (USA)}
\affiliation[b]{Fine Theoretical Physics Institute, University of Minnesota, Minneapolis, 55455 (USA)}
\affiliation[c]{Minnesota Institute for Astrophysics, University of Minnesota, Minneapolis, 55455 (USA)}
\affiliation[d]{Amherst Center for Fundamental Interactions, Department of Physics, University of Massachusetts, Amherst, MA 01003 (USA)}
\affiliation[e]{Department of Physics and Astronomy, Valparaiso University, Valparaiso, IN 46383 (USA)}

\abstract{Trapped inflation is a mechanism in which particle production from the moving inflaton is the main source of friction in the inflaton equation of motion.  The produced fields source inflaton perturbations, which dominate over the  vacuum ones. We review the computation of the perturbations performed in the original work and in a successive analysis that makes use of the effective field theory (EFT) of inflation. We verify the validity of several results obtained in these works, but we show that the final expression for the power spectrum is affected by one invalid approximation. This approximation also affects the computation of the bispectrum. We show that, once this approximation is replaced by an exact computation, the different schemes used in the literature to compute the perturbations at second order lead to different results for the bispectrum. 
}

\begin{document}

\begin{flushright}  ACFI-T16-07, UMN--TH--3520/16  \end{flushright}

\maketitle
\flushbottom

%%%%%%%%%%%%%%%%%%%%%%%%%%%%%%%%%%%%%%%%%%%%%%%%%
\section{Introduction}
\label{sec:intro}
%%%%%%%%%%%%%%%%%%%%%%%%%%%%%%%%%%%%%%%%%%%%%%%%%

A key difficulty  in inflationary model-building is to enforce the required flatness of the inflaton potential. Even the simplest Taylor expansion $V = \frac{m^2}{2} \, \varphi^2$, where $\varphi$ denotes the inflaton and $m^2$ the curvature of the potential at the minimum, is now ruled out by the data \cite{Ade:2015lrj}.  Moreover, the inflaton potential, as is typical for scalar potentials, is very sensitive to ultraviolet physics: unless some symmetry is at work, the potential will receive UV corrections that will spoil at the very least the predictability of the theory, and in the worst case the flatness of the potential and with it the viability of the model altogether. In the best case scenario the ultraviolet scale is associated to the reduced Planck mass $M_p \equiv \left( 8 \pi \, G_N \right)^{-1/2} \simeq 2.4 \cdot 10^{18}$~GeV (if such a scale is smaller, problems associated with UV-sensitivity become even stronger). Therefore, UV-sensitivity becomes much more dramatic for models for which $\varphi$ changes by an amount comparable to, or larger than, the Planck scale.

The above conclusions and considerations can drastically change if the inflaton evolves in a strongly dissipative regime, as was first considered in the context of warm inflation~\cite{Berera:1995ie,BasteroGil:2009ec,Bartrum:2013fia,Bastero-Gil:2014raa}. If the motion of the inflaton amplifies the fields coupled to it efficiently enough, the backreaction associated with this amplification can be the dominant source of friction for the inflationary dynamics. This may lead to successful slow roll inflation in potentials that may not be flat enough in absence of dissipation. Moreover, the increased friction can reduce the dynamical range scanned by the inflaton during inflation; then one needs to ensure flatness only in this more restricted range. 

The model of trapped inflation~\cite{Green:2009ds} provides a simple and minimal framework to test this idea. It is characterized by the Lagrangian
\begin{align}
\mathcal{L} = - \dfrac{1}{2} \partial_\mu \varphi \partial^\mu \varphi - V(\varphi) - \sum_i \dfrac{1}{2} \partial_\mu \chi_i \partial^\mu \chi_i - \sum_i \dfrac{g^2}{2} \left( \varphi - \varphi_{0i} \right)^2 \chi_i^2,
\label{trapped}
\end{align}
where  $\chi_i$ is a set of fields coupled to the inflaton $\varphi$.  The quantities $\varphi_{0i}$ are numerical values crossed by $\varphi$ during inflation. As the inflaton zero mode crosses the value $\varphi_{0i}$, quanta of $\chi_i$ are nonperturbatively produced. The production happens at the expense of the kinetic energy of $\varphi$ and acts as a friction term for the motion of the inflaton.~\footnote{The tower of couplings in (\ref{trapped}) may be a consequence of an intrinsic periodicity of the inflaton motion during inflation, broken by monodromy effects \cite{Silverstein:2008sg,Kaloper:2008fb} that generate the potential $V$.}  Prior to~\cite{Green:2009ds}, the idea that a coupling of the inflaton to auxiliary fields would lead to dissipation of the inflaton kinetic energy was also considered in the context of warm inflation, see e.g.~\cite{Berera:1998px} and~\cite{Graham:2008vu}.

Immediately after their production, the quanta of each $\chi_i$ species become non-relativistic, and they are rapidly diluted by the inflationary expansion. Therefore, they do not provide a direct contribution to the curvature perturbations after inflation. However, before being diluted, they source inflaton and metric perturbations. The resulting scalar modes were studied in \cite{Green:2009ds}, which directly computed and solved the integro-differential equations of motion following from (\ref{trapped}) under some approximations.~\footnote{
Mechanisms of particle production identical or similar to (\ref{trapped}) were studied in Refs.  \cite{Chung:1999ve,Kofman:2004yc,Romano:2008rr,Barnaby:2009mc,Barnaby:2009dd,Langlois:2009jp,Battefeld:2010sw,Battefeld:2011yj,D'Amico:2012ji,D'Amico:2013iaa,Battefeld:2013bfl,Amin:2015ftc}  either as a way to trap moduli on enhanced symmetry points, or as a way to induce a signature on the scalar power spectrum  from a single event of particle production.  Dissipative effects have been studied in the formalism of the effective field theory of inflation in~\cite{Barnaby:2011pe,LopezNacir:2011kk}.} 

Successively,  refs.~\cite{Lee:2011fj,Bugaev:2015npw} formulated a Langevin functional approach for the perturbations and solved it in a regime of weak coupling between the inflaton and the $\chi_i$ fields.~\footnote{Contrary to what obtained in  \cite{Green:2009ds} and in this work, a blue scalar power spectrum is obtained in~\cite{Lee:2011fj,Bugaev:2015npw}, in disagreement with the expectation of approximate scale invariance.} The production of gravitational waves in trapped inflation was studied in~\cite{Senatore:2011sp,Cook:2011hg}, where it was found to be subdominant to the scalar production. As shown in \cite{Barnaby:2012xt}, this is due to the non-relativistic nature of the $\chi_i$ quanta, which suppresses their quadrupole moment (see also \cite{Carney:2012pk} for a subsequent reanalysis of the production of gravitational waves).  

In this paper, we compare the results obtained from the computational scheme developed in the original work~\cite{Green:2009ds} with those of the effective field theory (EFT) study of~\cite{LopezNacir:2011kk}. The results for the power spectrum and bispectrum of the inflaton perturbations given in these two works agree with each other. Specifically, ref.~\cite{LopezNacir:2011kk} quotes and employs the power spectrum obtained in~\cite{Green:2009ds}. It then estimates the amplitude of the bispectrum using a three point function operator for the inflaton perturbations that is not derived from an explicit computation, but  is motivated by covariance. This operator, which is not present in the computations of \cite{Green:2009ds}, leads nonetheless to a bispectrum that is parametrically in agreement with that obtained by~\cite{Green:2009ds}. The mechanism of trapped inflation is extremely complicated, and an exact analytic computation of the inflaton perturbations is not possible. All the computations provided in the literature have some degree of approximation. The agreement between the results of~\cite{Green:2009ds,LopezNacir:2011kk}, if confirmed,  would corroborate the different computational schemes and approximations employed in these two works. 
 
To understand why the two computations lead to the same result, we critically reviewed the analysis of ~\cite{Green:2009ds}, which is more detailed and can be more easily reproduced. We verified a number of subtle approximations and assumptions made in that work. We  however find that one approximation used in~\cite{Green:2009ds} to estimate the correlation between the produced quanta is not sufficiently accurate. We show that, once this approximation is replaced by an exact computation, the terms evaluated by~\cite{Green:2009ds} give a bispectrum that is parametrically suppressed with respect to that of~\cite{LopezNacir:2011kk}. We then systematically study other diagrams for the bispectrum that emerge from the formalism of~\cite{Green:2009ds} but that were disregarded in that work. We find that also these diagrams lead to a bispectrum that is in disagreement with that of ~\cite{LopezNacir:2011kk}. We actually find that the leading operator of \cite{LopezNacir:2011kk} does not emerge in the computational scheme of~\cite{Green:2009ds}.  In fact, it is unclear to us whether this operator should appear at all in the full equations for the perturbations of trapped infaltion, and whether arguments of covariance can be applied straightforwardly to this problem. Particle production is a result of quantization, that clearly singles out the time direction. While a non-adiabatic variation in time of the frequency of a mode leads to particle production, this is not the case for a spatial variation. Namely, assume a coupling $\varphi^2 \chi^2$ between two species. A fast variation of $\varphi \left( t \right)$
leads to production of quanta of $\chi$. Instead, no particle production takes place if $\varphi$ has a spatially dependent profile  $\varphi \left( {\bf x} \right)$ which is constant in time. 

Moreover, we find that the same inaccurate approximation also affects the power spectrum obtained in~\cite{Green:2009ds}, which was used by~\cite{LopezNacir:2011kk}. Correcting this approximation results in a  power spectrum that differs by more than four orders of magnitude from that of \cite{Green:2009ds,LopezNacir:2011kk}. This has a significant impact on the region of parameter space of the model associated with any given phenomenological result. 

Our system contains many degrees of freedom, so that the curvature perturbation $\zeta$ will generally get contributions both from the fluctuations in the inflaton and by fluctuations in the $\chi_i$ fields. As a consequence, one would study perturbations in this model using the language of coupled adiabatic and isocurvature perturbations, with the additional complication that the contributions from the $\chi_i$ fields to energy perturbations starts at the quadratic level. In this work however we use the same working assumptions of~\cite{Green:2009ds} and~\cite{LopezNacir:2011kk}, i.e., that the metric perturbations are dominated by those in the inflaton, which leads to the simple identification  $\zeta=-H\,\delta\varphi/\dot\varphi$.  This choice is motivated by the fact that the energy density in $\chi_i$ modes is subdominant with respect to that in the inflaton, as we imposed in our computations. We follow~\cite{Green:2009ds} in making this choice in the present work, so to be able to study the results of~\cite{Green:2009ds}, and to compare them with those of~\cite{LopezNacir:2011kk}. Nevertheless, it would be important to perform an analysis where the role of the $\chi_i$ fields and invariance under time reparametrization are accounted for in a consistent way, as it is done in the context of warm inflation, for instance in Section III of~\cite{BasteroGil:2011xd}. We plan to return to this question in a future work.

The structure of the paper is as follows: In Section \ref{sec:equations}, we review the computational scheme of  \cite{Green:2009ds},  introducing the system of equations which governs the background solution of the inflaton field and its first and second order perturbations. In Section \ref{sec:background} we present the  slow-roll background solution for the model, as well as a set of conditions which are necessary either for trapped inflation to occur or for the analysis to be valid.  In section \ref{sec:perturbations} we then compute the first and second order scalar perturbations, as well as the tensor power spectrum, which are obtained from this scheme, and we compare the bispectrum obtained in this way with that obtained in \cite{LopezNacir:2011kk} . In Section  \ref{sec:pheno} we study the phenomenological consequences of these results.  Section \ref{sec:conslusions} contains our concluding remarks.

%%%%%%%%%%%%%%%%%%%%%%%%%%%%%%%%%%%%%%%%%%%%%%%%%
\section{Effective system of equations}
\label{sec:equations}
%%%%%%%%%%%%%%%%%%%%%%%%%%%%%%%%%%%%%%%%%%%%%%%%%

In this section, we present the background equations  following from (\ref{trapped}), as well as the equations for the first-order and second-order inflaton fluctuations. Their derivation is shown in more detail in Appendix \ref{app:equations}. As mentioned in the Introduction, our goal is to critically review the computations of~\cite{Green:2009ds} to be able to compare them with those of~\cite{LopezNacir:2011kk}.  Therefore, our derivation closely follows that of~\cite{Green:2009ds}. The full equation for the second order perturbations that we write below was not derived in~\cite{Green:2009ds}, which only focused on a subset of terms in that equation. In deriving the full second order equation, we follow the same ``rules'' used by~\cite{Green:2009ds} to derive the first order equation. We will find that the subset of terms considered in~\cite{Green:2009ds}, once correctly evaluated, provides the same parametric dependence of the bispectrum as the full set of terms included here.  

First, we expand the inflaton field as
\begin{equation}
\varphi \left( t ,\, \vec{x} \right) = \varphi_0 \left( t  \right) + \delta \varphi_1 \left( t ,\, \vec{x} \right) +  \delta \varphi_2 \left( t ,\, \vec{x} \right) + \dots ,
\label{eq:phi_expansion}
\end{equation} 
where $\varphi_0$ is the background solution, while $\delta \varphi_1$ and $\delta \varphi_2$ are, respectively, the first and second-order fluctuations of the inflaton field.  We decompose these into momentum modes via
\begin{equation}
\delta \varphi_i \left( t ,\, \vec{x} \right) = \int \frac{d^3 k}{\left( 2 \pi \right)^{3/2}} \, {\rm e}^{i \vec{k} \cdot \vec{x}} \, \delta \varphi_i \left( t ,\, \vec{k} \right) ,
\label{eq:mode_expansion}
\end{equation} 
with an identical decomposition for the $\chi_i$ fields.  The evolution of the background field is governed by 
\begin{align}
0 &= \ddot{\varphi}_0 + 3 H \dot{\varphi}_0 + V' + \int^t g^{5 \slash 2} \dfrac{\dot{\varphi}_0\,|\dot{\varphi}_0|^{3 \slash 2}}{\Delta(2\pi)^3} \dfrac{a(\tilde{t})^3}{a(t)^3} d\tilde{t},
\label{eq:background_integral_eq_motion}
\end{align}
which includes the backreaction on the inflaton motion  due to the production of $\chi_i$ quanta. The (mass dimension one) quantity $\Delta$ expresses the difference between the values of the inflaton field between two successive episodes of particle production, $\Delta = \vert \varphi_{0,i+1} - \varphi_{0,i} \vert$.   For brevity, we defined $V' \equiv  \dfrac{\partial V}{\partial \varphi} \bigg|_{\varphi = \varphi_0} $ (and an analogous definition holds below for $V''$). 

 The two other relevant background equations are the $(0,0)$ and $(i,i)$ Einstein equations that read, respectively, 
\begin{eqnarray}
3 M_p^2 \, H^2 &=&  \frac{\dot{\varphi}_0^2}{2} + V \left( \varphi_0 \right) +   \rho_{\chi} \vert_{\rm background} \;, \nonumber\\ 
6 M_p^2 \, H \dot{H} &=& - 3 H \left( \dot{\varphi}_0^2 +   \rho_{\chi} \vert_{\rm background} \right) \;, 
\label{einstein-bck}
\end{eqnarray} 
where we have used the fact that the $\chi_i$ quanta are non-relativistic, and where the total background energy in the produced quanta 
$ \rho_{\chi} \vert_{\rm background}  $ is given in eq. (\ref{rho-chi}).

To extend the computation to first and second order perturbations, it is convenient to define
\begin{equation}
g_i \equiv \frac{\delta \varphi_i}{x} = \frac{\delta \varphi_i}{- k \tau},
\label{g-dphi}
\end{equation} 
where we have introduced the variable $x = - k \tau$. In what follows, we will use a prime to denote differentiation with respect to $x$.  We also introduce the dimensionless  parameters 
\begin{equation}
\mu^2 \equiv \frac{g^{5/2} \, \vert \dot{\varphi}_0 \vert^{3/2}}{\Delta \, H^2 \left( 2 \pi \right)^3} \;\;\;,\;\;\; 
{\tilde \mu}^2 \equiv \frac{7 \mu^2}{2} \;\;,
\label{mu-mut} 
\end{equation} 
and we define the sources
\begin{align} 
& s_1 \equiv   - \sum_i \frac{g^2 \,  \left( \varphi_0 - \varphi_{0i} \right) }{H^2 x^3} \left( \chi_i^2 - \left\langle \chi_i^2 \right\rangle \right)_1 \;\;,\;\; 
{\hat s}_1 \equiv   -  \sum_i \frac{g^2}{H^2 x^2}  \,  \left( \chi_i^2 - \left\langle \chi_i^2 \right\rangle \right)_1 \;,  \nonumber\\ 
& s_2 \equiv  - \sum_i \frac{ g^2  \left( \varphi_0 - \varphi_{0i} \right)  }{H^2 x^3 }   \left( \chi_i^2 - \left\langle \chi_i^2 \right\rangle \right)_2,
\label{eq:sources}
\end{align} 
where the subscripts $1,2$ on $\left( \chi_i^2 - \left\langle \chi_i^2 \right\rangle \right)_i$ indicate the first and second order contributions to $\left( \chi_i^2 - \left\langle \chi_i^2 \right\rangle \right)$. As we review in Appendix  \ref{app:equations}, the computational scheme developed in~\cite{Green:2009ds} results in the following  master equations for the first  and second order inflaton perturbations 
\begin{align} 
\begin{cases}
&g_1'' + \left[ 1 + \frac{ {\tilde  \mu}^2 - 2 }{ x^2} \right] g_1  - 3 {\tilde \mu}^2 \,  \int_x  \frac{ d x' }{ x^{' 3}}  \, g_1 \left( x' \right)  =  s_1    \;, \\ 
&g_2'' + \left[ 1 + \frac{ {\tilde  \mu}^2 - 2 }{ x^2} \right] g_2  - 3 {\tilde \mu}^2 \,  \int_x  \frac{ d x' }{ x^{' 3}}  \, g_2 \left( x' \right)  =  s_2 + \frac{p}{2k} \, g_1 \star {\hat s}_1   +  {\hat s}_1  \star \frac{\vert \vec{k} - \vec{p} \vert}{2k} \, g_1 \\ 
& \qquad  -  \frac{ 2 H {\tilde \mu}^2}{7 \, \dot{\varphi}_0} \frac{p}{k} \frac{\vert \vec{k}-\vec{p}\vert}{k} \, 
\left( \frac{g_1 \star g_1'+g_1' \star g_1}{2} - \frac{27}{8} \frac{g_1 \star g_1}{x} \right) 
  - \frac{5 \, H \, {\tilde \mu}^2  }{4  \, \dot{\varphi}_0}   \frac{p}{k} \frac{\vert \vec{k}-\vec{p}\vert}{k} \,  \int_x 
 d x' \, \left( g_1' \star g_1'  + \frac{2}{x^{' 2}} \, g_1 \star g_1  \right)  \;, 
\end{cases} 
\label{eq:system-2nd}
\end{align} 
where in the source terms on the right hand side of the second equation we have used the convolution
\begin{equation}
\left( f \star g \right) \left( \vec{k} \right) \equiv \int \frac{d^3 p}{\left( 2 \pi \right)^{3/2}} \, f \left( \vec{p} \right) \, g \left( \vec{k} - \vec{p} \right) .
\label{eq:convolution}
\end{equation} 

The time variable ${\tilde t}$ in the integral of eq.~(\ref{eq:background_integral_eq_motion}) and the rescaled time variable $x'$ in the integrals of eqs.~(\ref{eq:system-2nd}) run over the production time of the different $\chi_i$ species. These continuous integrals replace  discrete sums (over the various species) as outlined in eq. (\ref{Delta}). This can be done if the integrand varies sufficiently slowly, namely if the conditions (\ref{sumtoint}) are met. We discuss these conditions, together with other conditions that must be satisfied for our results to be valid, in subsection 
\ref{sec:conditions}. 

Each master equation in \eqref{eq:system-2nd} is an integro-differential equation. We solve them by differentiating them, resulting in 
\begin{align} 
\begin{cases}
{\hat O} \, g_1 \left( x \right) = s_1'  \left( x \right) \;, \\ \\ 
{\hat O} \, g_2 \left( x \right)   =    \frac{  H {\tilde \mu}^2}{28 \,\dot{\varphi}_0}  \frac{p}{k} \,  \frac{\vert \vec{k} - \vec{p} \vert}{k}  \left( - 8 g_1 \star g_1'' + 27 \, g_1' \star g_1' + \frac{54 \, g_1 \star g_1'}{x} + \frac{43 \, g_1 \star g_1}{x^2} \right) \\ 
\quad\quad\quad\quad  \quad 
+ s_2' +  \frac{1}{2} \left( \frac{p}{k} g_1' \star {\hat s}_1 +    {\hat s}_1 \star  \frac{\vert\vec{k} - \vec{p} \vert}{k} g_1' 
+  \frac{p}{k}  g_1 \star {\hat s}_1 '  +   {\hat s}_1 '   \star   \frac{\vert\vec{k} - \vec{p} \vert}{k}  \, g_1 \right) \; ,
\end{cases}  
\label{eq:system-3rd}
\end{align} 
where the third-order operator is  
\begin{equation} 
{\hat O} \equiv \partial_x^3 + \left( 1 +  \frac{ {\tilde  \mu}^2 - 2 }{ x^2} \right)  \partial_x + \frac{ {\tilde \mu}^2 + 4 }{x^3} .
\label{eq:operator}
\end{equation} 
These differential equations can be solved using Green's functions; we then impose that the solutions to \eqref{eq:system-3rd} also satisfy \eqref{eq:system-2nd}.  This is done in section \ref{sec:green} below.

%%%%%%%%%%%%%%%%%%%%%%%%%%%%%%%%%%%%%%%%%%%%%%%%%
\section{Slow roll background solution and conditions}
\label{sec:background}
%%%%%%%%%%%%%%%%%%%%%%%%%%%%%%%%%%%%%%%%%%%%%%%%%

This section is divided in two parts. In subsection \ref{sec:slow} we solve the background equations presented above using the slow roll approximation. In subsection  \ref{sec:conditions} we list the various conditions that must be met for our results to be valid, and we review their origin.

%%%
\subsection{Slow roll background solution} 
\label{sec:slow}
%%%

We solve the background equations  (\ref{eq:background_integral_eq_motion}) and (\ref{einstein-bck}) in slow roll approximation, 
\begin{equation}
\epsilon \equiv - \frac{\dot{H}}{H^2} \ll 1 \;\;\;,\;\;\; \vert \ddot{\varphi}_0 \vert \ll 3 \, H \vert \dot{\varphi}_0 \vert \;. 
\label{slowroll-conds}
\end{equation} 
We then impose that the Hubble friction in the motion of the inflaton is subdominant to the friction from particle production (which is the core assumption of the mechanism of trapped inflation) 
\begin{equation} 
 3 H \vert \dot{\varphi}_0 \vert \ll \vert V' \vert \;. 
 \end{equation} 

In the slow roll approximation, the time dependence of the integrand of  (\ref{eq:background_integral_eq_motion}) is domainated by that of the scale factor, and we can therefore write 
\begin{equation}\label{eq:zeromode-local}
\ddot{\varphi}_0 + 3 H \dot{\varphi}_0 + V' + 
\frac{g^{5/2}\,\dot{\varphi}_0 \, \vert \dot{\varphi}_0 \vert^{3/2}}{3 \, H \, \Delta \, \left( 2 \pi \right)^3} \simeq 0 \;\;, 
\end{equation}
with the first two terms negligible, so that this equation is solved by 
\begin{equation}
\dot{\varphi}_0 \simeq - \frac{1}{g} \left( 24 \pi^3 \, H \, \Delta \, V' \right)^{2/5} \;\;, 
\label{phidot}
\end{equation}
where, without loss of generality, we have assumed that $V'  > 0$, and so $\dot{\varphi}_0 < 0$.

%%%
\subsection{Conditions} 
\label{sec:conditions}
%%%

Our results assume several conditions which we discuss here. The conditions are of various natures: some of them have a physical origin, on which the trapped inflation mechanism relies, while others are technical conditions that we impose for our approximations to be valid. This second class  of conditions could be evaded, and trapped inflation would still be in effect, but our phenomenological results cannot be trusted if they are violated. We label and list these conditions in Table \ref{tab:conditions} to be able to immediately refer to them in the following sections. 

We start with the three  physical conditions  written in the previous subsection: 
\begin{equation}
\epsilon \equiv - \frac{\dot{H}}{H^2} \ll 1 \;\;\;,\;\;\; \vert \ddot{\varphi}_0 \vert \ll 3 \, H \vert \dot{\varphi}_0 \vert \;\;\;,\;\;\;  3 H \vert \dot{\varphi}_0 \vert \ll \vert V' \vert \;. 
\label{C-physical}
\end{equation} 
These conditions are denoted, respectively, as  $C_1 ,\, C_2 ,\,$ and $C_3$ in Table  \ref{tab:conditions}. 

To evaluate the first slow roll condition we make use of the background equations (\ref{einstein-bck}), with $V$ dominating the right hand side (r.h.s.) of the first equation, and $ \rho_{\chi} \vert_{\rm background}$ dominating the r.h.s.\ of the second equation (one can verify that this last assumption holds once the third condition in (\ref{C-physical}) does).~\footnote{As $ \rho_{\chi} \vert_{\rm background} $ dominates $\dot{H}$, the condition $\epsilon \ll 1$ also ensures that 
$ \rho_{\chi} \vert_{\rm background} \ll V \left( \varphi \right)$.}  

To evaluate the second slow roll condition, we differentiate the result (\ref{phidot}), from which we obtain 
\begin{equation}
\ddot{\varphi}_0 = \frac{2}{5} \left( - \epsilon H \dot{\varphi}_0 + \dot{\varphi}_0^2 \frac{V''}{V'} \right) \;\;. 
\label{eq:ddotphi0}
\end{equation} 
When inserted into the left-hand-side of this condition, the first term in the parenthesis is clearly negligible.  Dropping the first term while keeping the second one gives the expression written in the third column of the table. 

The third condition in (\ref{C-physical}) is the defining condition for the trapped mechanism: namely the requirement that the Hubble friction term in the equation of motion for the background inflaton field (eq.~(\ref{eq:background_integral_eq_motion})) is subdominant to the friction from particle production.  If this is the case, the last two terms in (\ref{eq:background_integral_eq_motion}) approximately cancel each other, and the Hubble friction term is subdominant to either of them. 

The fourth condition $C_4$ in Table  \ref{tab:conditions} is obtained from the requirements that the mass of $\chi_i$ varies rapidly at the moment the quanta of $\chi_i$ are produced and that the expansion of the universe can be disregarded during the production.  This gives $H \delta t \sim \frac{H}{\sqrt{g \vert \dot{\varphi}_0}} \ll1$.

All the following conditions in  Table  \ref{tab:conditions} are imposed to ensure the validity of our results. The trapped inflationary mechanism may be in effect also when these conditions are violated, but our phenomenological results would be invalid. 

The fifth condition $C_5$ in Table  \ref{tab:conditions} is obtained from the requirement that the quanta $\chi_i$ have an inefficient annihilation into inflaton quanta via the $g^2 \varphi^2 \chi_i^2$ interaction in the Lagrangian.  If this annihilation was efficient, we would still expect the trapped mechanism to work (since energy is still extracted from the inflaton motion), but we would obtain different results for both the inflaton background evolution and its perturbations.  As can be seen from Appendix \ref{app:chi}, quanta of $\chi_i$ are produced with typical physical momentum $p \sim \sqrt{g \, \vert \dot{\varphi}_0 \vert}$ and with a number density $n \sim \left( g \, \vert \dot{\varphi}_0 \vert \right)^{3/2} $. A time $\delta t$ after they are produced, their mass is approximately $m_\chi \sim g \, \vert \dot{\varphi}_0 \vert \, \delta t$. The annihilation cross-section is therefore $\sigma \sim \frac{g^4}{m_\chi^2} \sim \frac{g^4}{g^2 \,  \dot{\varphi}_0^2 \, \delta t^2}$, corresponding to a rate 
\begin{equation}
\Gamma = \langle n \, \sigma \, v \rangle \sim \frac{g^4}{g \, \vert \dot{\varphi}_0 \vert \, \delta t^3} \sim g^4 \, \sqrt{g \vert \dot{\varphi}_0 \vert} \,, 
\label{gamma-annihilation}
\end{equation} 
where the final estimate is obtained by evaluating the rate immediately after the quanta are produced. As discussed in Appendix \ref{app:chi}, the production takes place in the time $\delta t \sim \frac{1}{\sqrt{ g \, \vert \dot{\varphi}_0 \vert}}$. Condition  $C_5$ is obtained by imposing that this rate is much smaller than $H$.  We note that if this is the case at the time  $\delta t \sim \frac{1}{\sqrt{ g \, \vert \dot{\varphi}_0 \vert}}$, then it will also be the case at later times, due both to the decrease of $\Gamma$ explicitly written in eq. (\ref{gamma-annihilation}) and to the dilution of the $\chi_i$ quanta due to the expansion of the universe, which we have not included in  (\ref{gamma-annihilation}).

The next two conditions in the Table, namely $C_6$ and $C_7$, are obtained by requiring that the sum over the $\chi_i$ species can be transformed into the integral in eqs. (\ref{eq:background_integral_eq_motion}) and (\ref{eq:system-2nd}). The replacement is discussed in eq.~(\ref{Delta}), resulting in the two conditions 
\begin{equation}\label{con:sum-to-integral-exact}
\left\vert \frac{\ddot{\varphi} \, \Delta}{\dot{\varphi}^2} \right\vert \ll 1 \;\;\;,\;\;\; \left\vert \frac{ H \, \Delta}{\dot{\varphi}} \right\vert \ll 1 \;\;. 
\end{equation}
Expanding this at zeroth and first order in the perturbations results in
\begin{align}
\begin{cases}
|\ddot{\varphi}_0| \Delta  \ll  \dot{\varphi}_0^2, \qquad H\, \Delta  \ll |\dot{\varphi}_0| & \qquad \mathrm{0th} \; \mathrm{order}\,, \\
\Delta \, \delta \ddot{\varphi_1} \ll \dot{\varphi}_0 \; \delta \dot{\varphi}_1  & \qquad \mathrm{1st} \; \mathrm{order}\,.
\end{cases}
\label{con:sum-to-integral}
\end{align}
Condition $C_6$ is given by the second of the 0th order equations in eqs.~\eqref{con:sum-to-integral} above. The first of those conditions is identically satisfied since, by using \eqref{eq:ddotphi0}, it turns out to be equivalent to $\Delta\ll V'/V''\simeq \varphi_0$ for monomial potentials like the ones we will consider. 

Condition $C_7$ emerges from the 1st order equation in eqs.~\eqref{con:sum-to-integral}, once we note that the largest first order perturbations are generated when the physical momentum of the modes equals $\mu\,H$, so that  $|\delta\ddot{\varphi}_1|\sim \mu^2\,H^2\,|\delta\varphi_1|$ and $|\delta\dot{\varphi}_1|\sim \mu\,H\,|\delta\varphi_1|$. 

The final condition in  Table  \ref{tab:conditions}, $C_8$,  results from an approximation made when calculating the sourced inflaton power spectrum and bispectrum below; specifically, we impose that the internal loop momentum is much greater than that of the perturbations in the correlators.

\begin{table}[t]
\begin{center}
\begin{tabular}{|c||c|c|} \hline 
 Label & Physical condition & Rewrites as    \\ \hline
 ${\rm C}_1$ & 
$\epsilon \ll 1 $ & $ \frac{2^{1/5} \, \pi^{6/5} \vert V' \vert^{7/5} \Delta^{2/5}}{3^{3/5} \, g \, H^{13/5} \, M_p^2} \ll 1 $ \\ \hline 
  ${\rm C}_2$ & 
$ \vert \ddot{\varphi}_0 \vert \ll 3 \, H \, \vert \dot{\varphi}_0 \vert   $ & 
$\frac{2^{6/5} \pi^{6/5} \vert V'' \vert \, \Delta^{2/5}}{3^{3/5} \, g \, H^{3/5} \, \vert V' \vert^{3/5}} \ll1 $   \\ \hline 
 ${\rm C}_3$ & 
$ 3 \, H \, \vert \dot{\varphi}_0 \vert \ll \vert V' \vert $ & $\frac{2^{6/5} \, 3^{7/5} \pi^{6/5} H^{7/5} \Delta^{2/5}}{g \,  \vert V' \vert^{3/5} } \ll 1 $   \\ \hline 
 ${\rm C}_4$ & 
$H^2 \ll g \, \vert \dot{\varphi}_0 \vert $ & $ \frac{H^{8/5}}{6^{6/5} \, 3^{2/5} \, \pi^{6/5} \, \vert  V' \vert^{2/5} \, \Delta^{2/5}} \ll 1 $ \\ \hline 
 ${\rm C}_5$ &
$g^4 \, \sqrt{ g \, \vert \dot{\varphi}_0 \vert } \ll H $ & $\frac{2^{3/5} 3^{1/5} \pi^{3/5} \, g^4 \, \vert V' \vert^{1/5} \, \Delta^{1/5}}{H^{4/5}} \ll 1 $  \\ \hline 
 ${\rm C}_6$ & 
$H \, \Delta \ll  \vert \dot{\varphi}_0 \vert $ & $ \frac{g \, H^{3/5} \, \Delta^{3/5}}{2^{6/5} 3^{2/5} \pi^{6/5}  \vert  V' \vert^{2/5} } \ll 1$ \\ \hline 
 ${\rm C}_7$ & 
$\Delta \vert \ddot{\varphi} \vert \ll \vert \dot{\varphi}^2 \vert \; \Rightarrow \mu \, H \, \Delta \ll \vert \dot{\varphi}_0 \vert $  & 
$\frac{g^{3/2} \, \Delta^{2/5}}{2^{9/5} \, 3^{1/10} \, \pi^{9/5} \, H^{1/10}   \vert  V' \vert^{1/10} } \ll 1$ \\ \hline 
 ${\rm C}_8$ & 
$ 21 \, \pi \, \mu^2 \, H^2 \ll 2 \, g \, \vert \dot{\varphi}_0 \vert$ &   $\frac{3^{1/5} \, 21 \, g \, H^{1/5} \, \vert V' \vert^{1/5}}{2^{2/5} 8 \pi^{7/5} \, \Delta^{4/5}} \ll 1 $ \\ \hline 
\end{tabular}
\end{center}
\caption{List of conditions for the validity of the trapped mechanisms ($C_1-C_4$) and of our computational scheme ($C_5-C_8$). The conditions are discussed in the text.  
}\label{tab:conditions}
\end{table}

%%%%%%%%%%%%%%%%%%%%%%%%%%%%%%%%%%%%%%%%%%%%%%%%%
\section{The perturbations}
\label{sec:perturbations}
%%%%%%%%%%%%%%%%%%%%%%%%%%%%%%%%%%%%%%%%%%%%%%%%%

In this section we solve the differential equations \eqref{eq:system-3rd}, using the background solution \eqref{phidot}.  In the first subsection below, we discuss the Green's function for the operator appearing in \eqref{eq:operator} and we provide the formal solution for the first and second order inflaton perturbations.  We then show that these solutions also automatically satisfy the starting second-order integro-differential equations. In subsection  \ref{sec:correlators} we provide formal expressions for the correlators of the inflaton perturbations.  Starting from these formal solutions, in subsection \ref{sec:twopoint} we provide the explicit solution for the first order perturbations, and we compute the inflaton power spectrum. In subsection \ref{sec:threepoint} we then obtain an explicit approximate solution for the second order perturbations, and evaluate the equilateral bispectrum of the inflaton perturbations. In subsection  \ref{sec:gradient} we show that, contrary to the solution of the same equations obtained in ref.~\cite{Green:2009ds}, the bispectrum obtained from our solution of those equations in in parametrical disagreement with that of  \cite{LopezNacir:2011kk}. We further show that the operator that governs the bispectrum of \cite{LopezNacir:2011kk} does not emerge in the computational scheme of~\cite{Green:2009ds}.  In subsection \ref{sec:GW} we review the computation~\cite{Cook:2011hg} of the amplitude of the tensor modes in this scenario. 

%%%
\subsection{Green's function, and formal solutions}
\label{sec:green}
%%%

The Green's function for the operator \eqref{eq:operator} appearing in the differential equations reads
\begin{align}
G(x,x^\prime) = \Theta(x^\prime - x) \;  \left[ c_1(x^\prime) \, f_1(x) + c_2(x^\prime)  \, f_2(x)  + c_3(x^\prime)  \, f_3(x) \right] \,, 
\label{eq:green}
\end{align}
where
\begin{align} 
f_1 \left( x \right) &=  \frac{1}{x} \; _1F_2 \left[ \left\{ -\frac{1}{2}  \right\} , \left\{  - \frac{1}{2} - i \frac{\tilde \mu}{2} ,\, 
 - \frac{1}{2} + i \frac{\tilde \mu}{2} \right\} ,\,  - \frac{x^2}{4} \right] \,, \nonumber\\ 
f_2 \left( x \right)  &= x^{2+i {\tilde \mu}}  \; _1F_2 \left[ \left\{ 1 + \frac{i \, {\tilde \mu}}{2}  \right\} , \left\{ 
\frac{5 + i {\tilde \mu}}{2} ,\, 1 + i \, {\tilde \mu}  \right\} , - \frac{x^2}{4} \right] \;\;,\;\; f_3 \left( x \right)= f_2^* \left( x \right) \;\;, 
\label{eq:homogeneous_solns}
\end{align} 
and
\begin{align} 
c_1 \left( x' \right) &=  \frac{ f_2 \left( x' \right) \, f_3' \left( x' \right) -  f_2' \left( x' \right) \, f_3 \left( x' \right) }{  2 i {\tilde \mu} \left( 9 + {\tilde \mu}^2 \right)} \,, \nonumber\\ 
c_2 \left( x' \right) &=  \frac{ f_3 \left( x' \right) \, f_1' \left( x' \right) -  f_3' \left( x' \right) \, f_1 \left( x' \right) }{  2 i {\tilde \mu} \left( 9 + {\tilde \mu}^2 \right)} \;\;,\;\; 
c_3 \left( x' \right) = c_2^* \left( x' \right) \;, 
\label{ci-sol}
\end{align} 
where $\phantom{}_1F_2$ is a generalized hypergeometric function.  We introduce the notation $G(x,x^\prime) = \Theta(x^\prime-x)\, \tilde{G}(x,x^\prime)$ to denote the Green's function without the step function.  The functions $f_i(x)$ have the following small-$x$ behavior 
\begin{align}
f_1 \left( x \right) &= \frac{1}{x} \left[ 1 + \frac{x^2}{2 \left( 1 + {\tilde \mu}^2 \right)} +  {\rm O } \left( x^4 \right) \right] \,, \nonumber\\ 
f_2 \left( x \right) &= x^2 \, x^{i \, {\tilde \mu}}  \left[ 1 + {\rm O } \left( x^2 \right) \right] \,,  
\end{align}  
which can be used when $x \ll x' ,\, 1$.  Consequently, in this regime, 
\begin{equation}
x \, G \left( x ,\, x' \right) = \left( 1 + \frac{x^2}{2 \left( 1 + {\tilde \mu}^2 \right)} \right)  c_1 \left( x' \right) + {\rm O } \left( x^3 \right) \, .
\label{eq:xG}
\end{equation}

\paragraph{}The functions $c_i(x^\prime)$  can be expressed exactly in terms of Bessel functions,
\begin{align}
c_1 \left( x' \right) &= - i \frac{\pi}{4} \, \frac{1+{\tilde \mu}^2}{{\tilde \mu} \, \cosh \left( \frac{\pi \, {\tilde \mu}}{2} \right)} \, x' \left[ \left(1 + i \, {\tilde \mu} \right) J_{\frac{1+i \, {\tilde \mu}}{2}} \left( \frac{x'}{2} \right)  J_{-\frac{1+i \, {\tilde \mu}}{2}} \left( \frac{x'}{2} \right) 
- \left(1 - i \, {\tilde \mu} \right) J_{\frac{1-i \, {\tilde \mu}}{2}} \left( \frac{x'}{2} \right)  J_{-\frac{1-i \, {\tilde \mu}}{2}} \left( \frac{x'}{2} \right) \right]  \,, \nonumber\\ 
c_2 \left( x' \right)  &=   \frac{i \, \left( 1 + i \, {\tilde \mu} \right) \left( 3 + i \, {\tilde \mu} \right)}{2^{7+2 i {\tilde \mu}} \, {\tilde \mu}} \, 
\Gamma^2 \left( - \frac{3+ i {\tilde \mu}}{2} \right) \, x' \left[ \left( 1 + i \, {\tilde \mu} \right) J^2_{-\frac{1+i \, {\tilde \mu}}{2}} \left( \frac{x'}{2} \right) +  \left( 1 - i \, {\tilde \mu} \right) J^2_{\frac{1-i \, {\tilde \mu}}{2}} \left( \frac{x'}{2} \right) \right]  \,. 
\label{eq:cs}
\end{align}

In terms of this Green's function, the solutions to eqs.~\eqref{eq:system-3rd} can be expressed as
\begin{align}
g_i(x) &= [g_{i,h}(x) + h.c.] + \int_x^\infty dx^\prime \,\tilde{G}(x,x^\prime)\, S_i' (x^\prime) \,, 
\label{eq:soln_form}
\end{align}
where the index $i = 1,2$ labels the first order and second order inflaton perturbations, $S_i'$ is the expression on the r.h.s. of  \eqref{eq:system-3rd}, and $g_{i,h}$ denotes the homogeneous vacuum solution.  We recall that eqs.~\eqref{eq:system-3rd} are the derivatives of the actual master equations  \eqref{eq:system-2nd} that must be satisfied by the perturbations.  These equations have the form
\begin{equation}
g_i'' + \left[ 1 + \frac{ {\tilde  \mu}^2 - 2 }{ x^2} \right] g_i  - 3\, {\tilde \mu}^2 \,  \int_x^M  \frac{ d x' }{ x^{' 3}}  \, g_i \left( x' \right)  =  S_i \left( x \right) \,, 
\label{formal-mastereq}
\end{equation}
where $S_i$ ($i=1,2$) are the sources on the right hand side of equations \eqref{eq:system-2nd}. 

We are interested in the limit $M  = k \slash a_\mathrm{in} H\rightarrow \infty$ (where $a_\mathrm{in}$ is the scale factor at the beginning of inflation; we note that $M \gg 1$ means that the mode is initially deep inside the horizon). Since by construction eqs.~\eqref{eq:soln_form} solve the derivatives of \eqref{formal-mastereq} (that is, equations \eqref{eq:system-3rd}), we can be sure that they solve \eqref{formal-mastereq} at all values of $x$ if they solve them at one given value of $x$.  We verify that they solve them at $x=M$; namely we ensure that 
\begin{equation}
\lim_{M \rightarrow \infty} \; g_{i,h}'' \left( M \right) + \left[ 1 + \frac{ {\tilde  \mu}^2 - 2 }{ M^2} \right] g_{i,h}  \left( M \right) 
= \lim_{M \rightarrow \infty} \; g_{i,h}'' \left( M \right) +  g_{i,h}  \left( M \right)  = 0 \,, 
\label{eq:satisfy_integroDE}
\end{equation}
where we have used the fact that the sourced solution vanishes in the UV (assuming that the source itself vanishes in the UV limit).  In fact, since the source must vanish in the UV limit, we are guaranteed that the sourced part of (\ref{eq:soln_form})  satisfies the master equation. We disregard the homogeneous second order solution $g_{2,h}$, since we are only interested in the second order solutions for their contribution to the bispectrum and the vacuum solution provides a negligible departure from Gaussianity. We also impose that the homogeneous solution approaches the  adiabatic vacuum solution at early times; namely that 
\begin{equation}
\lim_{x \rightarrow \infty} g_{1,h} \left( x \right) = \frac{H}{\sqrt{2}\, k^{3/2}} \, {\rm e}^{i x} \,  \hat{a}_{\vec{k}} \,, 
\label{adiabatic-in}
\end{equation} 
where $\hat{a}_{\vec{k}}$ is the free inflaton annihilation operator. We see that, once (\ref{adiabatic-in}) is enforced, the condition (\ref{eq:satisfy_integroDE}) is automatically satisfied.  This guarantees also that the master equation \eqref{eq:system-2nd} is satisfied for both the sourced and homogeneous contributions. 

%%%
\subsection{Correlators}
\label{sec:correlators} 
%%%

We start from the relation
\begin{equation}
\zeta_i \left( \tau ,\, \vec{k} \right) = - \frac{H}{\dot{\varphi}_0} \, \delta \varphi_i \left( \tau ,\, \vec{k} \right)  \,,  
\label{zeta}
\end{equation} 
between the first ($i=1$) and second ($i=2$) order inflaton perturbations and the curvature perturbation on uniform density hypersurfaces.  We are interested in the two- and three-point correlation functions
\begin{eqnarray} 
\left\langle \zeta \left( \tau ,\, \vec{k} \right)  \zeta \left( \tau ,\, \vec{k}' \right)  \right\rangle &\equiv& \frac{2 \pi^2}{k^3} \, \delta^{(3)} \left( \vec{k} + \vec{k}' \right) \, P_{\zeta} \left( \tau ,\, k \right) \,, \nonumber\\ 
\left\langle \zeta \left( \tau ,\, \vec{k}_1 \right)  \zeta \left( \tau ,\, \vec{k}_2 \right)   \zeta \left( \tau ,\, \vec{k}_3 \right)  \right\rangle &\equiv&  \delta^{(3)} \left( \vec{k_1} + \vec{k}_2 + \vec{k}_3 \right) \, B_{\zeta} \left( \tau ,\, k_i \right)  \,. 
\label{def-PB}
\end{eqnarray} 

The homogeneous and sourced components of the perturbations are uncorrelated,  and therefore the two- and higher-order correlators can be written as the sum of the vacuum (homogeneous) and the sourced correlators. 

Disregarding higher order corrections, we write the power spectrum as the sum of the vacuum  plus the sourced power spectrum 
\begin{equation}
P_\zeta = \frac{k^3}{2 \pi^2} \frac{H^2}{\dot{\varphi}_0^2}  \left[ \left\langle \delta \varphi_{1,h} \left( \tau ,\, \vec{k} \right)  \delta \varphi_{1,h} \left( \tau ,\, - \vec{k} \right)  \right\rangle' +   \left\langle \delta \varphi_{1,s} \left( \tau ,\, \vec{k} \right)  \delta \varphi_{1,s} \left( \tau ,\, - \vec{k} \right)  \right\rangle' \right] \equiv P_{\zeta,h} + P_{\zeta,s} \,, 
\label{PS} 
\end{equation} 
where the prime denotes the correlator without the delta function, while the suffix ``{\em s}'' denotes the sourced component in  (\ref{eq:soln_form}). Disregarding the vacuum contribution and higher order corrections, we write the bispectrum as 
\begin{eqnarray} 
&& 
%\!\!\!\!\!\!\!\!  \!\!\!\!\!\!\!\! 
B_\zeta = - \frac{H^3 }{\dot{\varphi}^3} \Bigg[ \left\langle \delta \varphi_{1,s} \left( \tau ,\, \vec{k}_1 \right)  \delta \varphi_{1,s} \left( \tau ,\, \vec{k}_2 \right)  \delta \varphi_{1,s} \left( \tau ,\, \vec{k}_3 \right) \right\rangle'  \nonumber\\ 
&& + \left(  \left\langle \delta \varphi_{1,s} \left( \tau ,\, \vec{k}_1 \right)  \delta \varphi_{1,s} \left( \tau ,\, \vec{k}_2 \right)  \delta \varphi_{2,s} \left( \tau ,\, \vec{k}_3 \right) \right\rangle' 
+ 2 \; {\rm permutations} \right) \Bigg] \equiv B_\zeta^{(1,1,1)} +  B_\zeta^{(1,1,2)} \,.  \nonumber\\ 
\label{BS} 
\end{eqnarray} 
%

%%%
\subsection{First order perturbations, and their $2-$ and $3-$point correlation functions}
\label{sec:twopoint}
%%%

We now consider $g_1(x)$, which describes the first order perturbation in the inflaton field (we recall that 
$\delta \varphi_i = x \, g_i$, where $x = - k \, \tau$). First, we discuss the vacuum (homogeneous) solution.  Using the Green's function found above, the most general homogeneous solution of eq.~\eqref{eq:system-2nd} is 
\begin{equation}
g_{1,h} (x) = c_{h1} \, f_1(x) + c_{h2} \, f_2(x) + c_{h3} \, f_3(x) \,, 
\end{equation}
where the function $f_i$ are given in (\ref{eq:homogeneous_solns}). The requirement (\ref{adiabatic-in}) fixes the integration constants to 
\begin{align}
c_{h1} &= \dfrac{i(1+ \tilde{\mu}^2)}{\cosh \left( \frac{\pi \tilde{\mu}}{2} \right)} \dfrac{H \hat{a}_{\vec{k}} }{\sqrt{2} k^{3 \slash 2}} \,, 
\nonumber \\
c_{h2} &= - \dfrac{e^{\pi \tilde{\mu} \slash 2} \pi}{2^{3 + 2 i \tilde{\mu}} \cosh^2 \left( \frac{\pi \tilde{\mu}}{2} \right) \Gamma \left( \frac{1}{2} + \frac{i \tilde{\mu}}{2} \right) \Gamma \left( \frac{5}{2} + \frac{i \tilde{\mu}}{2} \right) } \dfrac{H \hat{a}_{\vec{k}} }{\sqrt{2} k^{3 \slash 2}} \,, 
\nonumber \\
c_{h3} &= - \dfrac{e^{-\pi \tilde{\mu} \slash 2} \pi}{2^{3 - 2 i \tilde{\mu}} \cosh^2 \left( \frac{\pi \tilde{\mu}}{2} \right) \Gamma \left( \frac{1}{2} - \frac{i \tilde{\mu}}{2} \right) \Gamma \left( \frac{5}{2} - \frac{i \tilde{\mu}}{2} \right) } \dfrac{H \hat{a}_{\vec{k}} }{\sqrt{2} k^{3 \slash 2}} \,. 
\end{align}
We are interested in the power spectrum at asymptotically large scales ($x \rightarrow 0$). In this regime, $\vert   f_{2,3 }\left( x \right) / f_1 \left( x \right) \vert = {\rm O } \left( x^3 \right)$, and we just write 
\begin{equation}
\lim_{x\rightarrow 0} g_{1,h} (x) =  \dfrac{1}{x} \dfrac{i (1 + \tilde{\mu}^2)}{\cosh \left( \frac{\pi \tilde{\mu}}{2} \right) } \dfrac{H \hat{a}_{\vec{k}} }{\sqrt{2} k^{3 \slash 2}} \,.
\end{equation}
(For $x = e^{-60}$, the $c_{h2} \, f_2(x)  $ contribution would become relevant only for ${\tilde \mu} \ga 127$.) We therefore obtain the two point correlation function for the vacuum modes 
\begin{equation}
\left\langle \delta \varphi_{1,h} \left( \vec{k} ,\, \tau \right) \delta \varphi_{1,h} \left( \vec{k}' ,\, \tau \right) \right\rangle = 
\frac{\delta^{(3)} \left( \vec{k} + \vec{k}' \right)}{k^3} \,  \frac{H^2 \, \left( 1 + \tilde{\mu}^2 \right)^2}{2 \, \cosh^2 \left( \frac{\pi \tilde{\mu}}{2} \right)} \;. 
\label{dp1hdp1h}
\end{equation}

Let us now turn our attention to  the sourced modes. In Appendix \ref{app:dphi1n}, we calculate the correlation function  $\left< \delta \varphi_{1,s}^n \right>$ between an arbitrary number of the sourced first-order inflaton perturbations. Eq. (\ref{g1n-general-time}) gives 
\begin{eqnarray} 
\left\langle \delta \varphi_{1,s} \left( \vec{k} ,\, \tau \right) \delta \varphi_{1,s} \left( \vec{k}' ,\, \tau \right) \right\rangle &=& 0.21 \, \frac{\delta^{(3)} \left( \vec{k} + \vec{k}' \right)}{k_1^3} \, \frac{g^{7/2} \, { |\dot{\varphi}_0}|^{5/2}}{\pi^3 \, H^2 \, \Delta} \, 
\int_x^{\sqrt{\frac{g \, \vert \dot{\varphi}_0 \vert}{2 \pi }} \, \frac{1}{H}}  \frac{d x_0}{x_0^4} \, \left[ x \, G \left( x ,\, x_0 \right) \right]^2 \nonumber\\ 
& \simeq & 0.18 \, \mu   \, g \, |\dot{\varphi}_0| \,  \frac{\delta^{(3)} \left( \vec{k} + \vec{k}' \right)}{k^3} \,. 
\label{dpshdp1s}
\end{eqnarray} 

This expression is in parametric agreement with the power spectrum of~\cite{Green:2009ds,LopezNacir:2011kk}, but suppressed by a factor $\sim 15,000$ with respect to their result.  We stress that we obtained (\ref{dpshdp1s}) starting from the same  first order equation derived in~\cite{Green:2009ds}, so the difference arises from the way that this equation is solved. From a line-to-line comparison of our and their solutions, which we present in Appendix \ref{ap:comparison}, we  identified the precise reason of the discrepancy. It is the approximation (\ref{zz-us-converted}) done in~\cite{Green:2009ds} for the two point correlator of the produced quanta, which are assumed to be correlated only at equal times. Our computation does not rely on this assumption, but rather on the exact correlator (\ref{SS-us-converted}). The strong discrepancy between the two results indicates that the approximation  (\ref{zz-us-converted}) is not adequate. 

To obtain the final expression (\ref{dpshdp1s}), we started with eqs.~\eqref{eq:xG} and \eqref{eq:cs}, performed the integration in $x_0$ numerically at fixed values of $\mu$, and then fit the $\mu-$dependence numerically. We note that the integrand is peaked at $x_0 \approx \tilde{\mu}$. We verified that ${\tilde \mu} = {\rm O } \left( 10-1000 \right)$ in the region of parameter space that satisfies all the contraints (see Figure \ref{fig:bounds}). As we discuss in  Appendix \ref{app:dphi1n}, $x_0  \equiv  - k \, \tau_0$ where $\tau_0$ is the conformal time of particle production. Therefore, the sourced power of modes of wave number $k$ receives its greatest contribution from the events of particle production that occurred at the time $\tau_0 \approx - {\tilde \mu} / k$. 

We need to ensure that the peak of the integral is smaller than the upper extremum of integration in eq. (\ref{dpshdp1s}). This gives the condition 
\begin{equation}
g \vert \dot{\varphi}_0 \vert \gg 7 \, \pi \mu^2 \, H^2 \;. 
\label{C8-phiphi} 
\end{equation}

Let us now turn our attention to the three point correlator. We start by considering the first of the two terms defined in (\ref{BS}).  Again using the result (\ref{g1n-general-time}),  we obtain 
\begin{align} 
&\left\langle \delta \varphi_{1s} \left( \vec{k}_1 ,\, \tau \right) \delta \varphi_{1s} \left( \vec{k}_2 ,\, \tau \right) 
\delta  \varphi_{1s} \left( \vec{k}_3 ,\, \tau \right) \right\rangle 
 =  0.16 \,  {\rm sign } \left( \dot{\varphi}_0 \right) \,  \frac{\delta^{(3)} \left( \vec{k}_1 + \vec{k}_2 + \vec{k}_3 \right)}{k_1^3 \, k_2^3 \, k_3^3} \, \frac{g^{9/2} \, |\dot{\varphi}_0|^{5/2}}{\pi^{9/2} \, H \, \Delta} \, \nonumber\\ 
&
 \times \int_{\tau_{\rm min}}^\tau \frac{d \tau_0}{\left( - \tau_0 \right)^4} \, \left[ - k_1 \, \tau G \left( - k_1 \, \tau ,\, - k_1 \, \tau_0 \right) \right] \, 
 \left[ - k_2 \, \tau \, G \left( - k_2 \, \tau ,\, - k_2 \, \tau_0 \right) \right] \,  \left[ - k_3 \, \tau G \left( - k_3 \, \tau ,\, - k_3 \, \tau_0 \right) \right] \,, \nonumber\\ 
\end{align} 
with
\begin{equation}
\tau_{\rm min} \equiv - \frac{1}{H \, {\rm Max } \left( k_1 ,\, \dots ,\, k_n \right)  } \, \sqrt{\frac{g \, \vert \dot{\varphi}_0 \vert}{3 \pi}} \;. 
\end{equation} 

We again use equations \eqref{eq:xG} and \eqref{eq:cs} and perform the last integral numerically at fixed $\mu$.  Fitting the $\mu$-dependence numerically, we find 
\begin{align}
&\left\langle : 
\delta \varphi_{1s} \left( \tau ,\, \vec{k}_1 \right) 
 \delta \varphi_{1s} \left( \tau ,\, \vec{k}_2 \right) 
  \delta \varphi_{1s} \left( \tau ,\, \vec{k}_3 \right) 
: \right\rangle \bigg|_{k_1 = k_2 = k_3 = k} = 0.017 \, \mu^2    \, g^2 \, H \, \dot{\varphi}_0 \, \frac{ \delta^{(3)} \left( \vec{k}_1 + \vec{k}_2 + \vec{k}_3 \right)  }{ k^6 }     \,.  
\label{dph1dph1dph1}
\end{align} 
In this case the integrand is also peaked at $\tau_0 \simeq {\tilde \mu}$, giving rise to the condition $g \vert \dot{\varphi}_0 \vert \gg \frac{21}{2} \, \pi \mu^2 \, H^2$ (which is parametrically equal to (\ref{C8-phiphi}), but slightly stronger). This is the condition $C_8$ in Table~\ref{tab:conditions}. 

We conclude this subsection by studying how the general $n-$point function  of the sourced first order inflaton perturbations, $\langle \delta \varphi_{1,s}^n \rangle = x^n \, \langle g_1^n \rangle$,  scales with the model parameters.  The $x^n$ factor ensures  that the modes are frozen at large scales, but it does not affect the parametric dependence. We can therefore understand this dependence by studying the expression  (\ref{g1n-general-time}). We set $H=1$ in this discussion, and we then restore it in the final expression through dimensional analysis. 

From the $g^2 \left( \varphi - \varphi_{0i} \right)^2 \chi_i^2$, interaction, we see that each mode $\chi$ sources the inflaton perturbation through  $\delta \varphi_{1,s} \propto \sum_i g^2  \left( \varphi - \varphi_{0i} \right) \int d^3 p \, \chi_i^2  \sim \sum_i g \,  m_{\chi_i}  \, {\rm sign } \left( \dot{\varphi}_0 \right) \, \int d^3 p \, \chi_i^2 $.

We therefore have 
\begin{equation}
\left\langle \delta \varphi_{1,s}^n \right\rangle \propto  \left( g \, {\rm sign } \left( \dot{\varphi}_0 \right) \right)^n \int d^3 p_1 \dots d^3 p_n \sum_i \left\langle   m_{\chi_i}  \left(  \chi_i \star \chi_i \right)_{\vec{k}_1} \dots    m_{\chi_i}  \left(  \chi_i \star \chi_i \right)_{\vec{k}_n}   \right\rangle \;. 
\end{equation} 
We note the final correlator is proportional to a single sum, since different $\chi$ species are uncorrelated. The correlator splits in a product of $n$ two-point correlators, each of which produces one $\delta-$function that ``cancels'' against one of the integrals. An overall $\delta-$function of the external momenta and one overall internal momentum integration remain, 
\begin{align} 
\left\langle \delta \varphi_{1,s}^n \right\rangle &\propto   \delta^{(3)} \left( \vec{k}_1 + \dots + \vec{k}_n \right) \;  \left( g \, {\rm sign } \left( \dot{\varphi}_0 \right) \right)^n  \, \int d^3 p \, {\rm e}^{-c \frac{p^2}{g \, \vert \dot{\varphi}_0 \vert}} \nonumber\\ 
&  \propto \sum_i \delta^{(3)} \left( \vec{k}_1 + \dots + \vec{k}_n \right) \;  \left( g \, {\rm sign } \left( \dot{\varphi}_0 \right) \right)^n \, 
\left( g \, \vert \dot{\varphi}_0 \vert \right)^{3/2}  \;. 
\label{dfn-par}
\end{align} 
($c$ is an order one factor emerging from the product of the wave functions.) The final factor in this expression can be understood as the number density of  $\chi_i$ particles with internal momentum $p$ running in the one loop diagram that gives (\ref{dfn-par}). The sum in (\ref{dfn-par}) is transformed into an integral over the time $\tau_0$ at which quanta are produced through    $\sum_i \rightarrow \int \frac{\vert \dot{\varphi}_0 \vert}{\Delta}\,dt$; see eq. (\ref{sumtoint}).  

The discussion so far explains all the parametric dependence in the first line of  (\ref{g1n-general-time}) as well as the integral in the second line. The remaining factors $G^n$ emerge from the time integral of the Green's function that relates each $\delta \varphi_{s,1}$ to its source. Each $G$ is dimensionless, and the remaining power of time in the integral ensures the correct scale invariant dependence given by the external  momenta. We find that the functions $G$ reach their maximum $G \sim \mu$ at the time $\tau_0 \sim - \mu / k$. Therefore the second line of   (\ref{g1n-general-time}) produces a factor $k^3 \, \mu^{n-3}$. This gives 
\begin{eqnarray}
\left\langle \delta \varphi_{1,s}^n \right\rangle &\sim& \frac{\delta^{(3)} \left( \vec{k}_1 + \dots + \vec{k}_n \right)}{k^{3 \left(  n-1 \right)}} 
 \times  \left( g \, {\rm sign } \left( \dot{\varphi}_0 \right) \right)^n \, g^{3/2} \,  \vert \dot{\varphi}_0 \vert^{3/2} \times \frac{\vert \dot{\varphi}_0 \vert}{\Delta} \times \mu^{n-3} \times  H^{n-4} \nonumber\\
 &\sim& \left[ {\rm sign} \left( \dot{\varphi}_0 \right) \right]^n \frac{\vert \dot{\varphi}_0 \vert}{H} \left( g \, \mu \, \frac{H}{k^3}\right)^{n-1} \, \delta^{(3)} \left( \vec{k}_1 + \dots + \vec{k}_n \right) \,, 
 \end{eqnarray} 
using eq. \eqref{mu-mut} to obtain the second line, and where where we have restored the proper $H$ dependence via dimensional analysis. We see that, in the special cases $n=2,3$, this expression reproduces the parametric dependence of eqs.~\eqref{dpshdp1s} and \eqref{dph1dph1dph1}.

%%%
\subsection{Second order perturbations and bispectrum in the scheme of~\cite{Green:2009ds}}
\label{sec:threepoint} 
%%%

We now turn our attention to the second order perturbations and to their contribution to the bispectrum  (\ref{BS}).  The goal of this and of the next subsection is not to provide a conclusive result for the bispectrum of trapped inflation, but to explicitly identify the disagreement between the results of the computational schemes of~\cite{Green:2009ds} and  \cite{LopezNacir:2011kk}. 

The sourced solution of the second order equation  \eqref{eq:system-3rd} can be written as 
\begin{align}
g_{2s} &= \int_x^\infty dx^\prime \,\tilde{G}(x,x^\prime) \Bigg[   \frac{  H {\tilde \mu}^2}{28 \,  \dot{\varphi}_0 }  \frac{p}{k} \,  \frac{\vert \vec{k} - \vec{p} \vert}{k} 
\left( - 8 g_1 \star g_1'' + 27 \, g_1' \star g_1' + \frac{54 \, g_1 \star g_1'}{x} + \frac{43 \, g_1 \star g_1}{x^2} \right)   \nonumber \\
& \qquad  \qquad  \qquad 
+  s_2' + 
 \frac{1}{2} \left( \frac{p}{k} g_1' \star {\hat s}_1 +    {\hat s}_1 \star  \frac{\vert\vec{k} - \vec{p} \vert}{k} g_1' 
+  \frac{p}{k}  g_1 \star {\hat s}_1 '  +   {\hat s}_1 '   \star   \frac{\vert\vec{k} - \vec{p} \vert}{k}  \, g_1 \right)
\Bigg],
\label{eq:2nd_order_solution_gen}
\end{align}
where $g_1$ is the solution to the first-order master equation (see \eqref{eq:g1_soln}), the sources are given by \eqref{eq:sources}, and the convolution denoted by $\star$ is defined in \eqref{eq:convolution}.  We see that several terms contribute to the source of the second order perturbations.  We denote by a suffix I and II, respectively, the contributions from the terms in the first and in the second line of \eqref{eq:2nd_order_solution_gen}.  In agreement with~\cite{Green:2009ds}, we expect also that the $s_2'$ contribution will be small, and we disregard it in our computations. 

As we show in Appendix \ref{ap:second_order}, the sources in the   first line of (\ref{eq:2nd_order_solution_gen}) lead to eq. \eqref{eq:B112-Id-res}, which reads 
\begin{align}
& \left\langle \delta \varphi_{1s} \left( \vec{k}_1 ,\, \tau \right)  \delta \varphi_{1s} \left( \vec{k}_2 ,\, \tau \right)  \delta \varphi_{2s} \left( \vec{k}_3 ,\, \tau \right)  \right\rangle_{\rm I, {\rm equil}} + 2 \, {\rm permut.} \nonumber\\
& \quad\quad \quad\quad \quad\quad \quad\quad \quad\quad \quad\quad
 \approx  -0.0037 \, \mu^2    \, g^2 \, H \,  \dot{\varphi}_0  \, \frac{ \delta^{(3)} \left( \vec{k}_1 + \vec{k}_2 + \vec{k}_3 \right)  }{ k^6 }     \,.   
\label{eq:threepoint2L1}
\end{align}
The sources in the second line lead instead to eq. \eqref{d1d1-d2IId-res}, which gives 
\begin{align}
& \!\!\!\!\!\!\!\!\!\!  \!\!\!\!\!\!\!\!\!\! 
 \left\langle \delta \varphi_{1s} \left( \vec{k}_1 ,\, \tau \right)  \delta \varphi_{1s} \left( \vec{k}_2 ,\, \tau \right)  \delta \varphi_{2s} \left( \vec{k}_3 ,\, \tau \right)  \right\rangle_{II,{\rm equil}}  + 2 \, {\rm permut.} \nonumber\\
  & \quad\quad \quad\quad \quad\quad \quad\quad \quad\quad \quad\quad
  \approx - 0.025 \, \mu^2 \, g^{2} \, H  \,  \dot{\varphi}_0  \, \ln \left( \frac{H \, \Delta}{\vert \dot{\varphi}_0 \vert} \right) \;  \frac{ \delta^{(3)} \left( \vec{k}_1 + \vec{k}_2 + \vec{k}_3  \right) }{ k^6 }   \;.  
\label{eq:threepoint2L2}
\end{align} 

If we exclude the logarithmic enhancement of the second term, the two contributions (\ref{eq:threepoint2L1}) and (\ref{eq:threepoint2L2}) have the same parametric dependence. Ref.~\cite{Green:2009ds} did not derive and evaluate the full term (\ref{eq:2nd_order_solution_gen}), but they considered a subset of terms that appear in the first line of  (\ref{eq:2nd_order_solution_gen}). They obtained a result for the bispectrum that is parametrically enhanced, by a factor ${\tilde \mu}^2$, with respect to (\ref{eq:threepoint2L1}). As we did for the power spectrum, in Appendix \ref{ap:comparison} we perform a line-to-line comparison of our and their solution, and in this case we also find that the discrepancy is due to their use of the inadequate approximation (\ref{SS-them}).

%%%
\subsection{Second order perturbations and bispectrum  in the EFT approach \cite{LopezNacir:2011kk}}
\label{sec:gradient}
%%%

The bispectrum of ref. \cite{LopezNacir:2011kk} arises from an EFT operator that contains spatial gradients of the perturbations. 
This presence of this operator arises from arguments of covariance, and ref. \cite{LopezNacir:2011kk} obtained their results by replacing the quantity $  \dot{\varphi}_0 $ that appears in the background equation of motion (\ref{eq:zeromode-local}) with the covariant expression $\left( g^{\mu \nu} \partial_\mu \varphi  \partial_\nu \varphi \right)^{1/2}$, with the spatial gradients acting on the perturbations. 

In the computational scheme of  \cite{Green:2009ds} that we are studying in this paper,  the covariantization $\dot\varphi\to\frac{1}{a}\,\sqrt{\dot\varphi^2-(\vec\nabla\varphi)^2}$ should be realized by replacing $\dot{\varphi}_0(\tilde{t}_0 + q\, \Delta t_1 + q^2\, \Delta t_2) + q\,\delta \dot{ \varphi_1}(\tilde{t}_0 + q \Delta t_1) + q^2\,\delta \dot{\varphi_2}(\tilde{t}_0  )$ by its covariantization in eq.~(\ref{time_shift_2order}). However, in this case, one can see that eqs.~(\ref{Delta_t}) imply that the gradient term $\vec\nabla\left[{\varphi}_0(\tilde{t}_0 + q\, \Delta t_1 + q^2\, \Delta t_2) + q\,\delta { \varphi_1}(\tilde{t}_0 + q \Delta t_1) + q^2\,\delta {\varphi_2}(\tilde{t}_0  )\right]$ identically vanishes (as it should, since by construction in this computational scheme $\Delta t$ is chosen in such a way that $\varphi(t_0+\Delta t)=\varphi_{0i}=$constant). This proves that the leading gradient operator of \cite{LopezNacir:2011kk} is not generated in this scheme. 

If the covariance arguments of \cite{LopezNacir:2011kk} are correct, we must conclude that this term will appear once some approximations intrinsic in this method are removed. To study the effect of this gradient term, we make use of the covariance argument of  \cite{LopezNacir:2011kk} in the context our second order equation in (\ref{eq:system-3rd}), and add the term  
\begin{equation}
+ \frac{5}{4} \, \frac{H \, {\tilde \mu}^2}{ \dot{\varphi}_0 } \, \frac{p}{k} \, \frac{ \vert \vec{k} - \vec{p} \vert }{ k } \, \vec{p} \cdot \left( \vec{k} - \vec{p} \right) \, g_1 \star g_1 \,, 
\label{gradient-operator}
\end{equation}
at the right hand side of that equation.  The $ \vec{p} \cdot \left( \vec{k} - \vec{p} \right) $ factor in this expression arises from the spatial derivatives acting on the $g_1$ perturbations inside the convolution, and it is obtained from the covariantization of the terms in  (\ref{eq:system-3rd}) with two time derivatives acting on the perturbations. This procedure is essentially the same one adopted in  \cite{LopezNacir:2011kk}, and indeed, adding this term to the second order equation of the perturbations we find  a result in parametric agreement with that of \cite{LopezNacir:2011kk} (as ref. \cite{LopezNacir:2011kk} provides only the parametric dependence, we cannot cross check the numerical coefficient); as shown in Appendix \ref{ap:BS-gradient}, we obtain
\begin{align}
& \left\langle \delta \varphi_1 \, \delta \varphi_1 \, \delta \varphi_2 \right\rangle_{\rm grad,1PR,equil} + 2 {\rm perm.} \simeq 
-0.0082 \, g^2 \,  \mu^4 \, H \,  \dot{\varphi}_0   \frac{\delta^{(3)} \left( \vec{k}_1 + \vec{k}_2 + \vec{k}_3 \right)}{k^6}    \;. 
\label{BS-grad}
\end{align} 
This contribution is ${\rm O } \left( \mu^2 \right)$ enhanced with respect to those evaluated in the previous subsection. The technical reason for this is that the operators considered there have two time derivatives rather than two gradients, and each time derivative results in a $1/\mu$ suppression of the result, as can be seen from the discussion after eq. (\ref{eq:B112-Id-res}).

%%%
\subsection{Gravitational waves}
\label{sec:GW}
%%%

The population of $\chi$ particles, besides affecting the evolution of the zero mode of the inflaton and sourcing its perturbations, is also a source of gravitational waves. The  power spectrum for gravitational waves in trapped inflation has been computed in Ref.~\cite{Cook:2011hg}.  Using Eqs.~\eqref{Delta} to convert the sum of the events of creation of gravitational waves into an integral, Ref.~\cite{Cook:2011hg} yields the tensor power spectrum
\begin{equation}
P^T \simeq \frac{2 H^2}{\pi^2 M_p^2} \left[ 1 + 4.8 \cdot 10^{-4} \, \frac{H^2}{M_p^2} 
\, \left( \frac{g \vert \dot{\varphi}_0 \vert}{H^2} \right)^{3/2} \, 
\ln^2 \left( \frac{\sqrt{g \, \vert \dot{\varphi}_0 \vert}}{H} \right)  
  \frac{\vert \dot{\varphi}_0 \vert}{\Delta \, H} \, \int \frac{  d x_0 }{  x_0 } 
\frac{\left( \sin x_{0} - x_{0} \cos x_{0} \right)^2}{x_{0}^3}  \right]\,,
\end{equation} 
where the first term originates from usual amplification of vacuum fluctuations of the tensors in  de Sitter space, while the second term is sourced by the $\chi_i$ particles. The integrand is peaked at $x_0 \simeq 2$, and the integral evaluates to $\frac{\pi}{6}$, so that
\begin{equation}
P^T \simeq \frac{2 H^2}{\pi^2 M_p^2} \left[ 1 +  .062 \, \mu^2\,\frac{\vert \dot{\varphi}_0 \vert}{g\,M_p^2}  \, 
\ln^2 \left( \frac{\sqrt{g \, \vert \dot{\varphi}_0 \vert}}{H} \right)   \right] .
\label{PT}
\end{equation} 
In section \ref{sec:pheno} we will determine the tensor-to-scalar ratio in the available parameter space for the model.

%%%%%%%%%%%%%%%%%%%%%%%%%%%%%%%%%%%%%%%%%%%%%%%%%
\section{Phenomenology}
\label{sec:pheno}
%%%%%%%%%%%%%%%%%%%%%%%%%%%%%%%%%%%%%%%%%%%%%%%%%

In the previous sections, we have introduced (Section \ref{sec:equations}) and solved (Sections \ref{sec:background} and \ref{sec:perturbations}) the master equations which govern the background inflaton solution and its first and second order perturbations in the computational scheme of  \cite{Green:2009ds}, showing explicitly how the result of \cite{Green:2009ds} for the bispectrum differs from that of  \cite{LopezNacir:2011kk}.  We have also reviewed the amount of gravitational waves produced during inflation (Section \ref{sec:GW}). Moreover,  we have collected the various constraints from our analysis, along with the constraints inherent to trapped inflation, in Table \ref{tab:conditions}. These conditions were discussed in Subsection \ref{sec:conditions}.  We now explore the phenomenology associated with these results. Similarly to~\cite{Green:2009ds}, we focus on a linear or a quadratic monomial potential, $V = M^3 \, \varphi$ and $V = \frac{1}{2} m^2 \varphi^2$, respectively (the explicit background evolutions in these two cases are presented in Appendix \ref{app:P_inflaton}). In Subsection \ref{sec:region}  we evaluate the constraints for these two models and obtain the viable parameter space. We show how the viable parameter space changes once the  power spectrum normalization of  \cite{Green:2009ds,LopezNacir:2011kk} is replaced by our correct expression.  In Subsection \ref{sec:ns} we discuss in more detail the power spectrum of the scalar perturbations in these models. In Subsection \ref{sec:fnl} we discuss how the relation between non-gaussianity and the model parameters is  affected by the change of normalization of the power spectrum that we have found. In Subsection \ref{sec:r} we study the tensor-to-scalar ratio.  

%%%
\subsection{Parameter space for a linear and quadratic inflaton potential}
\label{sec:region} 
%%%

For a monomial inflaton potential, the model is described by four parameters: the scale in the potential (denoted by $M$ or $m$ for linear and quadratic potentials, respectively), the coupling constant $g$ of the $\varphi^2 \,\chi_i^2$ interaction, the distance $\Delta$ between successive instances of particle production, and the number of e-folds $N$ before the end of inflation at which the largest CMB scales left the horizon. We fix this last parameter to $N=60$ in the present discussion; we have verified that similar results are obtained for $N=50$. 

The amplitude of the power spectrum (to be discussed in the next subsection) provides one relation between the three remaining parameters, which we use to express $\Delta$ in terms of the other two parameters. Therefore the phenomenology of the model can be discussed in the two dimensional plane $\left\{ g ,\, M \right\}$ in the case of linear potential, or  $\left\{ g ,\, m \right\}$, in the case of quadratic potential.  Table \ref{tab:conditions} summarizes the conditions that must be imposed for our results to be valid. The first two constraints in the Table are  slow-roll inflationary condition. 
The constraint $C_2$ is subdominant to $C_1$, which we use to determine the end of inflation at: 
\begin{eqnarray}
&& {\rm Linear \; potential} \;:\;\; \varphi_{\rm end} \simeq 5.9 \cdot 10^4 \, g^{2/3} \, M \;\;, \nonumber\\ 
&& {\rm Quadratic \; potential} \;:\;\; \varphi_{\rm end} \simeq 1.2 \cdot 10^8 \, g \, m \;\;. 
\end{eqnarray} 

\begin{figure}[ht!]
\centerline{
\includegraphics[width=0.5\textwidth,angle=0]{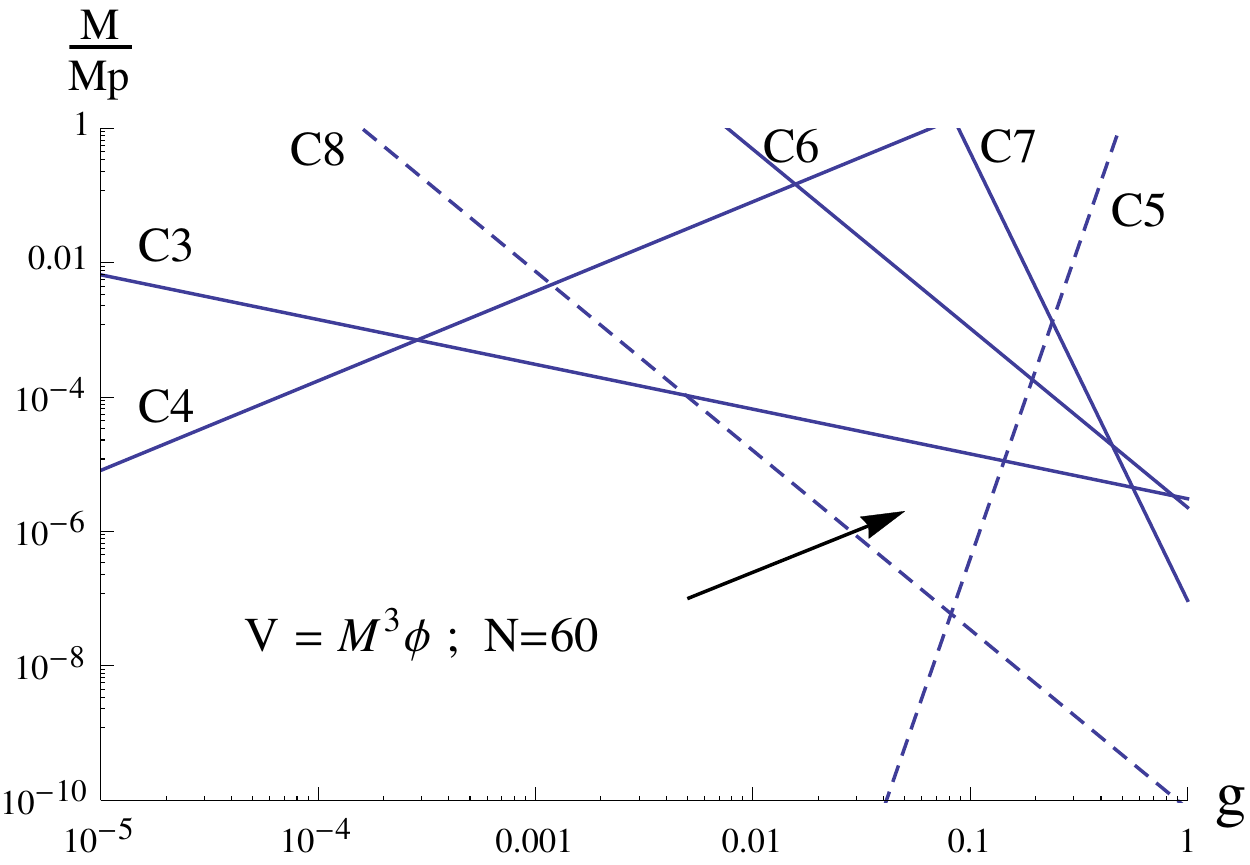}
\includegraphics[width=0.5\textwidth,angle=0]{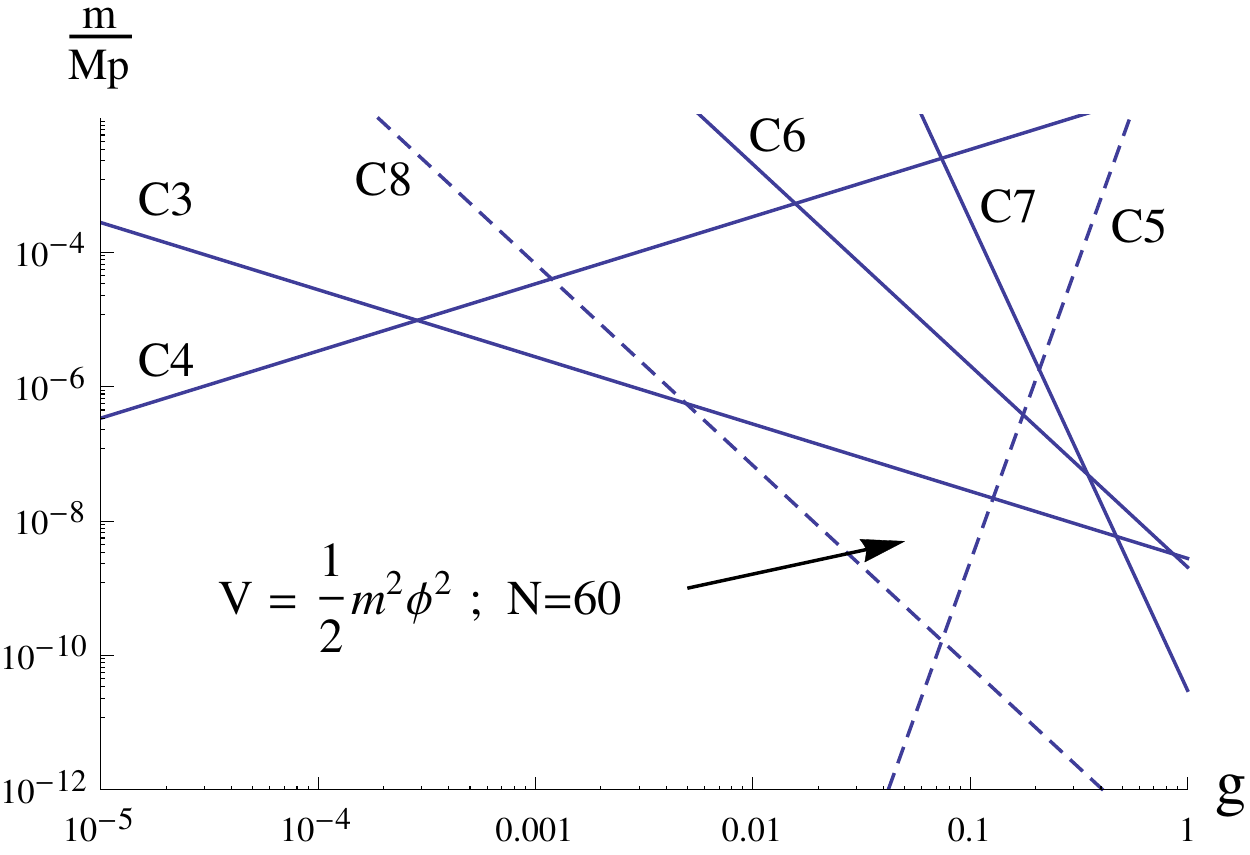}
}
\caption{
Conditions for trapped inflation to work, and for our results to be valid, in the case of a linear (left panel) and quadratic (right panel) inflaton potential. Solid (Dashed) lines are upper (lower) bounds. The triangle indicated by the arrows is the area that satisfies all these constraints.
}
\label{fig:bounds}
\end{figure}

The remaining conditions $C_3-C_8$ can be expressed as upper or lower bounds on the mass scale in the potential as a function of $g$. We do so in Figure \ref{fig:bounds}, where solid (dashed) lines represent upper (lower) bounds. The constraints are immediately obtained by using the relations written in Appendix \ref{app:P_inflaton} in the third column of Table  \ref{tab:conditions}.

As we can see from the figure, all the constraints are satisfied only in  a small triangular region in parameter space, delimited by the lower $C_5 ,\, C_8$ bounds, and by the upper $C_3$ bound. We recall that the constraint $C_3$ is the defining constraint for the trapped mechanism to work (the friction in the motion of the inflaton is dominated by particle production). The constraint $C_4$ enforces a non adiabatic variation of the mass of the quanta $\chi_i$ at their production. 
The remaining $C_5-C_8$ constraints ensure the validity of the approximations needed to obtain our results.

\begin{figure}[ht!]
\centerline{
\includegraphics[width=0.5\textwidth,angle=0]{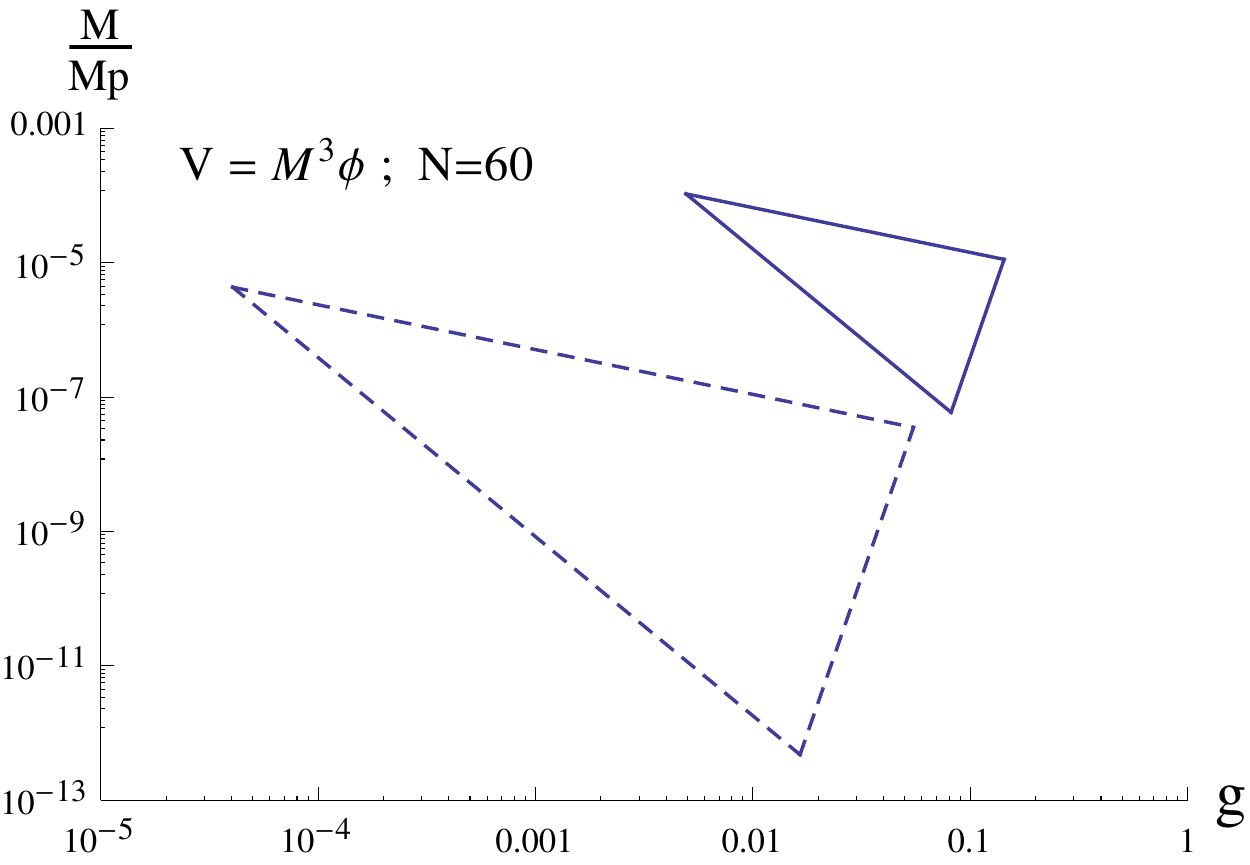}
\includegraphics[width=0.5\textwidth,angle=0]{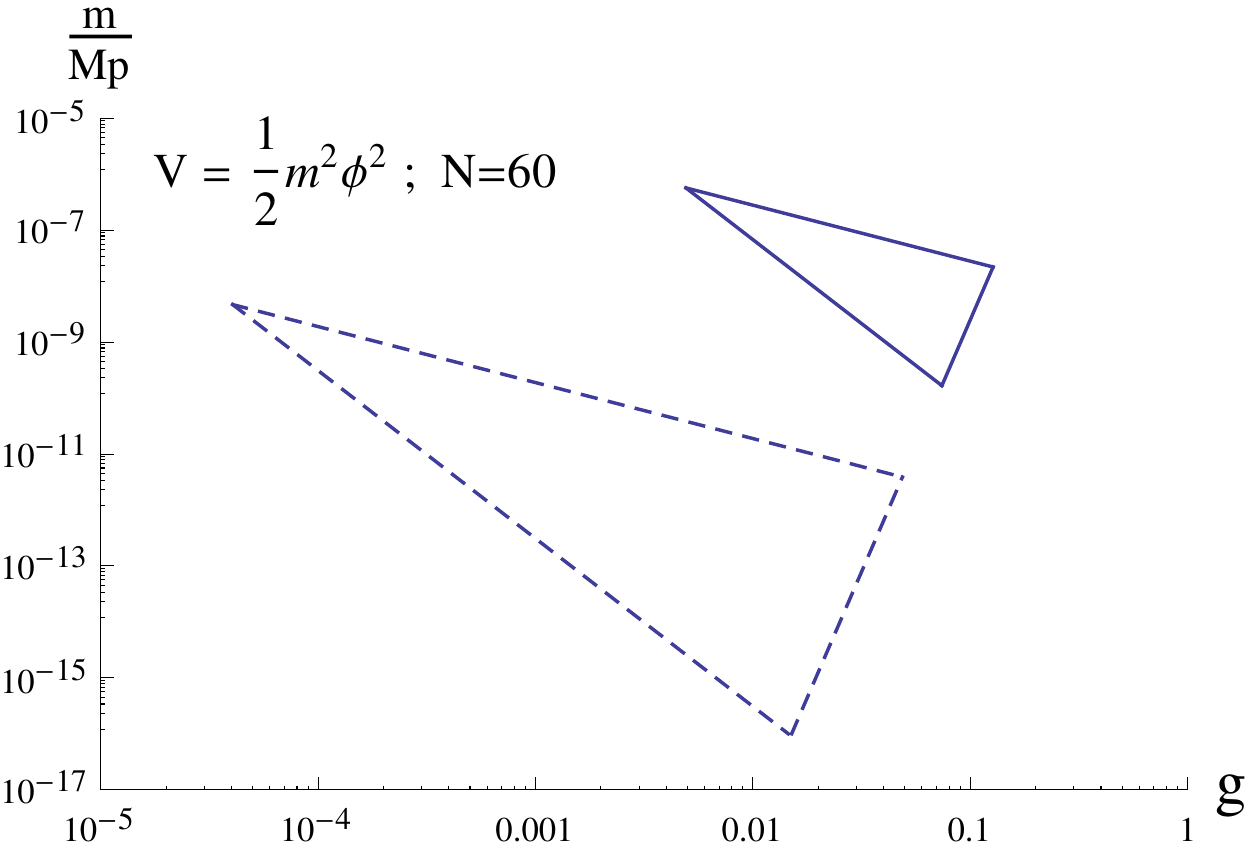}
}
\caption{Comparison of the validity region in parameter space obtained from our power spectrum normalization (solid lines) and the one of 
  \cite{Green:2009ds,LopezNacir:2011kk} (dashed lines). The left (right) panel is for a linear (quadratic) inflaton potential, and for $N=60$ e-folds of inflation.
}
\label{fig:compare}
\end{figure}

We stress that having the correct normalization of the power spectrum is crucial to relate the parameter space of the models to any phenomenological result and to identify the validity region in parameter space. Our result for the power spectrum disagrees by more than four order of magnitudes with that of \cite{Green:2009ds,LopezNacir:2011kk}. In Figure \ref{fig:compare} we show with solid lines the perimeter of the validity region  obtained with our normalization of the power spectrum. This is the same region shown in Figure \ref{fig:bounds}, and delimited by the three conditions $C_{3,5,8}$. We then evaluate the same three conditions with the   power spectrum normalization of  \cite{Green:2009ds,LopezNacir:2011kk}; the dashed lines show the valid parameter spaice if one uses that normalization. The difference between the two regions quantifies the impact of our power spectrum result.

\begin{figure}[ht!]
\centerline{
\includegraphics[width=0.5\textwidth,angle=0]{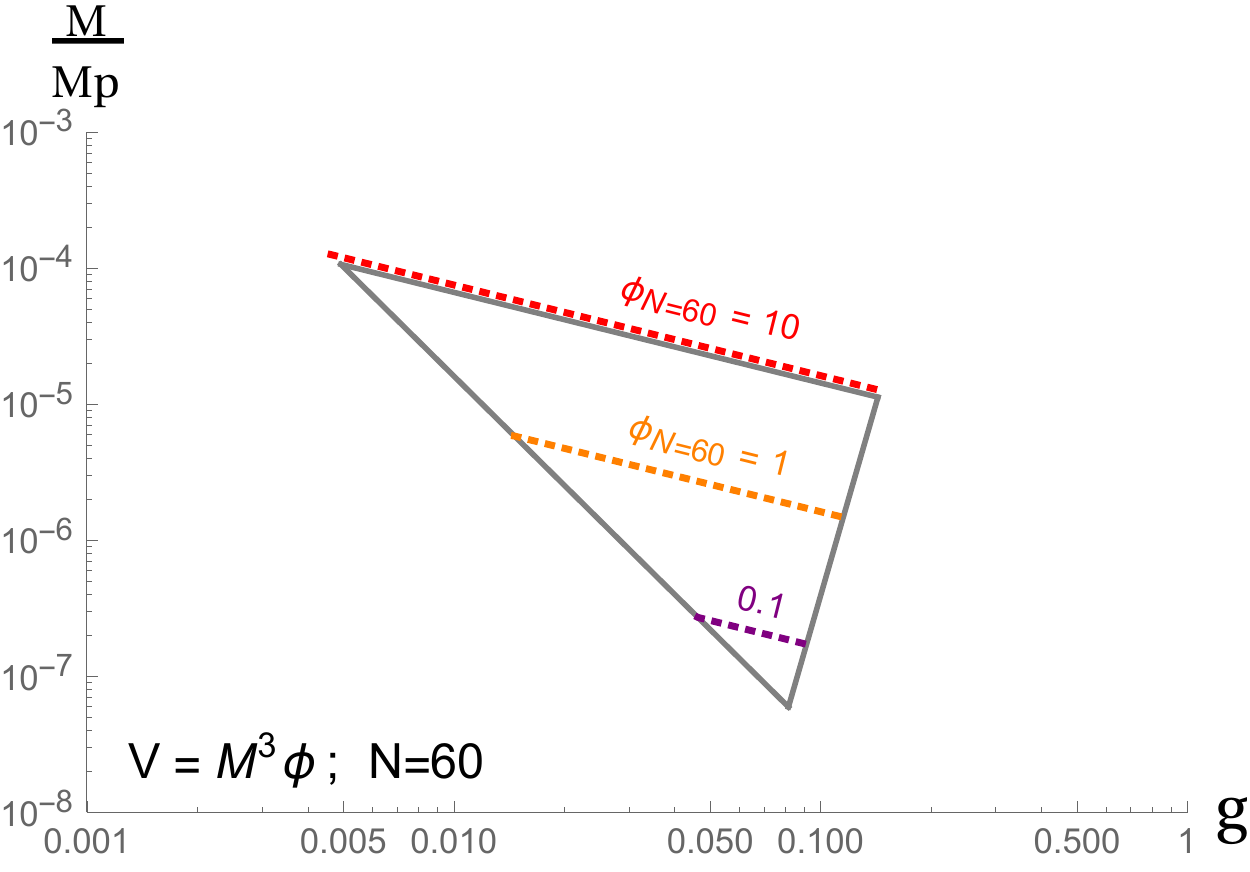}
\includegraphics[width=0.5\textwidth,angle=0]{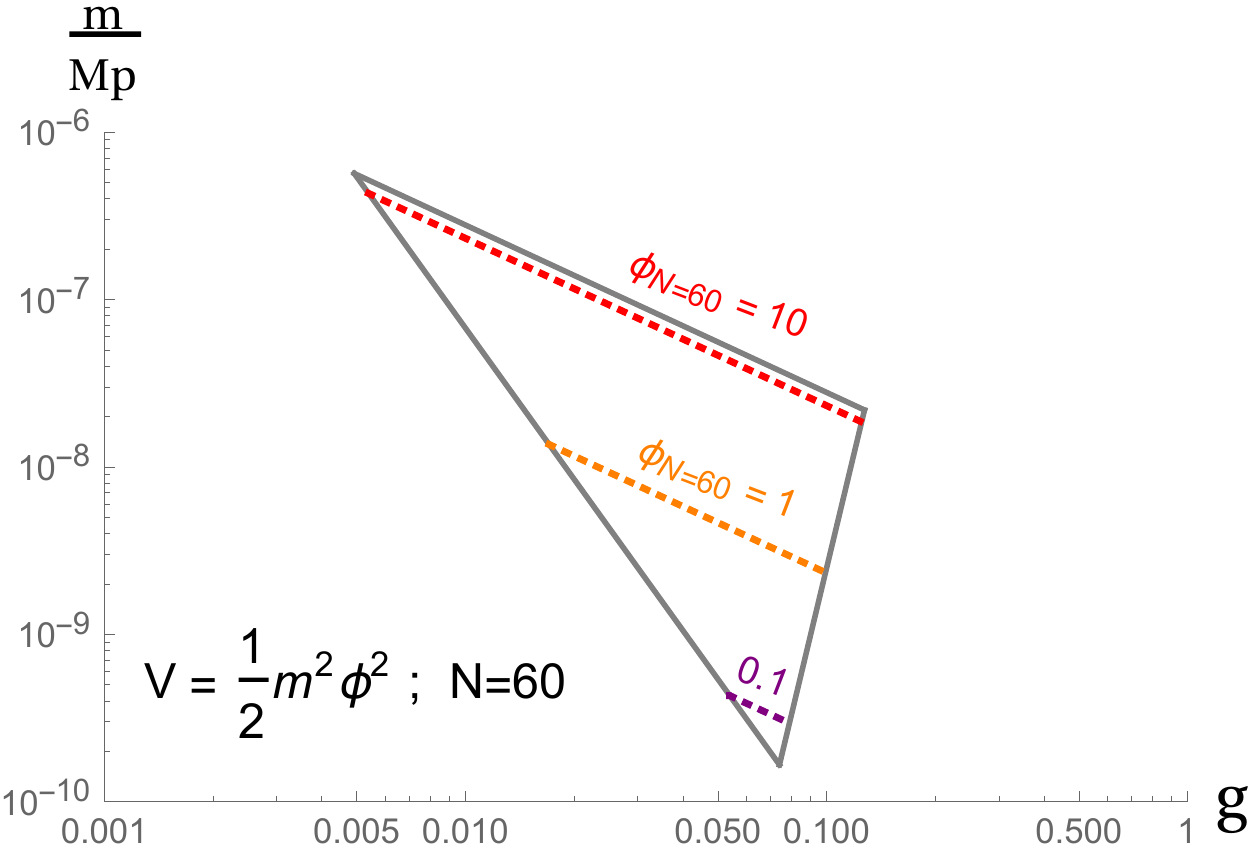}
}
\caption{
The dotted lines are contour lines indicating the value assumed by the inflaton at $N=60$ e-folds before the end of inflation, for any given choice of $g$ and of the potential mass parameter.  The labels on the lines indicate the numerical value $\varphi_{N=60}$ in units of the reduced Planck mass $M_p$. The result is shown only in the validity region identified in Figure \ref{fig:bounds}. 
}
\label{fig:phi60}
\end{figure}

The value of the inflaton at $N=60$ e-folds before the end of inflation is 
\begin{eqnarray}
&& {\rm Linear \; potential} \;:\;\; \varphi_{N=60} \simeq 2.9 \cdot 10^6 \, g^{2/3} \, M \;\;, \nonumber\\ 
&& {\rm Quadratic \; potential} \;:\;\; \varphi_{N=60} \simeq 4.3 \cdot 10^9 \, g \, m \;\;. 
\label{phi-60}
\end{eqnarray} 
We plot these values in Figure \ref{fig:phi60}, against the bounds that determine the triangular viable region. We see that in a significant portion of this region the inflation can have a sub-Planckian evolution all throughout inflation.  Not surprisingly, the lines with $\varphi_{N=60} \simeq 10\,M_p$ are close to the boundary determining the $C_3$ (slow roll) constraint, above which the Hubble friction controls the motion of the inflaton field. We recall that in this case  $\varphi_{N=60} = {\rm O } \left( 10 \right)$.  More precisely, we recall that the standard slow roll relations (no particle production) give  $\varphi_{N=60} \simeq 11 \, M_p$ and  $\varphi_{N=60} \simeq 15.5 \, {M_p}$ for a linear and quadratic inflaton potential, respectively. Instead, the more we enter in the region below the $C_3$ line, the more particle production is effective in slowing the inflaton field and in reducing the range scanned by the inflaton during inflation.

%%%
\subsection{Power spectrum and spectral tilt}
\label{sec:ns} 
%%%

From eqs. (\ref{zeta}), (\ref{PS}), (\ref{dp1hdp1h}), and (\ref{dpshdp1s}), we obtain the super-horizon power spectrum of the vacuum and sourced perturbations 
\begin{eqnarray}
P_{\zeta,h} \left( k \right) \simeq \dfrac{H^4}{4 \pi^2 \, \dot{\varphi}_0^2} \, \dfrac{\left( 1 + \tilde{\mu}^2 \right)^2}{\cosh^2 \left( \frac{\pi \tilde{\mu}}{2} \right)} \;\;\;,\;\;\; 
P_{\zeta,s} \left( k \right)  \simeq    \dfrac{0.18}{2\pi^2} \,   g \, \frac{\mu \, H^2}{|\dot{\varphi}_0|}  \;. 
\end{eqnarray} 
The ratio $\frac{P_{\zeta,h}}{P_{\zeta,s}}$ is a function of $\mu$ times $\frac{H^2}{g \vert \dot{\varphi}_0 \vert}$. The function of $\mu$ is smaller than one for $\mu \ga 1$, which we verified is always true in the viable region of parameter space. We then have $\frac{H^2}{g \vert \dot{\varphi}_0 \vert} \ll 1$ due to the condition $C_4$. We therefore see that the power spectrum is strongly dominated by the sourced modes, and so we have 
\begin{equation}
P_{\zeta} \left( k \right) \simeq 
P_{\zeta,s} \left( k \right) \simeq    \dfrac{0.18}{2\pi^2} \,   g \, \frac{\mu \, H^2}{|\dot{\varphi}_0|}  \simeq 
5.7 \cdot 10^{-4} \, \frac{g^{9/4} \, H}{\Delta^{1/2} \, \vert \dot{\varphi}_0 \vert^{1/4}} \;\;. 
\label{PS-res}
\end{equation}
We impose that the  power of the modes that left the horizon at $N=60$ matches with the observed one, $P_\zeta \simeq 2.2 \cdot 10^{-9}$ \cite{Ade:2015lrj}. This  fixes the parameter $\Delta$ in terms of the other parameters of the model. This relation is assumed in the results shown in our figures. 

The result (\ref{PS-res}) is about $10^{-4}$ smaller than the one found by \cite{Green:2009ds,LopezNacir:2011kk}. In Appendix \ref{ap:comparison} we discuss the origin of this discrepancy, and we clarify the specific approximation done in  \cite{Green:2009ds} on which we disagree. 

Let us now discuss the spectral tilt of the scalar power spectrum, 
\begin{equation}
n_s - 1 \equiv \frac{k}{ {\cal P}_\zeta } \, \frac{d \, {\cal P}_\zeta }{d k}  =  \frac{1}{H \, {\cal P}_\zeta} \, \frac{d \, {\cal P}_\zeta}{d t} \, \Big\vert_{a H = k} \;. 
\end{equation} 

As we will see shortly, if we impose that $\Delta$ is strictly constant, we obtain a value for $n_s$ which is greater than the range allowed by observations. To avoid this, we consider the case in which $\Delta$ depends on the value of the inflaton, so that it is slowly evolving during inflation. Reference~\cite{Flauger:2014ana} provides a setting where such an evolution might happen in the context of models of monodromy: as the inflaton rolls down its potential, the axion periodicity slowly changes, leading, in the context of trapped inflation, to a weak $\varphi$-dependence of $\Delta$. For definiteness, we consider a power law dependence 
\begin{equation}
\Delta \propto \varphi_0^\delta \;\;\;,\;\;\; \delta = {\rm constant} \;. 
\end{equation} 
This gives 
\begin{equation}
n_s-1 \simeq \frac{9}{10} \, \frac{\dot{H}}{H^2} - \frac{\dot{\varphi}_0 \, V''}{10 \, H \, V'} - \frac{\dot{\varphi}_0 \, \delta}{2 \, H \, 
\varphi_0} \;. 
\end{equation} 
where eq. (\ref{eq:ddotphi0}) has been used. For inflaton potential $V \propto \varphi^p$, one has $\frac{\dot{H}}{H} = \frac{p}{2} \, \frac{\dot{\varphi}_0}{\varphi_0} $, and $\frac{V''}{V} = \frac{p-1}{\varphi_0}$. We arrive to the compact expression 
\begin{equation}
V \propto \varphi^p \;\;\Rightarrow\;\; n_{s-1} \simeq \left( \frac{\delta}{p} - \frac{1}{5 p} - \frac{7}{10} \right) \epsilon \;, 
\end{equation} 
where we recall that $\epsilon \equiv - \frac{\dot{H}}{H^2}$. Using the background solutions of Appendix \ref{app:P_inflaton}, we obtain $\epsilon \simeq \frac{0.39}{N}$ for a linear potential and $\epsilon \simeq \frac{0.83}{N}$ for a quadratic potential. 

For $\delta  = 0$ (namely, for constant $\Delta$), and for $N=60$, we obtain $n_s - 1 \simeq -0.006$ (resp., 
$n_s - 1 \simeq -0.01$) for a linear (resp., quadratic) inflaton potential, in disagreement with the observed value 
 $n_s-1 = -0.0323 \pm 0.0060$ \cite{Ade:2015lrj}. For nonvanishing $\delta$, we instead find 
\begin{eqnarray}
&& {\rm Linear \; potential} \;:\;\; \delta = 0.9 + 2.6 \, N \left( n_s-1 \right) \;, \nonumber\\ 
&& {\rm Quadratic \; potential}  \;:\;\;  \delta = 1.6 + 2.4 \, N \left( n_s-1 \right) \;.   
\label{smalldelta}
\end{eqnarray} 
For $N=60$, the observed value of $n_s$ enforces  $-6 \la \delta \la -2.3$ in the case of a linear inflaton potential and  $-4.8 \la \delta \la -1.3$ in the case of quadratic inflaton potential.

%%%
\subsection{$f_{\rm NL}$} 
\label{sec:fnl} 
%%%

The bispectrum of trapped inflation has been computed in refs.~\cite{Green:2009ds} and \cite{LopezNacir:2011kk}. Both works obtain a result in agreement with each other, which is parametrically given by  (\ref{BS-grad}). In the previous section we showed that the agreement is due to an inaccurate approximation made in  \cite{Green:2009ds} (more details are given in Appendix \ref{ap:comparison}), and we further demonstrated that the leading EFT operator introduced in  \cite{LopezNacir:2011kk} does not appear in the computational scheme of  \cite{Green:2009ds}. Therefore, the results of  \cite{Green:2009ds} do not corroborate the method of  \cite{LopezNacir:2011kk}, or vice versa\footnote{In this respect, it is also worth noting that in the context of warm inflation the amplitude of the bispectrum computed in~\cite{Moss:2007cv,Bastero-Gil:2014raa} shows a weak, logarithmic dependence of the dissipation parameter, whereas~\cite{LopezNacir:2011kk} shows a stronger, linear dependence.} the bispectrum for warm inflation obtained in refs. shows a weak warm inflation. 

\begin{figure}[ht!]
\centerline{
\includegraphics[width=0.5\textwidth,angle=0]{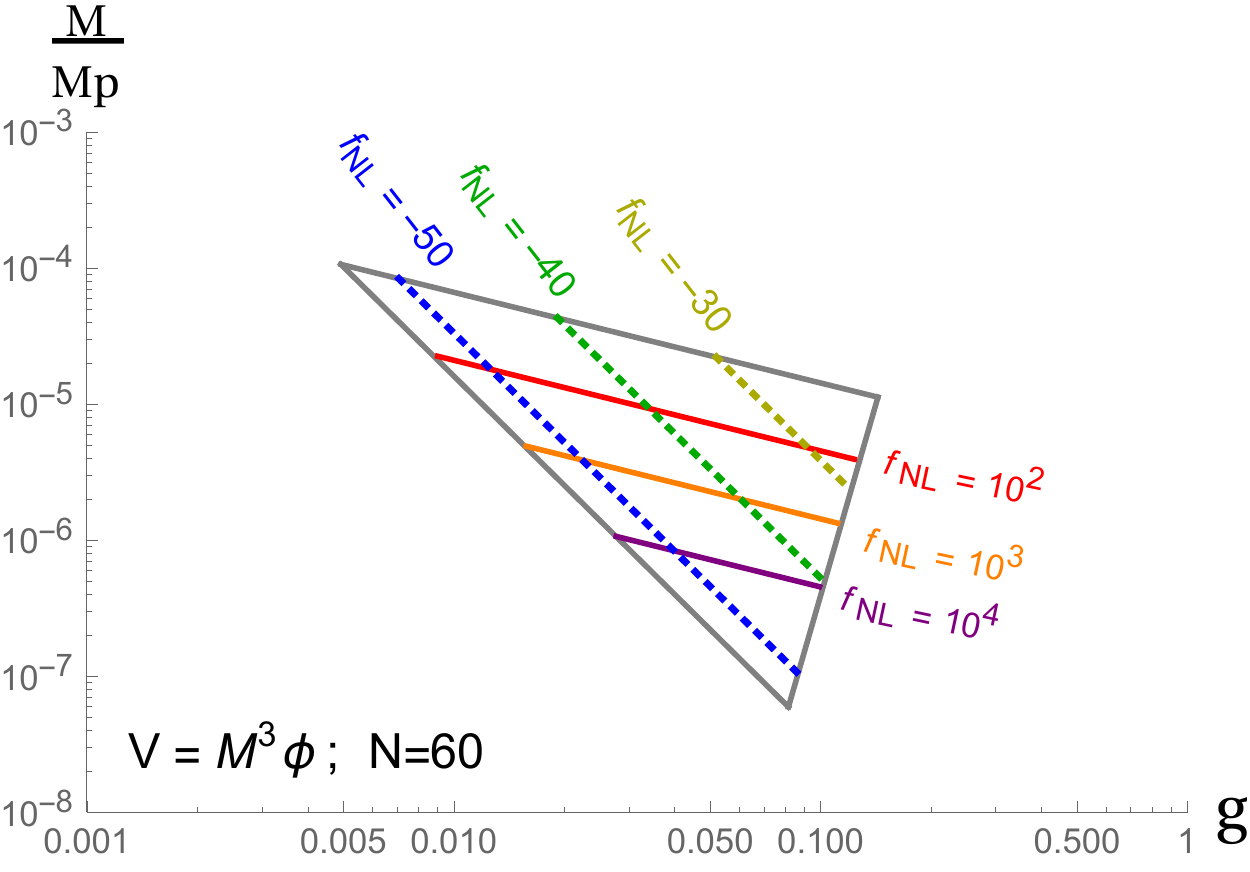}
\includegraphics[width=0.5\textwidth,angle=0]{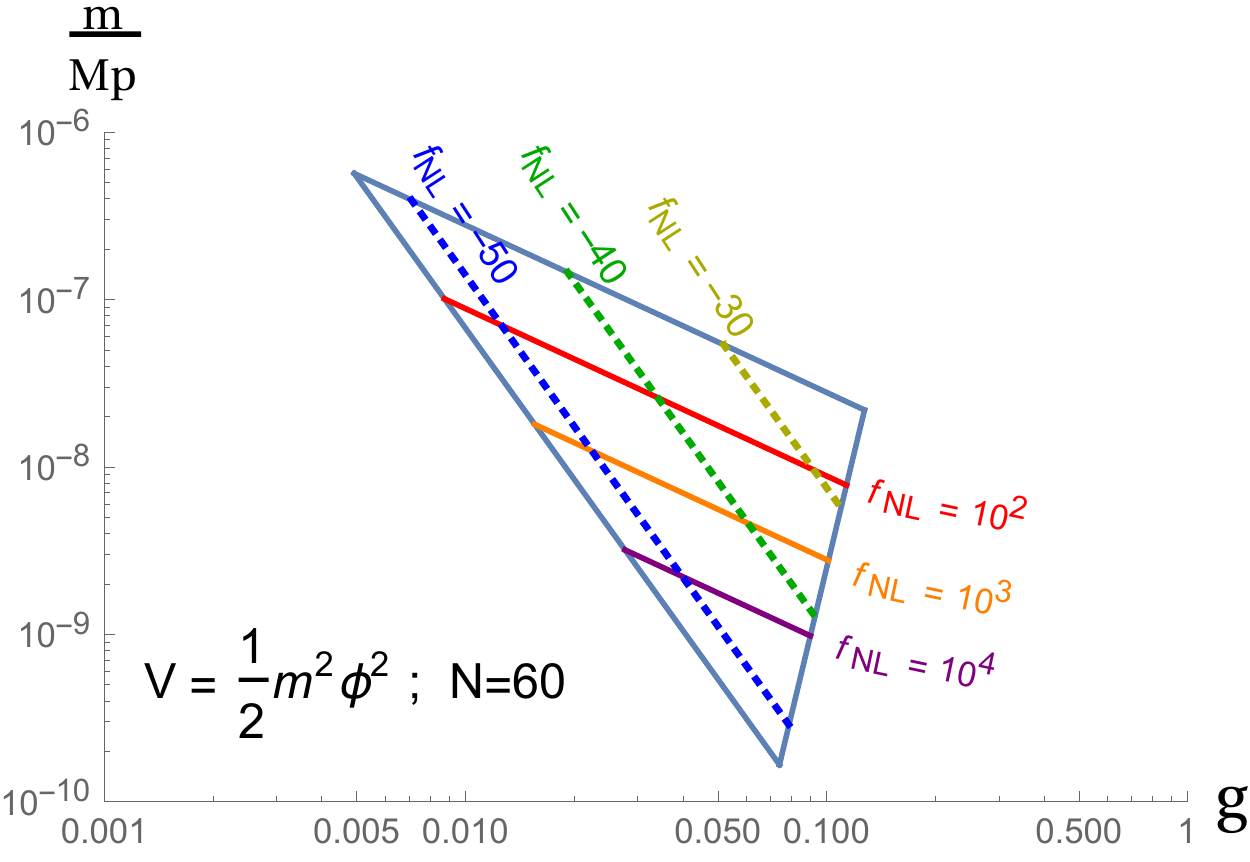}
}
\caption{
Comparison of the nonlinearity parameter obtained with the scheme of  \cite{Green:2009ds} (dashed lines) and with the EFT operator of 
\cite{LopezNacir:2011kk} (solid lines). The results are shown in the validity region obtained using the amplitude of the power spectrum computed in the present work. The left (right) panel is for a linear (quadratic) inflaton potential. 
}
\label{fig:fNL}
\end{figure}

To quantify the difference between the two results, in Figure \ref{fig:fNL} we show the nonlinearity parameter 
\begin{equation}
B_\zeta \left( k_1 ,\, k_2 ,\, k_3 \right) = \frac{3}{10} \left( 2 \pi \right)^{5/2} \, 
f_{\rm NL} \left( k_1 ,\, k_2 ,\, k_3 \right) \, P_\zeta^2  \, \frac{\sum_i k_i^3}{\prod_i k_i^3} \, \;,
\label{fNL-def}
\end{equation} 
obtained from the two different bispectra. From the results (\ref{eq:threepoint2L2}) and (\ref{BS-grad}), we obtain, respectively, 
\begin{equation}
f_{\rm NL,II,equil} \simeq  -  3.3 \, \ln \left( \frac{\vert \varphi_0 \vert}{H \, \Delta} \right) \;\;\;,\;\;\; 
f_{\rm NL,grad,equil} \simeq 1.1 \, {\tilde \mu}^2 \;. 
\label{fNL-result}
\end{equation}

As is typical for non-gaussianity from particle production, we expect the bispectrum to be maximal for equilateral configurations. This is because modes of a given species $\chi_i$ are only correlated with modes of the same species, $\langle \chi_i \chi_j \rangle \propto \delta_{ij}$. Quanta of a given species $\chi_i$ are produced only at the given (conformal) time $\tau_{0i}$. As discussed after eq. (\ref{dpshdp1s}), they mostly source inflaton modes of (comoving) momentum $k \sim - {\tilde \mu} / \tau_{0i}$. Therefore, sourced modes of the inflation are mostly correlated with modes of parametrically the same size. For this reason, the nonlinearity parameter shown in the figure is evaluated on exactly equilateral triangles. 

We stress that the relation between $f_{\rm NL}$ and the model parameters is strongly impacted by our correction of the amplitude of the power spectrum (cf. Figure \ref{fig:compare}).  The current bound on equilateral non-gaussianity is   $f_{\rm equil} = -4 \pm 43$ ($68 \, \%$ CL statistical) \cite{Ade:2015ava}. We see from the figure that only a narrow region of parameter space is viable. If a non-gaussianity of this type is observed, the correction of the power spectrum that we have obtained would lead to very different values of parameters than those that would be pointed out from the power spectrum of  \cite{Green:2009ds,LopezNacir:2011kk}.

%%%
\subsection{Tensor-to-scalar ratio}
\label{sec:r} 
%%%

\begin{figure}[ht!]
\centerline{
\includegraphics[width=0.5\textwidth,angle=0]{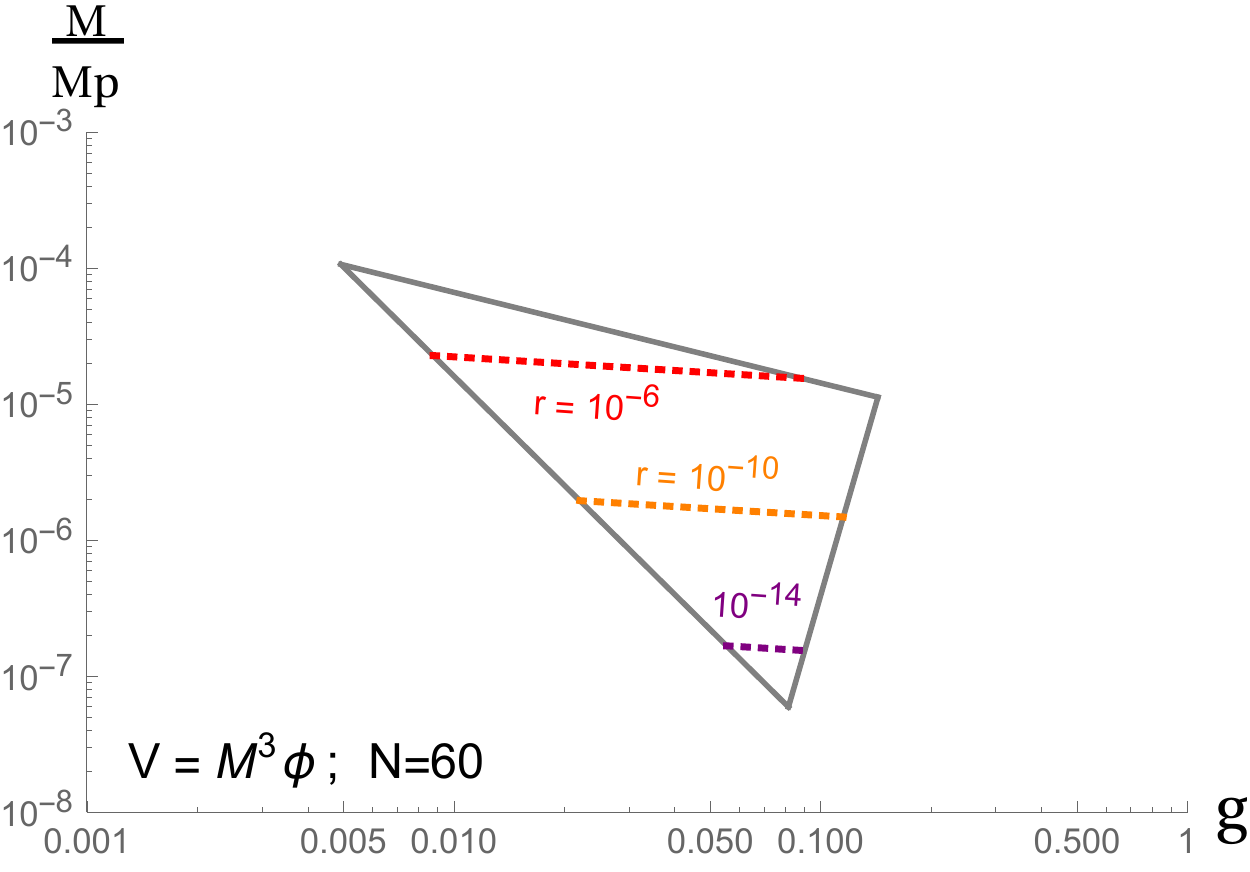}
\includegraphics[width=0.5\textwidth,angle=0]{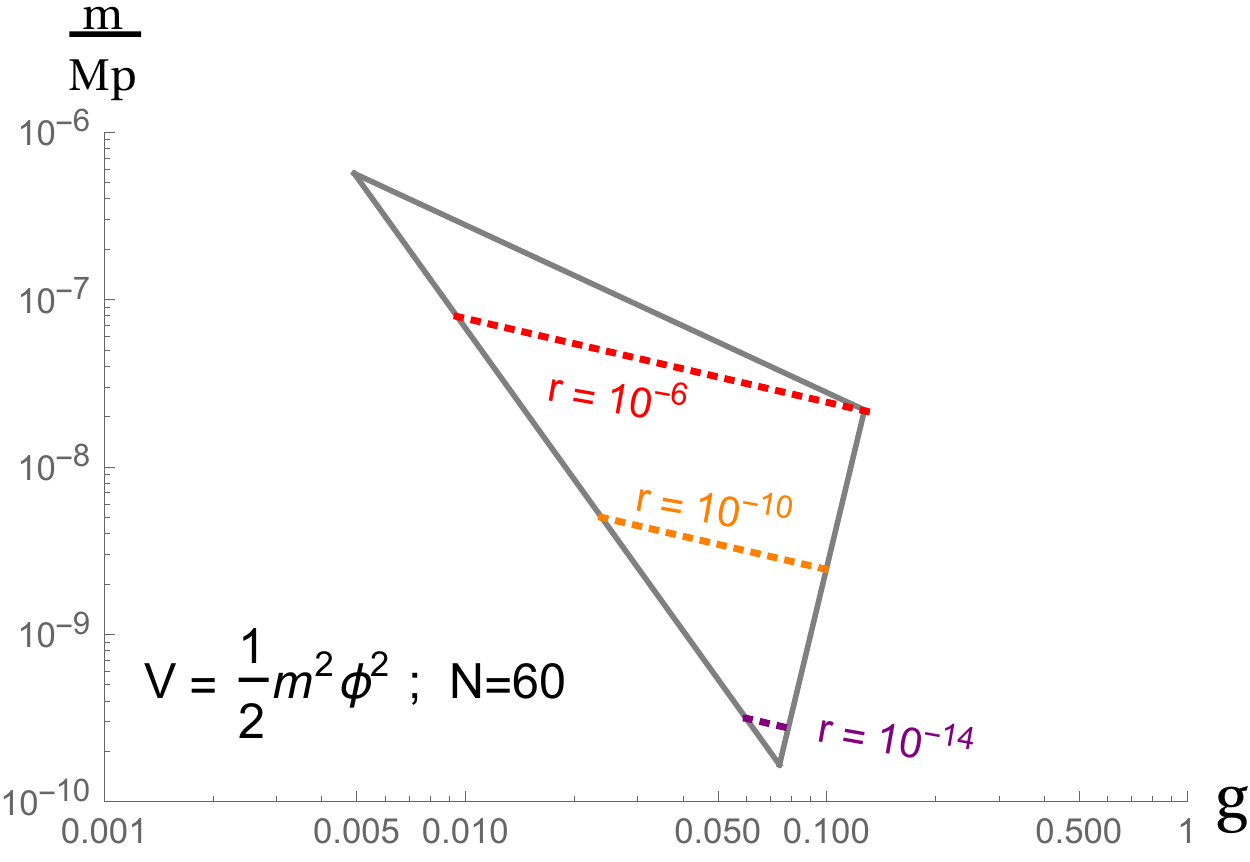}
}
\caption{
The dotted lines are contour lines indicating the value assumed by the tensor-to-scalar ratio $r$, for any given choice of $g$ and of the potential mass parameter in the  validity region. 
}
\label{fig:bounds+gw}
\end{figure}

Eq. (\ref{PT}) gives the power spectrum of tensor modes produced in trapped inflation. Disregarding the logarithmic correction in that expression, and using eq. (\ref{phidot}),  we find 
\begin{equation}
\frac{P_{T,{\rm sourced}}}{P_{T,{\rm vacuum}}} \sim  2.5 \cdot 10^{-4} \, \frac{g^{3/2} \, \vert \dot{\varphi}_0 \vert^{5/2}}{M_p^2 \, H^2 \, \Delta}  \simeq \frac{0.2 \, \vert V' \vert}{g \, H \, M_p^2} \simeq \frac{0.6 \, H \, \vert V' \vert}{g \, V} \;,  
\end{equation} 
and we have verified that the sourced part is highly subdominant in the viable region of parameters. From the vacuum part in (\ref{PT}), and from  the measured ${\cal P}_\zeta \simeq 2.2 \cdot 10^{-9}$, we obtain the tensor-to-scalar ratio 
\begin{eqnarray}
&& {\rm Linear \; potential} \;:\;\; r \simeq g^{2/3} \left( \frac{N}{60} \right) \, \left( 3,060 \, \frac{M}{M_p} \right)^4 \;, \nonumber\\  
&& {\rm Quadratic \; potential}  \;:\;\; r \simeq g^2 \left( \frac{N}{60} \right)^3 \, \left( 4.11 \cdot 10^6 \, \frac{m}{M_p} \right)^4 \;. 
\end{eqnarray} 
From Figure  \ref{fig:bounds+gw}, we see that in the region of our interest the GW signal is too small to be observed in the foreseeable future.

%%%%%%%%%%%%%%%%%%%%%%%%%%%%%%%%%%%%%%%%%%%%%%%%%
\section{Discussion} 
\label{sec:conslusions}
%%%%%%%%%%%%%%%%%%%%%%%%%%%%%%%%%%%%%%%%%%%%%%%%%

In this work we have explored the phenomenology of trapped inflation, which is an inflationary model in which particle production induced by the rolling inflaton is the main source of friction in the inflaton equation of motion.  We are indebted to the work of Ref.~\cite{Green:2009ds}, which proposed the trapped inflation model, and provided a computational scheme for the  background and the inflaton perturbations. In this work we  have built on the study of~\cite{Green:2009ds}, by providing a precise solution of the first order equation (in contrast to the approximate solution given in~\cite{Green:2009ds}), and by providing and solving the full equation for the second order inflaton perturbations in this scheme (in contrast to the subset of terms considered in~\cite{Green:2009ds}). 

An important motivation for this reanalysis was to compare the findings of~\cite{Green:2009ds} with those of ~\cite{LopezNacir:2011kk}.
Ref.~\cite{LopezNacir:2011kk} provided a general discussion of dissipative mechanisms in inflation, and trapped inflation is only one of their applications.  The portion of~\cite{LopezNacir:2011kk} devoted to the trapped inflation mechanism  does not perform an explicit computation of the second order equation, but it introduces an operator for the perturbations in the spirit of EFT. This operator is  proportional to spatial derivatives of the perturbations, and it is motivated by general arguments of covariance. All comments that we make on~\cite{LopezNacir:2011kk} are restricted to this part of their study.  The two works \cite{Green:2009ds} and \cite{LopezNacir:2011kk} use the same power spectrum (first derived in  \cite{Green:2009ds}), and claim results for the bispectrum that are in parametric agreement with each other. The fact that these two very different approaches lead to such an agreement appears to strongly support their validity. One is the naturally led to conclude that (i) the approximations done in both works are valid, and therefore can be applied to similar contexts, and (ii) it must be possible to identify how precisely the leading EFT operator of  \cite{LopezNacir:2011kk} emerges in the explicit computation of  \cite{Green:2009ds}. 

With this in mind, we reanalyzed the computations of these two works, and we found that the parametric agreement of the bispectrum is due to an inadequate approximation used in  \cite{Green:2009ds} (we discuss this in full details in Appendix \ref{ap:comparison}). Once this aproximation is replaced by an exact computation, the bispectrum obtained with the scheme of   \cite{Green:2009ds} is parametrically, and numerically, smaller than that obtained with the EFT operator of  \cite{LopezNacir:2011kk}, see eqs. (\ref{fNL-result}) and Figures \ref{fig:fNL}. 

Equally importantly, we found that the same inadequate approximation also affects the power spectrum obtained in~\cite{Green:2009ds} and used in~\cite{LopezNacir:2011kk}. By correcting for this approximation, we obtain a power spectrum that is  parametrically equal to that of \cite{Green:2009ds,LopezNacir:2011kk}, but smaller by a factor of $\simeq 6.8 \cdot 10^{-5}$ (in Appendix \ref{ap:comparison} we show in full detail how this numerical difference arises). This difference of more than four orders of magnitude has a profound impacts on the relation between the phenomenology and the values of the parameters in the model, as we demonstrate in Figure \ref{fig:compare}. If phenomenological evidence for this mechanism will be found, model parameters should be chosen very differently from what one would conclude from the power spectrum of  \cite{Green:2009ds,LopezNacir:2011kk}. 

In this summary section, we also want to comment on other interesting conclusions that were obtained in \cite{Green:2009ds}, and that  
remain valid also in presence of our updated results.  Eq.  (\ref{phi-60}) provides the value $\varphi_{60}$ of the inflaton $60$ e-folds before the end of inflation. As we see from Figure \ref{fig:phi60}, the range scanned by the inflaton during these $60$ e-folds can be fully sub-Planckian in a subset of this region. As we discussed in the Introduction, this may be helpful in ensuring that the predictions of the model are robust under the effects of UV physics.

From the result (\ref{PS-res}) we can also compute the spectral tilt of the scalar modes. In agreement with  \cite{Green:2009ds}, we find that, for the monomial potentials considered in the present work, the scalar spectrum is red, although too close to scale invariant to be compatible with observation. An easy way to solve this problem,  already remarked in  \cite{Green:2009ds}, is to assume that the parameter $\Delta$ it not exactly constant, but it changes for different values of the inflaton field. Assuming $\Delta \propto \varphi^\delta$ leads to the interval for $\delta$ given in eq. (\ref{smalldelta}). The relation $\Delta \left[ M ,\, g \right]$ (or $\Delta \left[ m ,\, g \right]$) should be then understood as the value of $\Delta$ when $\varphi = \varphi_{60}$. The slow variation of $\Delta$ can lead to corrections to some of our other results, which will be of the order of the slow roll parameters. We disregard these small corrections. 

In agreement with the results and discussions of \cite{Green:2009ds,Cook:2011hg,Senatore:2011sp,Barnaby:2012xt} we do not find a significant gravitational wave background from this model. This is due to the fact that the quanta of $\chi$ are non relativistic, and they source scalar perturbations more than tensor ones. The value of the tensor-to-scalar ratio obtained in the model is shown in Figure \ref{fig:bounds+gw}. 

Returning to our bispectrum study, we must conclude that either  the leading operator of~\cite{LopezNacir:2011kk} is 
not present in the full second order equations for the perturbations, or, in order to derive it,  one needs to relax some other approximations that are inherent in the scheme of  \cite{Green:2009ds}. We are not aware of any explicit computational scheme that improves over the one of   \cite{Green:2009ds}, and therefore this remains an interesting open question. We hope to come back to this in some future work.

\vskip.25cm
\noindent{\bf Acknowledgements:} 

We thank Bart Horn and Eva Silverstein for useful comments. The work of L.P. and M.P. is partially supported from the DOE grant DE-SC0011842  at the University of Minnesota. The work of L.S. is partially supported by the NSF grant PHY-1205986.

%%%%%%%%%%%
%%%%%%%%%%%
\appendix%%
%%%%%%%%%%%
%%%%%%%%%%%

%%%%%%%%%%%%%%%%%%%%%%%%%%%%%%%%%%%%%%%%%%%%%%%%%
\section{Derivation of the effective system of equations}

\label{app:equations}
%%%%%%%%%%%%%%%%%%%%%%%%%%%%%%%%%%%%%%%%%%%%%%%%%

In this section we review the derivation of the  equations for the background and for the first  order inflaton perturbations 
obtained in Ref.  \cite{Green:2009ds}. We then extend their scheme to second order in the inflaton perturbations.  In this way we recover the terms considered in  Ref.  \cite{Green:2009ds} in their bispectrum evaluation, plus additional terms. 

We begin with the trapped inflation Lagrangian
\begin{align}
\mathcal{L} = - \dfrac{1}{2} \partial_\mu \varphi \partial^\mu \varphi - V(\varphi) - \sum_i \dfrac{1}{2} \partial_\mu \chi_i \partial^\mu \chi_i - \sum_i \dfrac{g^2}{2} (\varphi - \varphi_{0i})^2 \chi_i^2,
\end{align}
which gives the equations of motion
\begin{align}
0 &= \ddot{\varphi} + 3 H \dot{\varphi} + \dfrac{\partial V}{\partial \varphi} - \dfrac{\nabla^2 \varphi}{a^2} + \sum_i g^2 (\varphi - \varphi_{0i}) \chi_i^2, \nonumber \\
0 &= \ddot{\chi}_i + 3 H \dot{\chi}_i- \dfrac{\nabla^2 \chi_i}{a^2} + g^2 (\varphi - \varphi_{0i})^2 \chi_{i}.
\label{eq:eq_motion}
\end{align}
By adding and subtracting $\sum_i g^2 (\varphi - \varphi_{0i}) \left< \chi_i^2 \right>$, the top equation may be rewritten as 
\begin{align}
0 &= \ddot{\varphi} + 3 H \dot{\varphi} + \dfrac{\partial V}{\partial \varphi}  - \dfrac{\nabla^2  \varphi}{a^2} + \sum_i g^2 (\varphi - \varphi_{0i}) \left( \chi_i^2 - \left< \chi_i^2 \right> \right)  +   {\rm sign} \left( \dot{\varphi} \right) \, \sum_i \, g \, m_{\chi_i}  \left< \chi_i^2 \right> ,
\label{eqphi-tmp}
\end{align}
We then use  (see the discussion after eq. (\ref{sumtoint}))
\begin{align}
n_{\chi_i} & =  m_{\chi_i} \left< \chi_i^2 \right> = \dfrac{|g\,\dot{\varphi}(t_i)|^{3 \slash 2}}{(2\pi)^3} \dfrac{a(t_i)^3}{a(t)^3}
\label{eq:n_chi} \;, 
\end{align}
in (\ref{eqphi-tmp}), to obtain 
\begin{align}
0 &= \ddot{\varphi} + 3 H \dot{\varphi} + \dfrac{\partial V}{\partial \varphi}  - \dfrac{\nabla^2  \varphi}{a^2} + \sum_i g^2 (\varphi - \varphi_{0i}) \left( \chi_i^2 - \left< \chi_i^2 \right> \right) +  \int^t \dot\varphi(\tilde{t})\, \dfrac{g^{5 \slash 2}\,|\dot{\varphi}(\tilde{t})|^{3 \slash 2}}{\Delta(2\pi)^3} \dfrac{a(\tilde{t})^3}{a(t)^3} d\tilde{t} \,. 
\label{eq:sum-to-integral}
\end{align}

In the last term, the sum over species has been converted into an integral, through the replacement  changed the following 
\begin{equation}
\sum_i \rightarrow \int \frac{\left \vert d \varphi \right\vert}{\Delta} \rightarrow \int \frac{1}{\Delta} \, \left\vert \frac{d \varphi}{d {\tilde t}} \right\vert \, d {\tilde t} \,. 
\label{Delta}
\end{equation} 
where the (mass dimension one) quantity $\Delta$ expresses the difference between the values of the inflaton field between two successive episodes of particle production, $\Delta = \vert \varphi_{0,i+1} - \varphi_{0,i} \vert$.  The conversion (\ref{Delta}) requires that particle production is very frequent, so that the integrand does not change appreciably between two successive episodes of particle production. This requires $\left\vert \frac{\ddot{\varphi}}{\dot{\varphi}} \, \delta t  \right\vert \ll 1 $ and $H \, \delta t \ll 1$, where $\Delta t \simeq \left\vert \frac{\Delta}{\dot{\varphi}} \right\vert$ is the time inteval between successive productions. This gives the two conditions 
\begin{equation}
\left\vert \frac{\ddot{\varphi} \, \Delta}{\dot{\varphi}^2} \right\vert \ll 1 \;\;\;,\;\;\; \left\vert \frac{ H \, \Delta}{\dot{\varphi}} \right\vert \ll 1 \;\;. 
\label{sumtoint}
\end{equation}
These two conditions need to hold for eq. (\ref{eq:sum-to-integral}) to be valid. 

In using  (\ref{eq:n_chi}), we have inserted an approximate solution of the second equation in  (\ref{eq:eq_motion}) into the equation for the inflaton. This approximate solution is obtained from  (\ref{n-rho-chi-bck}), which gives the amount of $\chi_i$ quanta produced by the inflaton zero mode, in which we have replaced the background inflaton time derivative entering in that expression with the time derivative of the total inflaton field ($\dot{\varphi}_0 \rightarrow \dot{\varphi}$). 
The production of $\chi_i$ particles is a function of the inflaton field.  Inflaton perturbations perturb the amount of produced $\chi_i$, which in turn backreacts on the evolution of the inflation perturbations. 

Let us now proceed with expanding eq. (\ref{eq:sum-to-integral}) up to second order in the inflaton perturbations. 
 When we do so, we must account for the fact that the inflaton perturbations $\delta \varphi_1$ and $\delta \varphi_2$ impact the time at which $\varphi(t) = \varphi_{0i}$; that is, the time at which there is resonant production of the $\chi_i$ quanta.  In particular, we define 
\begin{equation}
\varphi_0 \left( t_{0i} \right) = \varphi_{0i} \;\;,\;\; 
\varphi \left( t_{0i} + \Delta t \right) =  \varphi_{0i} ,
\end{equation}
where $\Delta t = q\, \Delta t_1 + q^2\, \Delta t_2 + {\rm O} \left( q^3 \right)$ and $\varphi = \varphi_0 \left( t \right) + q \,\delta \varphi_1 \left( t \right) +  q^2\, \delta \varphi_2 \left( t \right) + {\rm O } \left( q^3 \right)$. We have introduced the redundant parameter $q$ for bookkeeping: in the following equations, quantities multiplied by $q^n$ are of $n-$th order in the inflaton perturbations.   From this, we find 
\begin{equation}\label{Delta_t}
\Delta t_1 = - \frac{\delta \varphi_1 \left( t_{0i} \right)}{\dot{\varphi}_0 \left( t_{0i} \right)} \;\;,\;\; 
\Delta t_2 = - \frac{\delta \varphi_2 \left( t_{0i} \right)}{\dot{\varphi}_0 \left( t_{0i} \right)} + \frac{\delta \varphi_1 \left( t_{0i} \right) \, \delta \dot{\varphi}_1 \left( t_{0i} \right)}{ \dot{\varphi}_0^2 \left( t_{0i} \right) } - \frac{ \ddot{\varphi}_0 \left( t_{0i} \right) \, \delta \varphi_1^2 \left( t_{0i} \right) }{2 \, \dot{\varphi}_0^3 \left( t_{0i} \right)} .
\end{equation} 
Taking into account the shift of the time variable, we have
\begin{align}\label{time_shift_2order}
&0 = \ddot{\varphi}_0 + q\,\delta \ddot{\varphi}_1 + q^2\,\delta \ddot{\varphi}_2 + 3\, H\, \dot{\varphi}_0 + 3\, H\, q\,\delta  \dot{\varphi}_1 +  3\, H\,q^2\, \delta\dot{\varphi}_2 + \dfrac{\partial V}{\partial \varphi} \bigg|_{\varphi_0 + q \,\delta \varphi_1 + q^2\, \delta \varphi_2}   \nonumber \\
& - \dfrac{q\, \nabla^2  \delta \varphi_1 + q^2\, \nabla^2 \delta \varphi_2}{a^2} + \sum_i g^2 (\varphi_0 + q \,\delta \varphi_1 + q^2 \,\delta \varphi_2 - \varphi_{0i}) \left( \chi_i^2 - \left< \chi_i^2 \right> \right) \nonumber \\
& + \int^{t-q \,\Delta t_1 - q^2\, \Delta t_2} g^{5 \slash 2} \dfrac{\vert \dot{\varphi}_0(\tilde{t}_0 + q\, \Delta t_1 + q^2\, \Delta t_2) + q\,\delta \dot{ \varphi_1}(\tilde{t}_0 + q \Delta t_1) + q^2\,\delta \dot{\varphi_2}(\tilde{t}_0  )\vert^{3 \slash 2}}{\Delta(2\pi)^3} \nonumber \\
&  \cdot 
(\dot{\varphi}_0(\tilde{t}_0 + q\, \Delta t_1 + q^2\, \Delta t_2) + q\,\delta \dot{ \varphi_1}(\tilde{t}_0 + q \Delta t_1) + q^2\,\delta \dot{\varphi_2}(\tilde{t}_0  )) \dfrac{a(\tilde{t}_0 + q \Delta t_1 + q^2 \Delta t_2)^3}{a(t)^3} d\tilde{t}_0 + {\rm O } \left( q^3 \right) \,. 
\end{align}

Next, we consider the $\left( \chi_i^2 - \left< \chi_i^2 \right> \right)$ term, which when expanded to second order gives

\begin{align}
&\sum_i g^2 (\varphi_0 + q \delta \varphi_1 + q^2 \delta \varphi_2 - \varphi_{0i}) \left( \chi_i^2 - \left< \chi_i^2 \right> \right) \nonumber \\
& \qquad = \sum_i \left[ q  \,  g^2 \left( \varphi_0 - \varphi_{0i} \right)  \, \left( \chi_i^2 - \left\langle \chi_i^2 \right\rangle \right)_1 + q^2 \,  g^2 \delta \varphi_1    \left( \chi_i^2 - \left\langle \chi_i^2 \right\rangle \right)_1 
+ q^2   \, g^2 \left( \varphi_0 - \varphi_{0i} \right)  \,  \left( \chi_i^2 - \left\langle \chi_i^2 \right\rangle \right)_2 
 \right]  + {\rm O } \left( q^3 \right) \,, 
\end{align} 
where  the subscripts $1,2$ on $\left( \chi_i^2 - \left\langle \chi_i^2 \right\rangle \right)_i$ indicate, respectively, the first and second order contributions to $\left( \chi_i^2 - \left\langle \chi_i^2 \right\rangle \right)$.  After expanding the integral in $q$, we single out the $q^0 ,\, q^1 ,\, q^2$ terms, to obtain, respectively,  the equations at zeroth, first, and second order inflaton perturbations:

\begin{equation}
\ddot{\varphi}_0 +3 H \dot{\varphi}_0 + V' + \int^t d {\tilde t}_0 \, \dot{\varphi}_0\,\frac{g^{5/2} \,|\dot{\varphi}_0|^{3/2} \, a^3}{\left( 2 \pi \right)^3 \Delta \,a^3 \left( t \right)}   = 0 \,,
\end{equation}
\begin{eqnarray} 
&& \delta \ddot{\varphi}_1 + 3 H \delta \dot{\varphi}_1 + V'' \, \delta \varphi_1 - \frac{\nabla^2 \delta \varphi_1}{a^2} +   g^2 \left( \varphi_0 - \varphi_{0i} \right)  \, \left( \chi_i^2 - \left\langle \chi_i^2 \right\rangle \right)_1  \nonumber\\ 
&& \quad\quad +  \frac{g^{5/2}}{\left( 2 \pi \right)^3 \, \Delta \, a^3 \left( t \right)} \int^t d {\tilde t}_0 \, a^3 \, |\dot{\varphi}_0|^{3/2} \left\{ \frac{5}{2} \, \delta \dot{ \varphi }_1  - 3 H\, \left( 1 + \frac{5 \ddot{\varphi}_0}{6 H \dot{\varphi}_0} \right) \delta \varphi_1 \right\} 
+ \frac{g^{5/2} |\dot{\varphi}_0|^{3/2} \, \delta \varphi_1}{\left( 2 \pi \right)^3 \, \Delta} = 0  \,, \nonumber\\  
\end{eqnarray} 
\begin{eqnarray} 
&& \delta \ddot{\varphi}_2 + 3 H \delta \dot{\varphi}_2 + V'' \, \delta \varphi_2 + \frac{1}{2} V''' \, \delta \varphi_1^2 - \frac{\nabla^2 \delta \varphi_2}{a^2} \nonumber\\ 
&& \quad\quad +  g^2 \delta \varphi_1    \left( \chi_i^2 - \left\langle \chi_i^2 \right\rangle \right)_1 +    g^2 \left( \varphi_0 - \varphi_{0i} \right)  \,  \left( \chi_i^2 - \left\langle \chi_i^2 \right\rangle \right)_2 \nonumber\\ 
&& \quad\quad +  \frac{g^{5/2}}{\left( 2 \pi \right)^3 \, \Delta \, a^3 \left( t \right)} \int^t d {\tilde t}_0 \, a^3 \, |\dot{\varphi}_0|^{3/2} \Bigg\{ 
\frac{5}{2} \, \delta \dot{ \varphi }_2  - 3 H\, \left( 1 + \frac{5 \ddot{\varphi}_0}{6 H \dot{\varphi}_0} \right) \delta \varphi_2 - \frac{5 \delta \varphi_1 \delta \ddot{\varphi}_1}{2 \dot{\varphi}_0} + \frac{15 \delta \dot{\varphi}_1^2}{8 \dot{\varphi}_0} \nonumber\\ 
&& \quad\quad\quad\quad  \quad\quad\quad\quad  \quad\quad\quad\quad  \quad\quad\quad\quad 
 - \frac{9 H}{2 \dot{\varphi}_0} \left( 1 + \frac{5 \ddot{\varphi}_0}{18 H \dot{\varphi}_0} \right) \delta \varphi_1 \delta \dot{\varphi}_1 \nonumber\\ 
&& \quad\quad\quad\quad  \quad\quad\quad\quad  \quad\quad\quad\quad  \quad\quad\quad\quad 
 + \frac{9 H^2}{2 \dot{\varphi}_0} \left[ 1  + \frac{\dot{H}}{3 H^2} + \frac{4 \ddot{\varphi}_0}{3 H \dot{\varphi}_0} + \frac{5 \ddot{\varphi}_0^2}{36 H^2 \dot{\varphi}_0^2} + \frac{ 5 \dddot{\varphi}_0 }{18 H^2 \dot{\varphi}_0 } \right] \delta \varphi_1^2   \Bigg\} \nonumber\\ 
  && \quad\quad +  \frac{g^{5/2} \, \vert \dot{\varphi}_0 \vert^{3/2}}{\left(2 \pi \right)^3 \Delta} \left[  \delta \varphi_2 + \frac{3}{2 \, \dot{\varphi}_0} \delta \varphi_1 \delta \dot{\varphi}_1 - \frac{3 \, H}{2 \, \dot{\varphi}_0} \left( 1 + \frac{\ddot{\varphi}_0}{2 H \dot{\varphi}_0} \right) \delta \varphi_1^2 \right]  = 0 .
\end{eqnarray} 

The zeroth order equation is identical to equation \eqref{eq:background_integral_eq_motion}, which is solved in section \ref{sec:background}; we now proceed to simplify the first and second order equations.  As discussed in section \ref{sec:background}, the higher derivatives of the background solution are small, which allows us to drop all terms involving $\ddot{\varphi}_0$ and $\dddot{\varphi}_0$; we therefore treat $\dot{\varphi}_0$ as a constant.  Consistently with  \cite{Green:2009ds}, we can also  drop $V^{\prime \prime}$ and $V^{\prime \prime \prime}$; then we have
\begin{eqnarray} 
&& \delta \ddot{\varphi}_1 + 3 H \delta \dot{\varphi}_1 - \frac{\nabla^2 \delta \varphi_1}{a^2}  + \frac{H^2 \mu^2}{ a^3 \left( t \right)}  \int^t d {\tilde t}_0 \, a^3 \,  \left\{ \frac{5 \, \delta \dot{ \varphi }_1 }{2 } - 3 H  \delta \varphi_1 \right\} 
+ H^2 \mu^2  \, \delta \varphi_1 \nonumber\\  
&& \quad\quad\quad\quad =  -   g^2 \left( \varphi_0 - \varphi_{0i} \right)  \,  \left( \chi_i^2 - \left\langle \chi_i^2 \right\rangle \right)_1 \;, 
\end{eqnarray} 
and 
\begin{eqnarray} 
&& \delta \ddot{\varphi}_2 + 3 H \delta \dot{\varphi}_2 - \frac{\nabla^2 \delta \varphi_2}{a^2}  + \frac{H^2 \mu^2}{ a^3 \left( t \right)}  \int^t d {\tilde t}_0 \, a^3 \,  \left\{ \frac{5 \, \delta \dot{ \varphi }_2 }{2 } - 3 H  \delta \varphi_2 \right\} 
+ H^2 \mu^2  \, \delta \varphi_2 \nonumber\\  
&& = -  g^2 \delta \varphi_1    \left( \chi_i^2 - \left\langle \chi_i^2 \right\rangle \right)_1 -   g^2 \left( \varphi_0 - \varphi_{0i} \right)  \,  \left( \chi_i^2 - \left\langle \chi_i^2 \right\rangle \right)_2  - \frac{H^2 \mu^2}{ \dot{\varphi}_0 }  \left[   \frac{3}{2 } \delta \varphi_1 \delta \dot{\varphi}_1 - \frac{3}{2} H  \delta \varphi_1^2 \right]    \nonumber\\      
&& \quad\quad -  \frac{H^2 \mu^2 }{ \dot{\varphi}_0   \, a^3 \left( t \right)} \int^t d {\tilde t}_0 \, a^3 \,  \Bigg\{ 
 - \frac{5 \delta \varphi_1 \delta \ddot{\varphi}_1}{2 } + \frac{15 \delta \dot{\varphi}_1^2}{8 }  - \frac{9 H}{2 }  \delta \varphi_1 \delta \dot{\varphi}_1 
 + \frac{9}{2} H^2 \,   \delta \varphi_1^2   \Bigg\},     
\end{eqnarray} 
where $\mu$ is defined in (\ref{mu-mut}).   After moving to conformal time and using the expansion in momentum modes given in \eqref{eq:mode_expansion}, we find 
\begin{eqnarray} 
&& \partial_\tau^2 \, \delta \varphi_1 + 2 a H  \, \partial_\tau  \delta \varphi_1  + k^2 \, \delta \varphi_1 
+ \frac{7}{2} a^2 H^2 \, \mu^2 \, \delta \varphi_1 
 - \frac{21}{2} \frac{H^3 \mu^2}{ a \left( t \right)}  \int^\tau  d {\tilde \tau}_0 \, a^4 \,  \delta \varphi_1  \nonumber\\ 
&& \quad\quad\quad\quad =  - a^2 \, g^2 \left( \varphi_0 - \varphi_{0i} \right)  \left( \chi_i^2 - \left\langle \chi_i^2 \right\rangle \right)_1 \;, 
\end{eqnarray} 
and 
\begin{eqnarray} 
&& \partial_\tau^2 \, \delta \varphi_2 + 2 a H  \, \partial_\tau  \delta \varphi_2  + k^2 \, \delta \varphi_2 
+ \frac{7}{2} a^2 H^2 \, \mu^2 \, \delta \varphi_2 
 - \frac{21}{2} \frac{H^3 \mu^2}{ a \left( t \right)}  \int^\tau  d {\tilde \tau}_0 \, a^4 \,  \delta \varphi_2  \nonumber\\ 
&& =    -  a^2 \, g^2 \delta \varphi_1   \star \left( \chi_i^2 - \left\langle \chi_i^2 \right\rangle \right)_1 - a^2 \,  g^2  \left( \varphi_0 - \varphi_{0i} \right)   \left( \chi_i^2 - \left\langle \chi_i^2 \right\rangle \right)_2 \nonumber\\ 
&& +  \frac{a H^2 \mu^2}{ \dot{\varphi}_0 }  \delta \varphi_1 \star \partial_\tau  \delta \varphi_1 - \frac{35 \mu^2 H^2}{8 \,  \dot{\varphi}_0  \, a \left( \tau \right) }  \int^\tau d {\tilde \tau} \, a^2 \, \left( \partial_{\tilde \tau } \delta \varphi_1 \right) 
\star  \left( \partial_{\tilde \tau } \delta \varphi_1 \right),   \nonumber\\ 
\end{eqnarray} 
having also integrated by parts and defined the convolution as in eq. (\ref{eq:convolution}). 

It is convenient to define the new variable $ x = - k \, \tau $; we use a prime to denote differentiation with respect to $x$. We also rescale the fields by $\delta \varphi_i \equiv x \, g_i$, which results in 
\begin{eqnarray}
&& g_1'' + \left[ 1 + \frac{\left( {\tilde  \mu}^2 - 2 \right)}{ x^2} \right] g_1  - 3\, {\tilde \mu}^2 \,  \int_x  \frac{ d x' }{ x^{' 3}}  \, g_1 \left( x' \right)  =  - \frac{g^2 \,  \left( \varphi_0 - \varphi_{0i} \right) }{H^2 x^3} \left( \chi_i^2 - \left\langle \chi_i^2 \right\rangle \right)_1 \;, 
\end{eqnarray} 
and 
\begin{align} 
& 
 g_2'' + \left[ 1 + \frac{\left( {\tilde  \mu}^2 - 2 \right)}{ x^2} \right] g_2  - 3\, {\tilde \mu}^2 \,  \int_x  \frac{ d x' }{ x^{' 3}}  \, g_2 \left( x' \right)   =   - \frac{ g^2  \left( \varphi_0 - \varphi_{0i} \right)  }{H^2 x^3 }   \left( \chi_i^2 - \left\langle \chi_i^2 \right\rangle \right)_2 \nonumber\\ 
&  -  \frac{g^2}{2 H^2 x^2}  \left[ \frac{p}{k}   g_1   \star \left( \chi_i^2 - \left\langle \chi_i^2 \right\rangle \right)_1 
+  \left( \chi_i^2 - \left\langle \chi_i^2 \right\rangle \right)_1 \star \frac{\vert \vec{k} - \vec{p} \vert}{k} \,   g_1 \right] 
\nonumber\\ 
&
 -  \frac{ 2 H {\tilde \mu}^2}{7  \dot{\varphi}_0 } \frac{p}{k} \, \frac{\vert \vec{k} - \vec{p} \vert}{k} \, 
\left( \frac{g_1 \star g_1'+g_1' \star g}{2} - \frac{27}{8} \frac{g_1 \star g_1}{x} \right) 
 - \frac{5 \, H \, {\tilde \mu}^2  }{4  \,  \dot{\varphi}_0    }   \frac{p}{k} \, \frac{\vert \vec{k} - \vec{p} \vert}{k} \, 
 \int_x  d x' \, \left( g_1' \star g_1'  + \frac{2}{x^{' 2}} \, g_1 \star g_1  \right)   \;, 
\end{align} 
after another integration by parts, and where we have also rescaled $\mu^2 = \frac{2}{7 } \, {\tilde \mu}^2$.  This gives equations \eqref{eq:system-2nd} (with the definitions in \eqref{eq:sources}) in section \ref{sec:equations} of the main text.

%%%%%%%%%%%%%%%%%%%%%%%%%%%%%%
\section{Explicit background solutions}
\label{app:P_inflaton}
%%%%%%%%%%%%%%%%%%%%%%%%%%%%%%

In this section, we apply the background solution (\ref{phidot}) and the constraint~\eqref{PS-res} on the normalization of the power spectrum to specific models of inflation. In particular, in the two subsections below we study the case of a linear and a quadratic monomial inflaton potential, respectively.  In this Appendix, to simplify the notation, $\varphi_0 > 0$ is assumed. 

\subsection{Linear Inflaton Potential}

We first consider the linear inflaton potential,
\begin{equation}
V(\varphi) = M^3 \, \varphi \, , 
\end{equation}
resulting in the Hubble rate 
\begin{equation}
H \simeq   \frac{V^{1/2}}{\sqrt{3} \, M_p}  \simeq \frac{M^{3/2} \, \varphi_0^{1/2}}{\sqrt{3} \, M_p} \,. 
\end{equation}
(We assume $\varphi_0 >0$ during inflation. We also assume that the potential is modified near the origin so to have a stable minimum at $\varphi_0 =0$. We assume that the modification takes place at values attained after inflation.) Using eqs. (\ref{phidot}) and (\ref{mu-mut}) we then obtain 
\begin{equation}
\dot{\varphi}_0  \simeq - 11.3 \frac{M^{9/5} \Delta^{2/5} \varphi_0^{1/5}}{g \, M_p^{2/5}}  \;\;\;,\;\;\; 
\mu = \frac{0.678 g^{1/2} M_p^{7/10}}{M^{3/20} \Delta^{1/5} \varphi_0^{7/20}} \;. 
\label{eq:mu_A}
\end{equation}

We note that our analysis has three independent parameters, $M$ (which describes the inflaton potential), $g$ (which describes the coupling to the $\chi$ field), and the spacing $\Delta$ between the value of the inflaton at two successive instances of particle production. We can relate one of these parameters to the other two by imposing that the normalization of the power spectrum (\ref{PS-res}) agrees with the measured value 
 ${\cal P}_\zeta \simeq 2.2 \cdot 10^{-9}$ \cite{Ade:2015lrj}:   
\begin{equation}
{\cal P} \simeq 0.000179 \, \frac{g^{5/2} M^{21/20} \varphi_0^{9/20}}{\Delta^{3/5} M_p^{9/10}} \;\;\;\Rightarrow\;\;\; 
\Delta\simeq 1.53 \cdot 10^{8} \, \frac{g^{25/6} M^{7/4} \varphi_N^{3/4}}{M_p^{3/2}} \;,  
\label{PA}
\end{equation}
and, combining the last two equations, 
\begin{equation}
\dot{\varphi}_0 = - 21,200 \frac{g^{2/3} M^{5/2} \varphi_0^{1/5} \varphi_N^{3/10}}{M_p} \;\;,\;\; 
\mu \simeq \frac{0.0156}{g^{1/3} } \, \frac{M_p}{ M^{1/2} \varphi_0^{7/20} \varphi_N^{3/20}} \;. 
\end{equation}

In these expression, $\varphi_N$ is the value assumed by the inflaton $\varphi_0$ when the CMB modes exited the horizon. We relate $\varphi_N$ to the number of e-folds $N \equiv   \ln \left( \frac{a_{\rm end}}{a} \right)$ (where  $a_{\rm end}$ is the value assumed by the scale factor $a$ at the end of inflation) through 
\begin{equation}
N = \int _{\varphi_{\rm end}}^{\varphi_N} \frac{H \, d \varphi_0}{-\dot{\varphi}_0} \simeq \frac{0.0000209  }{g^{2/3}} \, \frac{\varphi_N}{M} 
\;\;\Rightarrow \;\; \varphi_{N} \simeq 4.78 \cdot 10^4 \, g^{2/3} N \, M \,.
\end{equation}
We verified that $\Delta \ll \varphi_N$ in the validity region of the parameters. 

\subsection{Quadratic Inflaton Potential}

Next we consider a quadratic inflaton potential,
\begin{equation}
V(\varphi) = \frac{m^2}{2} \, \varphi^2 \,. 
\end{equation}
Proceeding exactly as in the previous subsection, the normalization of the power spectrum fixes 
\begin{equation}
\Delta \simeq 9.11 \cdot 10^{7} \, \frac{g^{25/6} m^{7/6} \varphi_N^{4/3}}{M_p^{3/2}} \;, 
\end{equation}
and we then find 
\begin{equation}
\dot{\varphi}_0 = - 15,000 \frac{g^{2/3} m^{5/3} \varphi_0^{4/5} \varphi_N^{8/15}}{M_p} \;\;,\;\; 
\mu \simeq \frac{0.0221 }{g^{1/3} } \, \frac{M_p}{ m^{1/3} \varphi_0^{2/5} \varphi_N^{4/15}} \;,
\end{equation}
with 
\begin{equation}
\varphi_{N} \simeq 9.28 \cdot 10^6 \,  g \, N^{3/2} \, m \; .
\end{equation}
We verified that $\Delta \ll \varphi_N$ in the validity region of the parameters. 

%%%%%%%%%%%%%%%%%%%%%%%%%%%%%%
\section{Sourced fields $\chi_i$}
\label{app:chi}
%%%%%%%%%%%%%%%%%%%%%%%%%%%%%%

In this Appendix we focus our attention to a single field $\chi_i$, whose Lagrangian is given by 
\begin{equation}
{\cal L}_i = - \frac{1}{2} \, \partial_\mu \, \chi_i \, \partial^\mu \, \chi_i  - \frac{g^2}{2} \left( \varphi - \varphi_{0i} \right)^2 \chi_i^2 \;. 
\end{equation} 
The classical value of the inflaton provides an effective time-evolving mass for $\chi_i$. By construction, $\varphi_{0i}$ is a value that is assumed by the inflaton zero mode during inflation. When the inflaton is equal to this value, the field $\chi_i$ is instantaneously massless. Around this moment, the frequency of $\chi_i$ varies non-adiabatically, and quanta of $\chi_i$ are non-perturbatively produced. 

To evaluate the production, we quantize the field $\chi_i$ as we did for the inflaton, 
\begin{equation}
\chi_i \left( \tau ,\, \vec{x} \right) =  \int \frac{d^3 k}{\left( 2 \pi \right)^{3/2} } \; {\rm e}^{i \vec{k} \cdot \vec{x}} 
\left( \hat{a}_i \left( \vec{k} \right) \chi_{i,k} \left( \tau \right)  + \hat{a}_i^\dagger \left( -\vec{k} \right) \, \chi_{i,k}^* \left( \tau \right) \right) \;, 
\label{chi-quantum}
\end{equation} 
where $ \hat{a}_i $ destroys one quantum of $\chi_i$. We rewrite the mode functions in terms of  Bogolyubov coefficients, as 
\begin{equation}
 \chi_{ki} \left( \tau \right) = \frac{1}{a \left( \tau \right) \, \sqrt{2 \, \omega_i \left( \tau \right)}} \left[ 
 \alpha_i \left( \tau, k \right) \, {\rm e}^{-i \int^\tau d \tau' \omega_i \left( \tau' \right) } + 
  \beta_i \left( \tau, k \right) \, {\rm e}^{i \int^\tau d \tau' \omega_i \left( \tau' \right) } \right] \;, 
\label{boloyubov}
\end{equation} 
where consistency of the quantization enforces $\vert \alpha_i \left( k \right) \vert^2 -  \vert \beta_i \left( k \right) \vert^2 = 1$, and where $\omega_i$ is the comoving frequency after canonical normalization, 
\begin{equation}
\omega_i = \sqrt{k^2 + g^2 a^2  \left( \varphi_0 \left( \tau \right) - \varphi_{0i} \right)^2 - \frac{a''}{a}} \;.
\end{equation} 
(The contribution from the last factor is negligible, and it is disregarded in the following.) Only the inflaton background value is retained in these expressions. The produced quanta of $\chi_i$ act as sources for the inflaton perturbations. Including first order inflaton perturbations in the computations performed in this Appendix modifies the mode functions obtained here.  We discuss this in Appendix~\ref{app:equations}. 

In the asymptotic past ($a \rightarrow 0$) the modes are in the adiabatic vacuum state, $\beta_i = 0$. The vacuum state remains a good approximation as long as $\varphi \left( t \right) \neq  \varphi_{0i} $, as the frequency varies adiabatically in this regime. As the inflaton crosses this value during inflation, a burst of production of quanta of $\chi_i$ takes place, and the Bogolyubov coefficients have a rapid transition from the vacuum value to \cite{Kofman:1997yn} 
\begin{eqnarray}
\alpha_i \left( \tau > \tau_{0i} ,\, k \right) = \sqrt{1+ {\rm e}^{-\pi \, \kappa_i^2}} \, {\rm e}^{i \, \alpha_{\kappa i}} \;\;,\;\; 
\beta_i \left( \tau > \tau_{0i} ,\, k \right) = {\rm e}^{- \frac{\pi}{2} \, \kappa_i^2} \;\;, 
\label{alpha-beta}
\end{eqnarray} 
where $\tau_{0i}$ is the (conformal) time at which $\varphi = \varphi_{0i}$, and where 
\begin{eqnarray}
\alpha_{\kappa i} &\equiv& {\rm Arg} \left[ \Gamma \left( \frac{1+i \kappa_i^2}{2} \right) \right] + \frac{\kappa_i^2}{2} \, \left( 1 - \log \frac{\kappa_i^2}{2} \right) \;, \nonumber\\ 
\kappa_i &\equiv& \frac{ k}{a \left( \tau_{0i} \right) \, \sqrt{g \,|\dot{\varphi}_{0i}|}} \;, 
\label{alpha-beta2}
\end{eqnarray}
and, finally $\dot{\varphi}_{0i}$ is $\dot{\varphi}_0$ evaluated at $\tau_{0i}$. The production takes place on a (physical) timescale $\delta t = {\rm O } \left( \frac{1}{ \sqrt{g \, \dot{\varphi}_{0i}} } \right)$, after which the constant values (\ref{alpha-beta}) are obtained. The results  (\ref{alpha-beta}) are obtained by disregarding the expansion of the universe in this time interval, and therefore they are valid under the assumption that 
\begin{equation}
H \, \delta t \simeq  \frac{H}{ \sqrt{g \, |\dot{\varphi}_{0i}|}} \ll 1 \,. 
\end{equation} 

The amplification encoded in (\ref{alpha-beta}) does not change the gaussian statistics obeyed by (\ref{chi-quantum}). This allows us to express 
 $n-$point correlation functions $\langle \chi^n \rangle$ in terms of products of  two-point correlation functions. The latter need to be regularized. 
The decomposition (\ref{boloyubov}) can be rephrased as a rotation in the space of annihilation / creation operators, from the original operators appearing in (\ref{chi-quantum}) to a pair of time-dependent operators (that diagonalize the free hamiltonian of $\chi_i$ at any time). The regularization is performed by normal ordering with respect to these time dependent annihilation operators, and leads to (see Section III of \cite{Barnaby:2012xt} for details) 
\begin{eqnarray}
 \left\langle :  \chi_{i,\vec{p}_1} \left( \tau_1 \right)  \,  \chi_{j,\vec{p}_2  } \left( \tau_2 \right) : \right\rangle &=& 
\delta_{ij} \, \frac{  \theta \left( \tau_1 - \tau_{0i} \right)  \theta \left( \tau_2 - \tau_{0i} \right) 
\; \delta^{(3)} \left( \vec{p}_1 + \vec{p}_2  \right)}{2 \, a \left( \tau_1 \right) \,  a \left( \tau_2 \right)  \sqrt{ \omega \left( \tau_1 \right)  \omega \left( \tau_2 \right) }}  \nonumber\\
&& \quad \times \Bigg[ \vert  \beta_i \vert^2 \, 
\Phi_i \left( \tau_1 \right) \, \Phi_i^* \left( \tau_2 \right)  +  \alpha_i \,  \beta_i^* \, 
\Phi_i^* \left( \tau_1 \right) \, \Phi_i^* \left( \tau_2 \right)  + {\rm c. \, c.}  \Bigg]_{p_1} \;,  \nonumber\\ 
\label{eq:chi_chi}
\end{eqnarray}
where the $\delta _{ij}$ term indicates that different ($i \neq j$) species $\chi_i$ and $\chi_j$ are uncorrelated, and 
where we have defined 
\begin{equation}
\Phi_i \left( \tau \right) \equiv {\rm e}^{  i \int_{\tau_{0i}}^{\tau} \,  d  \tau' \, \omega_i \left(  \tau' \right) } \;. 
\label{phase}
\end{equation}

We can now compute the background energy density in the $\chi_i$ field. The quanta of this field are produced at $\tau = \tau_{0i}$, and they rapidly become non-relativistic, namely $\omega_i \simeq  g \,  a \left( \tau \right) \vert \varphi_0 \left( \tau \right) - \varphi_{0i} \vert$. The phase-dependent terms in (\ref{eq:chi_chi}) exhibit very fast oscillations, and average to zero. Therefore, we can write 
\begin{equation}
\left\langle: \chi_i \left( \tau ,\, \vec{x} \right) \,  \chi_i \left( \tau ,\, \vec{x} \right) : \right\rangle \simeq \frac{\theta \left( \tau - \tau_{0i} \right)}{ a^2 \left( \tau \right) \, \omega_i \left( \tau \right)} \int \frac{d^3 p}{\left( 2 \pi \right)^3} \, \vert \beta_i \left( p \right) \vert^2 
= \frac{\theta \left( \tau - \tau_{0i} \right) g^{1/2} \, \vert \dot{\varphi}_{0i} \vert^{3/2}}{ \left( 2 \pi \right)^3 \, \vert \varphi_0 \left( \tau \right) - \varphi_{0i} \vert} \, \frac{a \left( \tau_{0i} \right)^3}{a \left( \tau \right)^3} \;,  
\label{chi-chi-bck}
\end{equation}
which corresponds to the total number density and background energy density in all $\chi_i$ quanta 
\begin{eqnarray}
&& n_\chi = \sum_i n_{\chi,i} = \sum_i \frac{\omega_i}{a} \left\langle: \chi_i \left( \tau ,\, \vec{x} \right) \,  \chi_i \left( \tau ,\, \vec{x} \right) : \right\rangle 
= \sum_i \frac{\theta \left( \tau - \tau_{0i} \right) g^{3/2} \, \vert \dot{\varphi}_{0i} \vert^{3/2}}{ \left( 2 \pi \right)^3 } \, \frac{a \left( \tau_{0i} \right)^3}{a \left( \tau \right)^3} \;,  \nonumber\\ 
&& \rho_\chi \vert_{\rm background} = \sum_i \frac{\omega_i}{a} \, n_{\chi,i} \simeq  \sum_i \frac{\theta \left( \tau - \tau_{0i} \right) g^{5/2} \, \vert \dot{\varphi}_{0i} \vert^{3/2} \vert  \varphi_0 \left( \tau \right) - \varphi_{0i} \vert}{ \left( 2 \pi \right)^3 } \, \frac{a \left( \tau_{0i} \right)^3}{a \left( \tau \right)^3} \;. 
\label{n-rho-chi-bck}
\end{eqnarray}

The contribution of each $\chi_i$ to the background energy density is maximal at its production, and it then gets diluted away by the expansion of the universe. Therefore, we can approximate $  \vert  \varphi_0 \left( \tau \right) - \varphi_{0i} \vert \simeq \vert \dot{\varphi}_0 \vert \, \left( t - t_{0i} \right)$ where the energy density counts the most. We then replace the sum over the various species with an integral, as done in (\ref{Delta}), obtaining 
\begin{equation}
 \rho_\chi \vert_{\rm background}  \simeq  \int^t d t' \, \frac{1}{\Delta} 
  \frac{ g^{5/2} \, \vert \dot{\varphi}_0 \vert^{7/2}  \, \left( t - t' \right) }{ \left( 2 \pi \right)^3 } \, \frac{a \left( t' \right)^3}{a \left( t \right)^3} 
 \simeq \frac{g^{5/2}  \, \vert \dot{\varphi}_0 \vert^{7/2}  }{9 H^2 \Delta \left( 2 \pi \right)^3}  \;. 
\label{rho-chi}
\end{equation} 
(Namely, at any moment the total energy density in the $\chi_i$ quanta is dominated by the species just produced.)

%%%%%%%%%%%%%%%%%%%%%%%%%%%%%%
\section{Correlators between first order inflaton perturbations}
\label{app:dphi1n}
%%%%%%%%%%%%%%%%%%%%%%%%%%%%%%

In this appendix, we derive the $n-$point correlator between sourced first-order inflaton perturbations,  $\left< \delta \varphi_{1,s}^n \right>$.  
As done above in section \ref{sec:equations} and appendix \ref{app:equations}, we  rescale the inflaton perturbation as $\delta \varphi_{1,s} \equiv x \, g_{1,s}$, where $x = - k \, \tau$, and so we want  to compute 
\begin{align}
\left< \delta \varphi_{1,s}(\vec{k}_1, \tau_1) \dots \delta \varphi_{1,s}(\vec{k}_n, \tau_n) \right> = (-k_1 \tau_1) \dots (-k_n \tau_n) \left\langle g_{1,s} \left( \vec{k}_1 ,\, \tau_1 \right) \dots  g_{1,s} \left( \vec{k}_n ,\, \tau_n \right) \right\rangle \,.  
\label{eq:phi1_to_g1_correlation_function}
\end{align}

Using (\ref{eq:soln_form}), we can write~\footnote{In this and in the next appendix $G$ denotes the Green's function without the theta function.} 
\begin{align} 
g_{1,s} \left( x \right) &= \int_x^\infty G \left( x ,\, x' \right) \partial_{x'} \, s_1 \left( x' \right) 
\equiv \int_{-\infty}^\tau d \tau' \left[ \partial_{\tau'} G \left( - k \, \tau ,\, - k \tau' \right) \right] \sum_i s_{1,i} \left( \vec{k} ,\, \tau' \right) \,,
\label{eq:g1_soln} 
\end{align}  
where $s_1 = \sum_i s_{1,i}$ is the source defined in (\ref{eq:sources}), and the individual $s_{1,i}$ read 
\begin{eqnarray} 
s_{1,i}  \left(  \vec{k} \neq 0 ,\, \tau \right) & = &   \frac{ g^2 \left[ \varphi_0 \left( \tau \right) - \varphi_{0i} \right]  }{H^2 k^3 \tau^3 } \,  \int \frac{d^3 p}{\left( 2 \pi \right)^{3/2} } \,   \chi_{i,\vec{p}} \left( \tau \right) \,   \chi_{i,\vec{k} - \vec{p}   }    \left( \tau \right) \nonumber\\ 
 & = &  - \frac{ g \, \omega_i \left( \tau \right)  {\rm sign} \left( \dot{\varphi}_0 \right)  }{H k^3 \tau^2 } \,  \int \frac{d^3 p}{\left( 2 \pi \right)^{3/2} } \,   \chi_{i,\vec{p}} \left( \tau \right) \,   \chi_{i,\vec{k} - \vec{p}   }    \left( \tau \right)  \,, 
 \label{eq:s_1i}
\end{eqnarray} 
where  $\omega_i \left( \tau \right) \simeq a \left( \tau \right) m_{\chi_i} \left( \tau \right) = \frac{g \, \left \vert \varphi_0 \left( \tau \right) - \varphi_{0i} \right \vert }{- H \tau}$.  
One can disregard the momentum of the produced quanta of $\chi_i$ as compared to their inflaton-dependent mass. This is not true at the exact moment these quanta are produced, but in this computation we disregard this very brief transient period during which the quanta of $\chi_i$ are being produced and their frequency is varying non-adiabatically, as is commonly done. 

We therefore want to compute 
\begin{align} 
&\left\langle g_1 \left( \vec{k}_1 ,\, \tau_1 \right) \dots  g_1 \left( \vec{k}_n ,\, \tau_n \right) \right\rangle \nonumber \\
&\qquad = \int_{-\infty}^{\tau_1} d \tau'_1 \left[ \partial_{\tau'_1} G \left( - k_1 \, \tau_1 ,\, - k_1 \, \tau'_1 \right) \right] \dots 
 \int_{-\infty}^{\tau_n} d \tau'_n \left[ \partial_{\tau'_n} G \left( - k_n \, \tau_n ,\, - k_n \, \tau'_n \right) \right] \times \mathcal{S}_n \;, 
\label{g1n-par} 
\end{align} 
where
\begin{align} 
{\cal S}_n &\equiv \sum_{i_1 ,\, \dots ,\, i_n} \, \left\langle   : s_{1,i_1} \left( \vec{k}_1 ,\, \tau_1' \right) \dots   s_{1,i_n} \left( \vec{k}_n ,\, \tau_n' \right) : \right \rangle 
=  \sum_{i_1 ,\, \dots ,\, i_n} \,  \frac{\left( - g \,  {\rm sign} \left( \dot{\varphi}_0 \right)  \right)^n \omega_{i_1} \left( \tau_1' \right) \dots  \omega_{i_n} \left( \tau_1' \right) }{H^n k_1^3 \dots k_n^3 
\tau_1^{' 2} \dots \tau_n^{' 2} } \nonumber \\
& \qquad \times \int \frac{d^3 p_1 \dots d^3 p_n}{\left( 2 \pi \right)^{3 n/2} } 
\left\langle :   \chi_{i_1,\vec{p}_1} \left( \tau_1' \right) \,  \chi_{i_1,\vec{k}_1 - \vec{p}_1   } \left( \tau_1' \right)  \dots   \chi_{i_n,\vec{p}_n} \left( \tau_n' \right) \,  \chi_{i_n,\vec{k}_n - \vec{p}_n   } \left( \tau_n' \right) : \right\rangle \nonumber\\ 
&\equiv {\cal S}_{n,1PI} +   {\cal S}_{n,1PR}  \,.
\label{Sc-Sd}
\end{align} 
In the last line we have remarked that the  correlator in the final expression receives 1-particle irreducible (1PI) and 1-particle reducible (1PR) contributions.  By 1PI contributions, we mean diagrams in which a unique species runs in a loop, $i_1 = i_2 = \dots = i_n$. The 1PR contributions are instead characterized by more than one loop, in which different species are running. These 1PR terms do not contribute to  $\langle g_1^n \left( k_i \right) \rangle$ (where $k_i$ are all external momenta), and so we disregard them in this appendix. 

Using the fact the $\chi_i$ is gaussian, we can write, for the 1PI contribution, 
\begin{eqnarray} 
{\cal S}_{n,1PI} \equiv && 2^{n-1} \left( n - 1 \right) ! \, \sum_{i_1 ,\, \dots ,\, i_n} \,  \frac{\left( - g \,  {\rm sign} \left( \dot{\varphi}_0 \right)  \right)^n \omega_{i_1} \left( \tau_1' \right) \dots  \omega_{i_n} \left( \tau_1' \right) }{H^n k_1^3 \dots k_n^3 
\tau_1^{' 2} \dots \tau_n^{' 2} } \,   \int \frac{d^3 p_1 \dots d^3 p_n}{\left( 2 \pi \right)^{3 n/2} } \nonumber\\ 
&& \!\!\!\!\!\!\!\!  \!\!\!\!\!\!\!\!  \!\!\!\!\!\!\!\!  \times \, 
 \left\langle :   \chi_{i_1,\vec{p}_1} \left( \tau_1' \right) \,  \chi_{i_2,\vec{k}_2 - \vec{p}_2} \left( \tau_2' \right) : \right\rangle \, \left\langle :   \chi_{i_2,\vec{p}_2} \left( \tau_2' \right) \,  \chi_{i_3,\vec{k}_3 - \vec{p}_3} \left( \tau_3' \right) : \right\rangle \, \dots \,  \left\langle :   \chi_{i_n,\vec{p}_n} \left( \tau_n' \right) \,  \chi_{i_1,\vec{k}_1 - \vec{p}_1} \left( \tau_1' \right) : \right\rangle,  \nonumber\\ 
 \label{eq:Sn_1}
\end{eqnarray} 
where the factor $2^{n-1} (n-1)!$ accounts for all the permutations that give the same result as the term singled out in the last expression, and that all equally contribute to ${\cal S}_n$. 

We then use the result (\ref{eq:chi_chi}) for the two-point correlation functions. After accounting for the $\delta$-functions, we are left with a single independent internal momentum within the integral.  To simplify the computation we assume that this internal  momentum is much greater than the one of the particles in the correlator. With this assumption, which we make explicit in eq.~\eqref{C8-in} below, we obtain 
\begin{eqnarray} 
{\cal S}_{n,1PI} &=& \sum_i \frac{ \left( n - 1 \right)! \, \left( - g \,  {\rm sign} \left( \dot{\varphi}_0 \right)  \right)^n \, H^n}{2 \; k_1^3 \dots k_n^3} \delta^{(3)} \left( \vec{k}_1 + \dots + \vec{k}_n \right)  \, \theta \left(  \tau_1' - \tau_{0i} \right) \dots  \theta \left(  \tau_n' - \tau_{0i} \right) \nonumber\\ 
&&  \quad\quad\quad\quad \times \, \int \frac{d^3 p}{\left( 2 \pi \right)^{3 n/2}} \, f_n \left( p ;\, \tau_i \right) \;, \nonumber\\ 
\end{eqnarray}
where
\begin{eqnarray}  
 f_n \left( p \right) & \equiv  &  \Bigg[ \vert  \beta_i \vert^2 \, 
\Phi_i \left( \tau_1' \right) \, \Phi_i^* \left( \tau_2' \right)  +  \alpha_i \,  \beta_i^* \, 
\Phi_i^* \left( \tau_1' \right) \, \Phi_i^* \left( \tau_2' \right)  + {\rm c. \, c.}  \Bigg]_{p} \nonumber\\ 
&& \times \Bigg[ \vert  \beta_i \vert^2 \, 
\Phi_i \left( \tau_2' \right) \, \Phi_i^* \left( \tau_3' \right)  +  \alpha_i \,  \beta_i^* \, 
\Phi_i^* \left( \tau_2' \right) \, \Phi_i^* \left( \tau_3' \right)  + {\rm c. \, c.}  \Bigg]_{p} \nonumber\\ 
&& \times \dots \times \Bigg[ \vert  \beta_i \vert^2 \, 
\Phi_i \left( \tau_n' \right) \, \Phi_i^* \left( \tau_1' \right)  +  \alpha_i \,  \beta_i^* \, 
\Phi_i^* \left( \tau_n' \right) \, \Phi_i^* \left( \tau_1' \right)  + {\rm c. \, c.}  \Bigg]_{p}. 
\end{eqnarray} 

The phases $\Phi_i$ are rapidly oscillating due to the fact that the quanta of $\chi_i$ are very massive as soon as $\varphi_0$ moves past $\varphi_{0i}$. Therefore the integral is dominated by the terms in this product for which all $\Phi_i$ factors cancel. An immediate evaluation of the products in the $n=2,3,4$ cases gives 
\begin{eqnarray}
f_2 \left( p \right) & = & 4 \vert \beta \left( p \right) \vert^4 + 2 \vert \beta \left( p \right) \vert^2 \;, \nonumber\\ 
f_3 \left( p \right) & = & 8 \vert \beta \left( p \right) \vert^6 + 6 \vert \beta \left( p \right) \vert^4 \;, \nonumber\\ 
f_4 \left( p \right) & = & 16 \vert \beta \left( p \right) \vert^8 + 16 \vert \beta \left( p \right) \vert^6  + 2 \vert \beta \left( p \right) \vert^4 \;,  
\label{f2f3f4}
\end{eqnarray}
where $\vert \alpha \vert^2 = 1 + \vert \beta \vert^2$ has been used. Higher $f_n$ can be evaluated in an equally straightforward manner. 

We then use (\ref{alpha-beta}) and evaluate the momentum integral, to obtain 
\begin{equation} 
{\cal S}_{n,1PI} = \sum_i \delta^{(3)} \left( \vec{k}_1 + \dots + \vec{k}_n \right)  \, \theta \left(  \tau_1' - \tau_{0i} \right) \dots  \theta \left(  \tau_n' - \tau_{0i} \right) \, \frac{1}{k_1^3 \dots k_n^3} \, {\cal C}_n 
\frac{ \left( -   {\rm sign } \left( \dot{\varphi}_0 \right)  \right)^n \, g^n \, H^n }{\pi^{3 n / 2}} \, \frac{g^{3/2} \, |\dot{\varphi}_{0}|^{3/2}}{H^3 \left( - \tau_{0i} \right)^3} \,, 
\label{Sn-result}
\end{equation}
with
\begin{alignat}{2}
{\cal C}_2 = \frac{2 + \sqrt{2}}{16} \;\;,\;\; 
{\cal C}_3 = \frac{27 \sqrt{2} + 16 \sqrt{3} }{288 \sqrt{2}} \;\;,\;\; 
{\cal C}_4 &= \frac{36 + 9 \sqrt{2} + 32 \sqrt{3} }{384 } \;. 
\label{C2-C3-C4}
\end{alignat} 

Inserting this in (\ref{g1n-par}), and  replacing the sum with an integral as outlined in \ref{app:equations}, we obtain 
\begin{eqnarray} 
\left\langle g_{1s} \left( \vec{k}_1 ,\, \tau_1 \right) \dots  g_{1s} \left( \vec{k}_n ,\, \tau_n \right) \right\rangle_{1PI} 
& = & {\cal C}_n \,  \frac{ \delta^{(3)} \left( \vec{k}_1 + \dots + \vec{k}_n \right) }{ k_1^3 \dots k_n^3 } 
 \frac{  g^n \, H^n }{\pi^{3 n / 2}} \,  \frac{g^{3/2} \, |\dot{\varphi}_0|^{3/2}}{H^3 } \, \frac{\left( {\rm sign } \left( \dot{\varphi}_0 \right)  \right)^n  \, \vert \dot{\varphi}_0  \vert }{H \, \Delta} \, \nonumber\\ 
& & \!\!\!\!\!\!\!\!  \!\!\!\!\!\!\!\!  \!\!\!\!\!\!\!\!  \!\!\!\!\!\!\!\!  \!\!\!\!\!\!\!\!   \times \int_{ - \frac{1}{H \, {\rm Max } \left( k_1 ,\, \dots ,\, k_n \right)  } \, \sqrt{\frac{g \, \vert \dot{\varphi}_0 \vert}{n \pi}} }^{{\rm min} \left( \tau_1 ,\, \dots ,\, \tau_n \right)} \frac{d \tau_0}{\left( - \tau_0 \right)^4} \, 
G \left( - k_1 \, \tau_1 ,\, - k_1 \, \tau_0 \right) \dots  G \left( - k_n \, \tau_n ,\, - k_n \, \tau_0 \right) \,, \nonumber\\ 
\label{g1n-general-time} 
\end{eqnarray} 
where the suffix $1PI$ indicates that this is only the term coming from the 1PI contribution to eq. \eqref{Sc-Sd}, and where the lower extremum of integration is due to the requirement that the internal momentum $p$ is greater than the external ones. The value of the typical internal momentum can be obtained from eqs. (\ref{alpha-beta}) and  (\ref{f2f3f4}), leading to 
\begin{equation}
k_1 ,\, k_2 ,\, \dots ,\, k_n  \ll a \left( \tau_{0i} \right) \, \sqrt{\frac{g \, \vert \dot{\varphi}_{0i} \vert}{n \, \pi}} \,. 
\label{C8-in}
\end{equation}
When we evaluate the integral in (\ref{g1n-general-time})  we need to verify that the integrand is peaked at times after this lower extremum. This is the origin of the condition $C_8$ in Table \ref{tab:conditions}.

\section{Second Order Perturbations in the scheme of  \cite{Green:2009ds}}
\label{ap:second_order}

In this appendix, we consider the sourced second-order fluctuation in the inflaton field, with the aim of finding its contribution to the bispectrum.  We formally write the second-order fluctuation, and the corresponding contribution to the bispectrum, as 
\begin{equation}
g_{2s} \equiv g_{2s,I} + g_{2s,II} \;\; \Rightarrow \;\;  \left\langle g_{1s} g_{1s} g_{2s} \right\rangle \equiv  \left\langle g_{1s} g_{1s} g_{2s} \right\rangle_{I}   +   \left\langle g_{1s} g_{1s} g_{2s} \right\rangle_{II}   \,.   
\label{g2-I-II}
\end{equation}
where $I$ and $II$ denote,  respectively, the contribution from the first and second line in the source term of eq.   \eqref{eq:2nd_order_solution_gen} for the second-order perturbation. We compute these contributions separately in Subsections \ref{ap:line-I} and \ref{ap:line-II}.

\subsection{$g_1 \, g_1 \, \rightarrow g_2$ contributions (Line I)}
\label{ap:line-I}

In this subsection we compute the contribution from the terms in the first line of 
eq. \eqref{eq:2nd_order_solution_gen} to the second order perturbation. This is formally given by 
\begin{align} 
&g_{2s,I} \left( \vec{k}_3 ,\, \tau \right) =  
   \frac{  H {\tilde \mu}^2}{28 \,  \dot{\varphi}_0  \, k_3} \int \frac{d^3 p}{\left( 2 \pi \right)^{3/2}}  \frac{p \, \vert \vec{k}_3 - \vec{p} \vert }{k_3^2} \,  \int_{-\infty}^\tau d \tau' G \left( - k_3 \, \tau ,\, - k_3 \, \tau'  \right) \nonumber\\ 
& \qquad 
 \times \; 
\left[ - 4 \, \partial_{\tau'}^2  - 4 \, \partial_{\tau''}^2 + 27 \, \partial_{\tau'} \,  \partial_{\tau''} + \frac{27}{\tau'} \, \partial_{\tau'} 
 + \frac{27}{\tau'} \, \partial_{\tau''} + \frac{43}{\tau^{' 2}} \right] g_1 \left( \vec{p} ,\, \tau' \right)  g_1 \left( \vec{k}_3 - \vec{p} ,\, \tau'' \right) 
\Big\vert_{\tau'' = \tau'} \;. 
\label{g2-I}
\end{align} 
which leads to the following contribution to the bispectrum 
\begin{align}
& \!\!\!\!\!\!\!\! 
 \left\langle g_{1s} \left( \vec{k}_1 ,\, \tau \right)  g_{1s} \left( \vec{k}_2 ,\, \tau \right)  g_{2s} \left( \vec{k}_3 ,\, \tau \right)  \right\rangle_{I}   \simeq   
   \frac{  H {\tilde \mu}^2}{28 \,  \dot{\varphi}_0  \, k_3} \int \frac{d^3 p}{\left( 2 \pi \right)^{3/2}}  \, \frac{p \, \vert \vec{k}_3 - \vec{p} \vert }{k_3^2}  \, \int_{-\infty}^\tau d \tau' G \left( - k_3 \, \tau ,\, - k_3 \, \tau'  \right) \nonumber\\ 
& \qquad 
\times \; \left[ - 4 \, \partial_{\tau'}^2  - 4 \, \partial_{\tau''}^2 + 27 \, \partial_{\tau'} \,  \partial_{\tau''} + \frac{27}{\tau'} \, \partial_{\tau'} 
 + \frac{27}{\tau'} \, \partial_{\tau''} + \frac{43}{\tau^{' 2}} \right]   \nonumber\\ 
&  \quad\quad  \quad\quad   \quad\quad  \quad\quad     \quad\quad        
\times \left\langle g_1 \left( \vec{k}_1 ,\, \tau \right)  g_1 \left( \vec{k}_2 ,\, \tau \right)   g_1 \left( \vec{p} ,\, \tau' \right) \, g_1 \left( \vec{k}_3 - \vec{p} ,\, \tau'' \right)  \right\rangle \Big\vert_{\tau'' = \tau'} \;. 
\label{g1g1g2-c+d}
\end{align} 

To evaluate this term, we use the two expressions \eqref{g1n-par} and  \eqref{Sc-Sd}. The correlator appearing in this expressions separates in a 1PR plus 1PI  contribution (see the discussion after eq. \eqref{Sc-Sd}),~\footnote{In version one of the present work the 1PR contribution was disregarded. We thank Bart Horn for emphasizing its relevance in a private communcation.}  and, consequently, we define 
\begin{equation}
 \left\langle g_{1s} g_{1s} g_{2s} \right\rangle_{I} \equiv  \left\langle g_{1s} g_{1s} g_{2s} \right\rangle_{I,1PR}   +   \left\langle g_{1s} g_{1s} g_{2s} \right\rangle_{I,1PI}   \,.   
 \label{g1g1g2-d-and-c}
\end{equation} 
where the 1PR term  $\left\langle g_{1s} g_{1s} g_{2s} \right\rangle_{I,1PR}$ takes contribution from ${\cal S}_{4,1PR}$, and the 1PI term    $\left\langle g_{1s} g_{1s} g_{2s} \right\rangle_{I,1PI}$  from ${\cal S}_{4,1PI}$.  The 1PR and 1PI  contributions are computed in Subsections \ref{ap:second_order-Id} and \ref{ap:second_order-Ic}, respectively. The sum is dominated by the 1PR contribution, and the final result of this contribution is given in eq. \eqref{eq:B112-Id-res}.

\subsubsection{$\left\langle  g_{1s} g_{1s} g_{2s} \right\rangle_{I,1PR}$ contribution} 
\label{ap:second_order-Id}

Let us  compute the 1PR contribution 
\begin{align}
& \left\langle g_{1s} \left( \vec{k}_1 ,\, \tau \right)  g_{1s} \left( \vec{k}_2 ,\, \tau \right)  g_{2s} \left( \vec{k}_3 ,\, \tau \right)  \right\rangle_{I,1PR}   \simeq   
   \frac{  H {\tilde \mu}^2}{28 \,  \dot{\varphi}_0  \, k_3}  \int_{-\infty}^\tau d \tau' G \left( - k_3 \, \tau ,\, - k_3 \, \tau'  \right) \nonumber\\ 
& \qquad 
\times \; \left[ - 4 \, \partial_{\tau'}^2  - 4 \, \partial_{\tau''}^2 + 27 \, \partial_{\tau'} \,  \partial_{\tau''} + \frac{27}{\tau'} \, \partial_{\tau'} 
 + \frac{27}{\tau'} \, \partial_{\tau''} + \frac{43}{\tau^{' 2}} \right]  {\cal D}  \Big\vert_{\tau'' = \tau'} \;, \nonumber\\ 
\end{align}
where we have defined 
\begin{align}
{\cal D} \equiv   \int \frac{d^3 p}{\left( 2 \pi \right)^{3/2}}  
\frac{p \, \vert \vec{k}_3 - \vec{p} \vert}{k_3^2} \, 
 \left\langle g_1 \left( \vec{k}_1 ,\, \tau \right)  g_1 \left( \vec{k}_2 ,\, \tau \right)   g_1 \left( \vec{p} ,\, \tau' \right) \, g_1 \left( \vec{k}_3 - \vec{p} ,\, \tau'' \right)  \right\rangle_{1PR} \;. 
\label{start-disc}
\end{align} 

Using eqs. \eqref{g1n-par} and \eqref{Sc-Sd},  we can write  
\begin{align}
\!\!\!\!\!\!\!\! \!\!\!\!\!\!\!\! \!\!\!\!\!\!\!\! \!\!\!\!\!\!\!\! 
& {\cal D}  \equiv \int  \frac{d^3 p}{\left( 2 \pi \right)^{3/2}} 
 \int_{-\infty}^\tau d \tau_1 \left[ \partial_{\tau_1} G \left( - k_1 \, \tau ,\, - k_1 \, \tau_1  \right) \right] \,  
    \int_{-\infty}^\tau d \tau_2 \left[ \partial_{\tau_2} G \left( - k_2 \, \tau ,\, - k_2 \, \tau_2  \right) \right] \nonumber\\ 
&   \int_{-\infty}^{\tau'} d \tau_3 \left[ \partial_{\tau_3} G \left( - p \, \tau' ,\, - p \, \tau_3  \right) \right] \,  
    \int_{-\infty}^{\tau''} d \tau_4 \left[ \partial_{\tau_4} G \left( - \vert \vec{k}_3 - \vec{p} \vert  \, \tau'' ,\, - \vert \vec{k}_3 - \vec{p} \vert  \, \tau_4  \right) \right] \nonumber\\ 
&  \sum_{a,b,c,d} \frac{g^4 \, \omega_a \left( \tau_1 \right) \, \omega_b \left( \tau_2 \right) \, \omega_c \left( \tau_3 \right) \, \omega_d \left( \tau_4 \right) }{H^4 k_1^3 \, k_2^3 \, k_3^2 \, p^2 \vert \vec{k}_3 - \vec{p} \vert^2 \, \tau_1^2 \, \tau_2^2 \, \tau_3^2 \, \tau_4^2 } 
\; \int \frac{d^3 p_1 \, d^3 p_2 \, d^3 p_3 \, d^3 p_4}{\left( 2 \pi \right)^6}
 \nonumber\\ 
&  \left\langle : 
\chi_{a,\vec{p}_1} \left( \tau_1 \right)  \chi_{a,\vec{k}_1-\vec{p}_1} \left( \tau_1 \right) 
\chi_{b,\vec{p}_2} \left( \tau_2 \right)  \chi_{b,\vec{k}_2-\vec{p}_2} \left( \tau_2 \right) 
\chi_{c,\vec{p}_3} \left( \tau_3 \right)  \chi_{c,\vec{p}-\vec{p}_3} \left( \tau_3 \right) 
\chi_{d,\vec{p}_4} \left( \tau_4 \right)  \chi_{d,\vec{k}_3 - \vec{p}-\vec{p}_4} \left( \tau_4 \right) 
: \right\rangle_{1PR} \,. 
\end{align}

The 1PR contribution is proportional to 
\begin{align}
& \left\langle : \chi_{a,\vec{p}_1} \left( \tau_1 \right)  \chi_{c,\vec{p}-\vec{p}_3} \left( \tau_3 \right) : \right\rangle 
\left\langle : \chi_{a,\vec{k}_1-\vec{p}_1} \left( \tau_1 \right) \chi_{c,\vec{p}_3} \left( \tau_3 \right)  : \right\rangle \nonumber\\ 
& \quad\quad \times 
\left\langle : \chi_{b,\vec{p}_2} \left( \tau_2 \right)  \chi_{d,\vec{k}_3 - \vec{p}-\vec{p}_4} \left( \tau_4 \right) : \right\rangle 
\left\langle : \chi_{b,\vec{k}_2-\vec{p}_2} \left( \tau_2 \right) \chi_{d,\vec{p}_4} \left( \tau_4 \right) :  \right\rangle   
 + 7 \; {\rm equivalent \; contractions} \,. 
\end{align}

We perform the contractions, using eq. \eqref{eq:chi_chi}, and we keep the dominant term in which the fast oscillating phases are not present, which gives   
\begin{align}
& \!\!\!\!\!\!\!\! \!\!\!\!\!\!\!\! 
{\cal D}  = \frac{2 \, H^4 \, g^4}{k_1^5 \, k_2^5 \, k_3^2} \,  \frac{\delta^{(3)} \left( \vec{k}_1 + \vec{k}_2 + \vec{k}_3 \right)}{\left( 2 \pi \right)^{15/2}} \times \nonumber\\ 
& \!\!\!\!\!\!\!\! \!\!\!\!\!\!\!\!   \sum_a
\;  \int_{\tau_{0a}}^\tau d \tau_1 \left[ \partial_{\tau_1} G \left( - k_1 \, \tau ,\, - k_1 \, \tau_1  \right) \right] \;  \int_{\tau_{0a}}^{\tau'} d \tau_3 \left[ \partial_{\tau_3} G \left( - k_1 \, \tau' ,\, - k_1 \, \tau_3  \right) \right]   \; \int d^3 p_3 \; 
\left\{  \vert \beta_a \left( p_3 \right) \vert^2 + 2  \vert \beta_a \left( p_3 \right) \vert^4   \right\} \nonumber\\ 
& \!\!\!\!\!\!\!\! \!\!\!\!\!\!\!\!   \sum_b \int_{\tau_{0b}}^\tau d \tau_2 \left[ \partial_{\tau_2} G \left( - k_2 \, \tau ,\, - k_2 \, \tau_2  \right) \right] 
\int_{\tau_{0b}}^{\tau''} d \tau_4 \left[ \partial_{\tau_4} G \left( - k_2 \, \tau'' ,\, -  k_2  \, \tau_4  \right) \right]  
 \; \int d^3 p_4 \; 
\left\{  \vert \beta_b \left( p_4 \right) \vert^2 +  2 \vert \beta_b \left( p_4 \right) \vert^4    \right\} \,. 
\end{align}
where we have disregarded external momenta in comparison with the internal ones. We see the presence of two  loops in which there are separate sources. 

We perform the momentum integrals using (\ref{alpha-beta}) and (\ref{alpha-beta2}). We then  replace the sums with integrals according to 
(\ref{Delta}).  Finally, we note that, in this approximation, nothing outside $G$ depends on $\tau_{i}$, and so we can perform the $\tau_i$ integrals,  evaluating $G$ at the boundary. We find  
\begin{align}
& {\cal D}  = \frac{128 \, {\cal C}_2^2  \, g^7 \, \vert \dot{\phi}_0 \vert^5}{H^4 \, \Delta^2 k_1^5 \, k_2^5 \, k_3^2} \,  \frac{\delta^{(3)} \left( \vec{k}_1 + \vec{k}_2 + \vec{k}_3 \right)}{\left( 2 \pi \right)^{15/2}} \; 
 \int_{-\infty}^{\tau'}  \frac{  d \tau_{0a}  }{ \left( - \tau_{0a} \right)^4} 
\;    G \left( - k_1 \, \tau ,\, - k_1 \, \tau_{0a}  \right)  \;   G \left( - k_1 \, \tau' ,\, - k_1 \, \tau_{0a}  \right)  \nonumber\\ 
& \quad\quad\quad\quad  \quad\quad\quad\quad  \times \;  \int_{-\infty}^{\tau''}   \frac{  d \tau_{0b}  }{ \left( - \tau_{0b} \right)^4} 
 G \left( - k_2 \, \tau ,\, - k_2 \, \tau_{0b}  \right)   G \left( - k_2 \, \tau'' ,\, -  k_2  \, \tau_{0b}  \right)    \;, 
\end{align}
with ${\cal C}_2$ given in \eqref{C2-C3-C4}. 

We insert this into (\ref{start-disc}), and we consider  the equilateral case $k_1=k_2=k_3 \equiv k$. In terms of the  dimensionless rescaled times 
\begin{equation}
- k \tau \equiv x \;\;,\;\; - k \tau' \equiv x' \;\;,\;\; - k \tau'' \equiv x'' \;\;,\;\; - k \tau_{0a} \equiv x_{0a} \;\;,\;\; - k \tau_{0b} \equiv x_{0b} \;, 
\end{equation}  
we obtain (the suffix on the Green functions  denotes how many derivatives act on each argument) 
\begin{align}
& \left\langle g_1 g_1 g_2 \right\rangle_{\rm I,1PR,equil} 
  \simeq   
 \frac{2^{49/6} \sqrt{\pi} C_2^2}{7^{11/3}} \, \frac{{\tilde \mu}^{22/3} \, g^{1/3} H^{7/3} \Delta^{2/3}}{k^6 }  {\rm sign} \left( \dot{\varphi}_0 \right)  \nonumber\\ 
& \qquad \Bigg\{ -8 \int_x^\infty d x' G \left( x ,\, x'  \right)  \,  \int_{x'}^\infty  \frac{  d x_{0a}  }{ \left(  x_{0a} \right)^4} 
\;    G \left(  x ,\,  x_{0a}  \right)  \;   G^{(2,0)} \left(  x' ,\,  x_{0a}  \right)  
\;  \int_{x'}^\infty   \frac{  d x_{0b}  }{ \left(  x_{0b} \right)^4} 
 G \left(  x ,\,  x_{0b}  \right)   G \left(  x' ,\,  x_{0b}  \right) \nonumber\\ 
& \qquad \;\;   +27  \int_x^\infty d x' G \left( x ,\, x'  \right)  \left[  \int_{x'}^\infty  \frac{  d x_{0a}  }{ \left(  x_{0a} \right)^4} 
\;    G \left(  x ,\,  x_{0a}  \right)  \;   G^{(1,0)} \left(  x' ,\,  x_{0a}  \right)  \right]^2 \nonumber\\ 
& \qquad \;\;   +54  \int_x^\infty d x' \frac{G \left( x ,\, x'  \right)}{x'}  \int_{x'}^\infty  \frac{  d x_{0a}  }{ \left(  x_{0a} \right)^4} 
\;    G \left(  x ,\,  x_{0a}  \right)  \;   G^{(1,0)} \left(  x' ,\,  x_{0a}  \right)   \int_{x'}^\infty  \frac{  d x_{0b}  }{ \left(  x_{0b} \right)^4} 
\;    G \left(  x ,\,  x_{0b}  \right)  \;   G \left(  x' ,\,  x_{0b}  \right)  \nonumber\\ 
& \qquad \;\;   +43  \int_x^\infty d x' \frac{G \left( x ,\, x'  \right)}{x^{'2}}  \left[  \int_{x'}^\infty  \frac{  d x_{0a}  }{ \left(  x_{0a} \right)^4} 
\;    G \left(  x ,\,  x_{0a}  \right)  \;   G \left(  x' ,\,  x_{0a}  \right)  \right]^2 \Bigg\} 
 \end{align} 
where eq. (\ref{mu-mut}) has been used.

We perform one integration by parts to eliminate the $\left(2 ,\, 0 \right)$ terms. We then use eqs. \eqref{g-dphi} and (\ref{eq:xG}) to obtain 
\begin{align}\label{eq:intT0T1}
& \!\!\!\!\!\!\!\!   \!\!\!\!\!\!\!\!  
\left\langle \delta \varphi_1 \delta \varphi_1 \delta \varphi_2 \right\rangle_{\rm I,1PR,equil}'  + 2 {\rm perm.} 
  \simeq   
 \frac{3 \times 2^{49/6} \sqrt{\pi} C_2^2}{7^{11/3}} \, \frac{{\tilde \mu}^{22/3} \, g^{1/3} H^{7/3} \Delta^{2/3}}{k^6 }   {\rm sign} \left( \dot{\varphi}_0 \right) \nonumber\\ 
&    
 \int_x^\infty d x' \left\{  35 \, c_1 \left(  x'  \right)  \, {\cal T}_1^2 \left( x' \right)   +   \left[ \frac{54 \, c_1 \left(  x'  \right)}{x'} + 8 \, c_1' \left(  x'  \right) \right] 
{\cal T}_0 \left( x' \right) \, {\cal T}_1 \left( x' \right)  + 43  \, \frac{c_1 \left(  x'  \right)}{x^{'2}}  {\cal T}_0^2 \left( x' \right) \right\}  \;, 
 \end{align} 
where we have defined 
\begin{align}\label{eq:defT0T1}
& {\cal T}_0 \left( x' \right) \equiv  \int_{x'}^\infty  \frac{  d x_{0a}  }{ \left(  x_{0a} \right)^4}   c_1 \left(    x_{0a}  \right)  \;   G \left(  x' ,\,  x_{0a}  \right)  \;, \nonumber\\ 
& {\cal T}_1 \left( x' \right) \equiv \int_{x'}^\infty  \frac{  d x_{0a}  }{ \left(  x_{0a} \right)^4}   c_1 \left(    x_{0a}  \right)  \;   G^{(1,0)} \left(  x' ,\,  x_{0a}  \right)  \;. 
\end{align}
We perform the integration numerically, and we find that the numerical result is well fitted by 
\begin{align}
& 
 \left\langle \delta \varphi_{1s} \left( \vec{k}_1 ,\, \tau \right)  \delta \varphi_{1s} \left( \vec{k}_2 ,\, \tau \right)  \delta \varphi_{2s} \left( \vec{k}_3 ,\, \tau \right)  \right\rangle_{\rm I,1PR,equil} + 2 \; {\rm permutations} & \nonumber\\ 
& \quad\quad \approx  {\tilde \mu}^{22/3} \, g^{1/3} H^{7/3} \Delta^{2/3} \, 
 \dfrac{\delta(\vec{k}_1 + \vec{k}_2 + \vec{k}_3)}{k^6}   {\rm sign} \left( \dot{\varphi}_0 \right) 
 \times \frac{-0.018}{{\tilde \mu}^4} 
 \approx  -0.0037  \, g^2 H \mu^2  \dot{\varphi}_0   \dfrac{\delta(\vec{k}_1 + \vec{k}_2 + \vec{k}_3)}{k^6}  \;. 
\label{eq:B112-Id-res}
\end{align}

The $\propto {\tilde \mu}^{-4}$ scaling of the integral in (\ref{eq:intT0T1}) can be understood as follows: we verified that the Green functions reach a maximum when their argument is of ${\rm O } \left( {\tilde \mu} \right)$, and that  this maximum value is  ${\tilde \mu}$-independent. We also verified that the function $c_1$ depends on ${\tilde \mu}$ 
(the ${\tilde \mu}$ dependence of $c_1$ was not written explicitly in the expressions so far, to keep the notation simpler) and $x'$ as 
\begin{equation}
c_1 \left( {\tilde \mu} ,\, x' \right) \approx {\tilde \mu} \, f_{c_1} \left( \frac{x'}{{\tilde \mu}} \right) \;, 
\label{c1-scaling}
\end{equation} 
(the scaling is not exact, but it is very accurate in the $x' \simeq {\tilde \mu}$ region which dominates the integrand). As a consequence, $c_1' = \partial_{x'} c_1$ can be approximated by a ${\tilde \mu}-$independent function of $x'/{\tilde \mu}$. It then immediately follows that 
\begin{align} 
& {\cal T}_0 \left( x' ,\, {\tilde \mu} \right) \approx \frac{1}{{\tilde \mu}^2} f_{{\cal T}_0} \left( \frac{x'}{\tilde \mu} \right) \;\;,\;\; 
{\cal T}_1 \left( x' ,\, {\tilde \mu} \right) \simeq \frac{1}{{\tilde \mu}^3} f_{{\cal T}_1} \left( \frac{x'}{\tilde \mu} \right) \;. 
\label{T0T1-scaling}
\end{align}
We then see that the integrand of (\ref{eq:defT0T1}) is approximately a function of $x' / {\tilde \mu}$, with magnitude $\propto {\tilde \mu}^{-5}$ in its maximum, located at $x'  / {\tilde \mu} = {\rm O } \left( 1 \right)$; we further observe that the three terms in the second line of  (\ref{eq:defT0T1}) present the same scaling. Therefore, the integral scales with ${\tilde \mu}$ as ${\tilde \mu}^{-4}$. This is in excellent agreement with the numerical integration that gives the result reported in eq. (\ref{eq:B112-Id-res}).

\subsubsection{$\left\langle  g_{1s} g_{1s} g_{2s} \right\rangle_{I,1PI}$ contribution} 
\label{ap:second_order-Ic}

Inserting the result \eqref{g1n-general-time} in \eqref{g1g1g2-c+d}, we obtain the contribution to the bispectrum from the 1PI ${\cal S}_{4,1PI}$ contraction,     
\begin{align} 
& \!\!\!\!  \!\!\!\! 
 \left\langle g_{1s} \left( \vec{k}_1 ,\, \tau \right)  g_{1s} \left( \vec{k}_2 ,\, \tau \right)  g_{2s} \left( \vec{k}_3 ,\, \tau \right)  \right\rangle_{I,1PI}   \simeq  \,  {\rm sign } \left( \dot{\varphi}_0 \right) \, 
 \frac{\sqrt{2} \, \mathcal{C}_4 g^3 H^3 {\tilde \mu}^4}{49 \pi^{9/2}} \, \frac{\delta^{(3)} \left( \vec{k}_1 + \vec{k}_2 + \vec{k}_3 \right) }{ k_1^3 k_2^3 k_3^3 } \, \int \frac{d^3 p}{p^2 \, \vert \vec{k}_3 - \vec{p} \vert^2} \nonumber\\ 
 &  \qquad 
\times   \int_{-\infty}^{\tau} d \tau' \;  G \left( - k_3 \, \tau ,\, - k_3 \tau' \right) \, \left[ - 4 \, \partial_{\tau'}^2  - 4 \, \partial_{\tau''}^2 + 27 \, \partial_{\tau'} \,  \partial_{\tau''} + \frac{27}{\tau'} \, \partial_{\tau'}  + \frac{27}{\tau'} \, \partial_{\tau''} + \frac{43}{\tau^{' 2}} \right]     \nonumber\\ 
&  \qquad 
\times \int_{-\infty}^{{\rm min} \left( \tau' ,\, \tau''  \right)} \frac{d \tau_0}{\left( - \tau_0 \right)^4} \, 
G \left( - k_1 \, \tau ,\, - k_1 \, \tau_0 \right) \, 
G \left( - k_2 \, \tau ,\, - k_2 \, \tau_0 \right) \, 
G \left( - p \, \tau' ,\, - p \, \tau_0 \right) \nonumber \\
& \qquad \qquad \times
G \left( - \vert \vec{k}_3 - \vec{p} \vert \, \tau'' ,\, - \vert \vec{k}_3 - \vec{p} \vert \, \tau_0 \right)  \Big\vert_{\tau'' = \tau'}. 
\end{align} 
Because $G(x,x)=0$, there is no contribution from the derivatives acting on the extremum of integration.  Explicitly evaluating this expression as in the previous subsection we obtain, in the equilateral $k_1  = k_2 = k_3 = k$ case, 
\begin{eqnarray} 
&&   \!\!\!\!\!\!\!\!  \!\!\!\!\!\!\!\!   \!\!\!\!\!\!\!\!  \!\!\!\!\!\!\!\! 
 \left\langle \delta \varphi_1  \left( \vec{k}_1 ,\, \tau \right)  \delta \varphi_1 \left( \vec{k}_2 ,\, \tau \right)  \delta \varphi_2 \left( \vec{k}_3 ,\, \tau \right)  \right\rangle_{\rm I,1PI,{\rm equil}}   =   \,  {\rm sign } \left( \dot{\varphi}_0 \right) \, 
 \frac{\sqrt{2} \, C_4 g^3 H^3 {\tilde \mu}^4}{49 \pi^{9/2}} \, \frac{\delta^{(3)} \left( \vec{k}_1 + \vec{k}_2 + \vec{k}_3 \right) }{ k^6  } \, \int \frac{d^3 q}{ q^2 \, \vert {\hat k}_3 - \vec{q} \vert^2 } \nonumber\\ 
 &&   \!\!\!\!\!\!\!\!  \!\!\!\!\!\!\!\! 
\times   \int_0 d x' \;  c_1 \left( x' \right) \,   \int_{x'} \frac{d x_0}{ x_0^4 } \, c_1^2 \left( x_0 \right)   \nonumber\\ 
& &   \!\!\!\!\!\!\!\!  \!\!\!\!\!\!\!\! 
\Bigg\{ -4 q^2  G^{(2,0)} \left( q \, x' ,\, q \, x_0 \right) \, G \left(  \vert {\hat k}_3 - \vec{q} \vert \, x' ,\,  \vert {\hat k}_3 - \vec{q} \vert \, x_0 \right) - 4 \, 
\vert {\hat k}_3 - \vec{q} \vert^2  \, G \left( q \, x' ,\, q \, x_0 \right) \,  G^{(2,0)} \left(  \vert {\hat k}_3 - \vec{q} \vert \, x' ,\,  \vert {\hat k}_3 - \vec{q} \vert \, x_0 \right)   \nonumber\\ 
& &   \!\!\!\!\!\!\!\!  \!\!\!\!\!\!\!\! 
 + 27 \, q \, \vert {\hat k}_3 - \vec{q} \vert \,  G^{(1,0)} \left( q \, x' ,\,  q \, x_0 \right) \,  G^{(1,0)} \left(  \vert {\hat k}_3 - \vec{q} \vert \, x' ,\,  \vert {\hat k}_3 - \vec{q} \vert \, x_0 \right)     \nonumber\\ 
& &   \!\!\!\!\!\!\!\!  \!\!\!\!\!\!\!\! 
 + \frac{27}{x'} \left[   q G^{(1,0)} \left( q \, x' ,\, q \, x_0 \right) \,  G \left(  \vert {\hat k}_3 - \vec{q} \vert \, x' ,\,  \vert {\hat k}_3 - \vec{q} \vert \, x_0 \right) + 
 \vert {\hat k}_3 - \vec{q} \vert  G \left( q \, x' ,\, q \, x_0 \right) \,  G^{(1,0)} \left(  \vert {\hat k}_3 - \vec{q} \vert \, x' ,\,  \vert {\hat k}_3 - \vec{q} \vert \, x_0 \right)  \right]   \nonumber\\ 
& &   \!\!\!\!\!\!\!\!  \!\!\!\!\!\!\!\! 
 + \frac{43}{x^{' 2}}  \, G \left(  q \, x' ,\,  q \, x_0 \right) \,  G \left(  \vert {\hat k}_3 - \vec{q} \vert \, x' ,\,  \vert {\hat k}_3 - \vec{q} \vert \, x_0 \right) \Bigg\} \;. 
 \label{d1d1d2-Ic-par} 
 \end{eqnarray} 

We do not perform the integration, but show that this result has a strong parametric suppression with respect to (\ref{eq:B112-Id-res}). To see this, it is convenient to simplify the integrand by disregarding ${\hat k}$ with respect to $\vec{q}$.  (This is justified for $q \gg 1$. As we now show, the integrand is peaked at $q = {\rm O } \left( 1 \right)$, so our estimate may be incorrect by a order one factor which cannot impact the strong parametric suppression that we found below.) Doing this, and rescaling $y' \equiv q \, x' \;,\; y_0 \equiv q \, x_0$, we find 
\begin{eqnarray} 
&&   \!\!\!\!\!\!\!\!  \!\!\!\!\!\!\!\!   \!\!\!\!\!\!\!\!  \!\!\!\!\!\!\!\! 
 \left\langle \delta \varphi_1  \left( \vec{k}_1 ,\, \tau \right)  \delta \varphi_1 \left( \vec{k}_2 ,\, \tau \right)  \delta \varphi_2 \left( \vec{k}_3 ,\, \tau \right)  \right\rangle_{\rm I,1PI,{\rm equil}}    \simeq    \,  {\rm sign } \left( \dot{\varphi}_0 \right) \, 
 \frac{\sqrt{2} \, C_4 g^3 H^3 {\tilde \mu}^4}{49 \pi^{9/2}} \, \frac{\delta^{(3)} \left( \vec{k}_1 + \vec{k}_2 + \vec{k}_3 \right) }{ k^6  } \nonumber\\ 
 &&   \!\!\!\!\!\!\!\!  \!\!\!\!\!\!\!\! 
\times 4 \pi \int  d q  \, q^2 \,  \int_0 d y' \;  c_1 \left( \frac{y'}{q} \right) \,   \int_{y'} \frac{d y_0}{ y_0^4 } \, c_1^2 \left( \frac{y_0}{q} \right)   \nonumber\\ 
& &   \!\!\!\!\!\!\!\!  \!\!\!\!\!\!\!\! 
\Bigg\{ -8  \, G^{(2,0)} \left(  y' ,\,  y_0 \right) \,  G \left(  y' ,\,  y_0 \right) \, 
 + 27 \,    G^{(1,0)} \left(  y' ,\,   y_0 \right)   G^{(1,0)} \left(  y' ,\,   y_0 \right)   \nonumber\\ 
& &   \!\!\!\!\!\!\!\!  \!\!\!\!\!\!\!\! 
 + \frac{54}{y'}   \, G^{(1,0)} \left(  y' ,\,  y_0 \right) \,   G \left(  y' ,\,  y_0 \right)  
 + \frac{43}{y^{' 2}}  \, G \left(   y' ,\,   y_0 \right) \,   G \left(   y' ,\,   y_0 \right)  \Bigg\} \,. 
\label{dph1-1-2-connected-largeq}
\end{eqnarray} 

We now use the same scaling arguments that we presented after eq. (\ref{eq:B112-Id-res}), and that reproduced the ${\tilde \mu}-$dependence of that result. The expression in the curly parenthesis indicates that the integrand is dominated by $y' ,\, y_0 = {\rm O } \left( {\tilde \mu} \right)$. The four terms in that parenthesis scale as ${\tilde \mu}^{-2}$. From the scaling of $c_1$, it then follows that the integrand is dominated by $q = {\rm O } \left( 1 \right)$. Therefore the last three lines of (\ref{dph1-1-2-connected-largeq}) can be estimated  as ${\tilde \mu}^{-1}$ times a numerical factor. Therefore, the ratio between the 1PI and 1PR contributions parametrically scales as 
\begin{equation}
\frac{ \left\langle \delta \varphi_1  \left( \vec{k}_1 ,\, \tau \right)  \delta \varphi_1 \left( \vec{k}_2 ,\, \tau \right)  \delta \varphi_2 \left( \vec{k}_3 ,\, \tau \right)  \right\rangle_{\rm I,1PI,{\rm equil}} }{ \left\langle \delta \varphi_1  \left( \vec{k}_1 ,\, \tau \right)  \delta \varphi_1 \left( \vec{k}_2 ,\, \tau \right)  \delta \varphi_2 \left( \vec{k}_3 ,\, \tau \right)  \right\rangle_{\rm I,1PR,{\rm equil}} } \propto \frac{{\tilde \mu}^3 \, g^3 \, H^3}{g^2 H {\tilde \mu}^2 \, \vert \dot{\varphi}_0 \vert} 
= \frac{{\tilde \mu}^2 H^2}{g \vert \dot{\phi}_0 \vert} \times  \frac{g^2}{{\tilde \mu}} \ll 1  \;. 
\label{ratio-c-d}
\end{equation} 

This first factor in the final expression is $\ll 1$, due to the condition ${\rm C}_8$ in Table \ref{tab:conditions}. The second factor is also $\ll 1$, due to the fact that $g \ll 1$ and ${\tilde \mu } > 1$ in the region of our interest. We therefore see that the 1PI contribution can be disregarded. We note that the 1PI contribution was not considered in the estimates of Ref. \cite{Green:2009ds}, which assumed that the 1PR term is indeed dominant. The computations of this subsection  confirms this.

\subsection{$g_1 \, {\hat s}_1 \, \rightarrow g_2$ contributions  (Line II)}
\label{ap:line-II}

In this subsection we compute the contribution from the terms in the second line of 
eq. \eqref{eq:2nd_order_solution_gen} to the second order perturbation (as already  commented above, we disregard the $s_2'$ contribution).  This is formally given by 
\begin{align}
g_{2s,II} \left( \vec{k}_3 ,\, \tau  \right) &= - \int^\tau d\tau^\prime \,G ( - k_3 \, \tau , - k_3 \tau') \partial_{\tau'}  \int \frac{d^3 p}{\left( 2 \pi \right)^{3/2}}  \frac{p}{k_3}  g_1 \left( \tau' ,\, \vec{p} \right)  {\hat s}_1 \left( \tau' ,\, \vec{k}_3 - \vec{p} \right) \;. 
\label{g2-s-II}
\end{align}
All these contributions were disregarded in the estimates of  Ref. \cite{Green:2009ds}. We show that such terms provide the dominant contribution to the total bispectrum in  the available region of parameter space. 

From \eqref{g2-s-II}, we obtain the  following contribution to the bispectrum 
\begin{align}
& \left\langle g_{1s} \left( \vec{k}_1 ,\, \tau \right)  g_{1s} \left( \vec{k}_2 ,\, \tau \right)  g_{2s,II} \left( \vec{k}_3 ,\, \tau \right)  \right\rangle   \simeq      \int_{-\infty}^\tau d \tau' \left[ \partial_{\tau'} \; G \left( - k_3 \, \tau ,\, - k_3 \, \tau'  \right) \right] \; \tilde{\cal D} \;, 
\label{g1g1g2-II-formal}
\end{align} 
with 
\begin{align}
\tilde{\cal D} \equiv  \int \frac{d^3 p}{\left( 2 \pi \right)^{3/2}}  
\frac{p}{k_3}  \,  \left\langle g_1 \left( \vec{k}_1 ,\, \tau \right)  g_1 \left( \vec{k}_2 ,\, \tau \right)   g_1 \left( \vec{p} ,\, \tau' \right) \, {\hat s}_1 \left( \vec{k}_3 - \vec{p} ,\, \tau' \right)  \right\rangle \;. 
\end{align} 

We then make use of the expression (\ref{eq:g1_soln}) for the first order mode $g_1$, and (\ref{eq:sources}) for the source ${\hat s}_1$. We end up with a series of terms each involving the contraction of eight operators ${\hat \chi}_i$. 
This is completely analogous to the correlator appearing in eq. \eqref{Sc-Sd} in the $n=4$ case. We perform the contractions exactly as outlined in Appendix \ref{app:dphi1n}. We end up with 
\begin{eqnarray} 
\tilde{\cal D} &=& \tilde{\cal D}_{1PR} +  \tilde{\cal D}_{1PI} \;, \nonumber\\ 
\tilde{\cal D}_{1PR} &=& \Bigg[ -  {\rm sign } \left( \dot{\varphi}_0 \right)  \,  \frac{64 \, {\cal C}_2^2}{ \left( 2 \pi \right)^{15/2}}  \sum_{a,b} \; 
 G \left( - k_1 \tau ,\, - k_1 \tau_{0a} \right) \; G \left( - k_2 \tau ,\, - k_2 \tau_{0b} \right)   \; G \left( - k_1 \tau' ,\, - k_1 \tau_{0a} \right)   \nonumber\\ 
&&   \quad \quad \quad \times \, 
\frac{g^8 H^3 \vert \dot{\varphi}_0 \vert^3}{ k_1^5 \, k_2^5 \, k_3  } 
 \delta^{(3)} \left( \vec{k}_1 + \vec{k}_2 + \vec{k}_3  \right) \; 
 a^3 \left( \tau_{0a} \right)   \frac{\theta \left( \tau' - \tau_{0b} \right)}{ \omega_b \left( \tau' \right) } 
 a^3 \left( \tau_{0b} \right)  \Bigg]  + \left( \vec{k}_1 \leftrightarrow \vec{k}_2  \;\;,\;\; a \leftrightarrow b \right) \;,   \nonumber\\ 
&& \!\!\!\!\!\!\!\!  \!\!\!\!\!\!\!\!  \!\!\!\!\!\!\!\!  \!\!\!\!\!\!\!\! 
\tilde{\cal D}_{1PI}  =  -  {\rm sign } \left( \dot{\varphi}_0 \right)  \,  \frac{64 \, {\cal C}_4}{\left(2 \pi \right)^{15/2}} \,  \delta^{(3)} \left(  \vec{k}_1 +  \vec{k}_2 + \vec{k}_3   \right) \, \sum_a  G \left( - k_1 \tau ,\, - k_1 \tau_{0a} \right) \;  G \left( - k_2 \tau ,\, - k_2 \tau_{0a} \right) \; 
\, \int d^3 p \,  G \left( - p \tau' ,\, - p \tau_{0a} \right) \; \nonumber\\ 
&&  \quad\quad\quad\quad \times \, \frac{H^3 \, g^{13/2} \,  \vert \dot{\varphi}_{0a} \vert^{3/2}  }{  k_1^3 \, k_2^3 \, k_3 \, p^2 \, \vert \vec{k}_3 -  \vec{p} \vert^2 } \, \frac{ \theta \left( \tau' - \tau_{0a} \right)  }{  \, \omega_a \left( \tau' \right) } 
a^3 \left( \tau_{0a} \right) \;. 
\label{Dtilde-d-c}
\end{eqnarray} 

The two terms give rise, respectively, to the 1PR and 1PI diagram contributions that we evaluate in 
 Subsections \ref{ap:second_order-IId} and \ref{ap:second_order-IIc}.  The sum is dominated by the  1PR contribution, the final result of which is given in eq. \eqref{d1d1-d2IId-res}

\subsubsection{$\left\langle  g_{1s} g_{1s} g_{2s} \right\rangle_{II,1PR}$ contribution} 
\label{ap:second_order-IId} 

Inserting the $\tilde{\cal D}_{1PR}$ result of eq. \eqref{Dtilde-d-c} into the expression \eqref{g1g1g2-II-formal}, and converting the sums over the species $a$ and $b$ into integrals over the production time (see equation \eqref{Delta}), we obtain in the equilateral case 
\begin{align}
& \left\langle g_{1s} \left( \vec{k}_1 ,\, \tau \right)  g_{1s} \left( \vec{k}_2 ,\, \tau \right)  g_{2s} \left( \vec{k}_3 ,\, \tau \right)  \right\rangle_{\rm II,1PR, equil}   \simeq     -   {\rm sign } \left( \dot{\varphi}_0 \right)  \, 
 \frac{128 \, {\cal C}_2^2}{ \left( 2 \pi \right)^{15/2}}   \, 
\frac{g^8  \vert \dot{\varphi}_0 \vert^5}{H^5 k^{6}  \, \Delta^2} \, 
 \delta^{(3)} \left( \vec{k}_1 + \vec{k}_2 + \vec{k}_3  \right) \; 
 \nonumber\\ 
& \quad\quad\quad\quad \int \frac{d x_{0a}}{x_{0a}^4}  G \left( x ,\, x_{0a} \right)  \int \frac{d x_{0b}}{x_{0b}^4}  G \left( x ,\, x_{0b} \right)  \;  \int_x^{x_{0b}}  \frac{  d x' }{ \omega_b \left( x' \right) }  \left[ \partial_{\tau'} \; G \left(  x   ,\,  x'  \right) \right]  \; G \left( x' ,\, x_{0a} \right)   \,. 
\end{align} 

 Using 
\begin{equation}
 \omega_i \left( \tau \right) \simeq a \left( \tau \right) m_i \left( \tau \right) = \frac{g \, \vert \varphi_0 \left( \tau \right) - \varphi_{0i} \vert }{- H \tau} \simeq  \frac{g \, \vert \dot{\varphi}_0 \vert  \left( t - t_{0i} \right)  }{- H \tau} = 
  - \frac{g \vert \dot{\varphi}_0 \vert}{H^2 \tau} \; \ln \left( \frac{\tau_{0i}}{\tau} \right) \,, 
\end{equation}

as well as the relations (\ref{mu-mut}), and the limit (\ref{eq:xG}), we arrive at 
\begin{align}
& \!\!\!\!\!\!\!\!\!\!  \!\!\!\!\!\!\!\!\!\! 
 \left\langle \delta \varphi_{1s} \left( \vec{k}_1 ,\, \tau \right)  \delta \varphi_{1s} \left( \vec{k}_2 ,\, \tau \right)  \delta \varphi_{2s} \left( \vec{k}_3 ,\, \tau \right)  \right\rangle_{\rm II,1PR, equil}   \simeq     {\rm sign } \left( \dot{\varphi}_0 \right)  \, 
\frac{2^{61/6} {\cal C}_2^2 \sqrt{\pi}}{7^{8/3}} \, g^{1/3} \, \Delta^{2/3} \, H^{7/3} \, {\tilde \mu}^{16/3} 
 \; \frac{ \delta^{(3)} \left( \vec{k}_1 + \vec{k}_2 + \vec{k}_3  \right) }{ k^6 }  \nonumber\\ 
& \quad\quad\quad\quad   \int \frac{d x_{0a}}{x_{0a}^4}  c_1 \left( x_{0a} \right)  \int \frac{d x_{0b}}{x_{0b}^4}  c_1 \left( x_{0b} \right)  \;  \int_0^{x_{0b}}  d x'  \;  \frac{x'}{ \ln \left( \frac{x_{0b}}{x'} \right) } \; c_1' \left( x'  \right)   \; G \left( x' ,\, x_{0a} \right)   \;. 
\label{d1d1-d2IId-par}
\end{align} 

Let us denote by ${\cal I}$ the second line of this expression. We decompose the remaining Green function $G \left( x' ,\, x_{0a} \right)$ in the three terms according to eq. \eqref{eq:green}, so that  each term can be written as a two dimensional integral (over $x_{0b}$ and $x'$) times a decoupled one-dimensional integral (over $x_{0a}$). 
\begin{eqnarray} 
{\cal I} &=& \sum_{i=1}^3 \int \frac{d x_{0a}}{x_{0a}^4}  c_1 \left( x_{0a} \right)   c_i \left( x_{0a} \right) \times  \int \frac{d x_{0b}}{x_{0b}^4}  c_1 \left( x_{0b} \right)  \;  \int_0^{x_{0b}}  d x'  \;  \frac{x'}{ \ln \left( \frac{x_{0b}}{x'} \right) } \; c_1' \left( x'  \right)   \; f_i \left( x' \right) \nonumber\\ 
&\simeq&  \frac{0.2}{\tilde \mu}  \int \frac{d x_{0b}}{x_{0b}^4}  c_1 \left( x_{0b} \right)  \;  \int_0^{x_{0b}\left( 1 - \frac{H  \Delta}{\vert \dot{\varphi}_0 \vert} \right)}  d x'  \;  \frac{x'}{ \ln \left( \frac{x_{0b}}{x'} \right) } \; c_1' \left( x'  \right)  
f_1 \left( x' \right)  \,. 
\end{eqnarray} 

In the second step, we used the fact that the one dimensional integral in the $i=1$ term is several orders of magnitude greater than the one dimensional integrals in the $i=2,3$ contributions, so we disregarded the latter. 
We note  that the one dimensional integral scales as ${\tilde \mu}^{-1}$, in agreement with the scaling (\ref{c1-scaling}) and with the discussion presented around that equation. Finally, we note the shift in the upper extremum of the $d x'$ integral. Without this shift, the integral diverges. This is however a spurious divergence due to the approximation \eqref{Delta} (sum into integral)  which fails at $x' = x_{0b}$. We remove this spurious singularity by resolving the time between two separate instances of particle production. 

We can then approximate 
\begin{eqnarray}
{\cal I} &\simeq&  \frac{0.2}{\tilde \mu}  \int \frac{d x_{0b}}{x_{0b}^4}  c_1 \left( x_{0b} \right)  \;  
x_{0b} \, c_1' \left( x_{0b}  \right) \, f_1 \left( x_{0b} \right)  \int_0^{x_{0b}\left( 1 - \frac{H  \Delta}{\vert \dot{\varphi}_0 \vert} \right)}  \frac{ d x' }{ \ln \left( \frac{x_{0b}}{x'} \right)  } \;. 
\end{eqnarray} 
The $d x'$ integral can be then performed analytically and evaluated in the upper extremum, where it is dominated. In the $H \, \Delta \ll \vert \dot{\varphi}_0 \vert$ regime (condition $C_6$ in Table \ref{tab:conditions})  it approximately evaluates to $-x_{0b} \, \ln \frac{H \, \Delta}{\vert \dot{\varphi}_0 \vert}$. The remaining integral over $d x_{0b}$ can be performed numerically, and we arrive to ${\cal I} \simeq   - \frac{0.08}{{\tilde \mu}^2} \;  \ln \frac{H \, \Delta}{\vert \dot{\varphi}_0 \vert} $. The scaling of this result with ${\tilde \mu}$ can be understood already from (\ref{d1d1-d2IId-par}) using the same arguments presented at the end of Subsection \ref{ap:second_order-Id}.  

Inserting the result for ${\cal I}$ in eq. \eqref{d1d1-d2IId-par}, we finally obtain 
\begin{align}
& \!\!\!\!\!\!\!\!\!\!  \!\!\!\!\!\!\!\!\!\! 
 \left\langle \delta \varphi_{1s} \left( \vec{k}_1 ,\, \tau \right)  \delta \varphi_{1s} \left( \vec{k}_2 ,\, \tau \right)  \delta \varphi_{2s} \left( \vec{k}_3 ,\, \tau \right)  \right\rangle_{\rm II,1PR, equil}+ 2 \, {\rm perm.} \nonumber\\ 
& \quad\quad  \simeq -  {\rm sign } \left( \dot{\varphi}_0 \right)  \, 0.12 \,  \, g^{1/3} \, \Delta^{2/3} \, H^{7/3} \, {\tilde \mu}^{10/3} \, 
 \ln \frac{H \, \Delta}{\vert \dot{\varphi}_0 \vert} \;  \frac{ \delta^{(3)} \left( \vec{k}_1 + \vec{k}_2 + \vec{k}_3  \right) }{ k^6 }   \;.  
\label{d1d1-d2IId-res}  
\end{align} 

When compared with the 1PR diagram contribution evaluated in Subsection \ref{ap:second_order-Id}, we see that  
\begin{align}
& \!\!\!\!\!\!\!\!   \!\!\!\!\!\!\!\!  
\frac{
\left\langle \delta \varphi_1 \delta \varphi_1 \delta \varphi_2^{\rm extra} \right\rangle_{\rm II,1PR, equil} 
}{ 
 \left\langle \delta \varphi_1 \delta \varphi_1 \delta \varphi_2 \right\rangle_{\rm I,1PR, equil}
} 
\simeq 7  \,  \ln \frac{H \, \Delta}{\vert \dot{\varphi}_0 \vert} \;.  
\label{ratio-II-Id}
\end{align} 
This stems from the difference between the sources $s_1$ and ${\hat s}_1$.

\subsubsection{$\left\langle  g_{1s} g_{1s} g_{2s} \right\rangle_{II,1PI}$ contribution} 
\label{ap:second_order-IIc} 

We insert the $\tilde{\cal D}_{1PI}$ result of eq. \eqref{Dtilde-d-c} and proceed as in the previous subsection. We obtain, in the equilateral case 
\begin{align} 
& \!\!\!\!\!\!\!\!  \!\!\!\!\!\!\!\! 
\left\langle \delta \varphi_{1s} \left( \vec{k}_1 ,\, \tau \right)  \delta \varphi_{1s} \left( \vec{k}_2 ,\, \tau \right)  \delta \varphi_{2s} \left( \vec{k}_3 ,\, \tau \right)  \right\rangle_{\rm II,1PI, equil}    \simeq    {\rm sign } \left( \dot{\varphi}_0 \right)  \, 
\frac{ \delta^{(3)} \left(  \vec{k}_1 +  \vec{k}_2 + \vec{k}_3   \right) }{ k^6 } \,  \frac{  64 \, {\cal C}_4  \, g^{11/2} \,  \vert \dot{\varphi}_{0} \vert^{3/2} H }{ \left( 2 \pi \right)^{15/2}  \, \Delta   } \nonumber\\ 
&  \int \frac{d^3 q}{ q^2 \vert {\hat k}_3 - \vec{q} \vert^2}  \;  \int \frac{ d x_{0a} }{   x_{0a}^4   }  \, c_1^2 \left(  x_{0a} \right)   \, \int_x^{x_{0a}}  \frac{ x' \, d x' }{  \ln \left( \frac{x_{0a}}{x'} \right) }   c_1' \left(  x'  \right) \,   G \left( q \, x' ,\,  q \, x_{0a} \right) \; .  
\end{align}  

Using the same arguments employed to estimate the integral in eq. \eqref{d1d1d2-Ic-par}, we estimate this result as 
\begin{align} 
& \!\!\!\!\!\!\!\!  \!\!\!\!\!\!\!\! 
\left\langle \delta \varphi_{1s} \left( \vec{k}_1 ,\, \tau \right)  \delta \varphi_{1s} \left( \vec{k}_2 ,\, \tau \right)  \delta \varphi \left( \vec{k}_3 ,\, \tau \right)  \right\rangle_{\rm II,1PI, equil}   \simeq   {\rm sign } \left( \dot{\varphi}_0 \right)  \, 
\frac{ \delta^{(3)} \left(  \vec{k}_1 +  \vec{k}_2 + \vec{k}_3   \right) }{ k^6 } \frac{  g^{11/2} \,  \vert \dot{\varphi}_{0} \vert^{3/2} H }{  \Delta   } \times {\tilde \mu} \,  \ln \frac{H \, \Delta}{\vert \dot{\varphi}_0 \vert} \; . 
\end{align} 

When compared with the result obtained in Subsection \ref{ap:second_order-Ic}, we obtain the same scaling 

\begin{align}
& \!\!\!\!\!\!\!\!   \!\!\!\!\!\!\!\!  
\frac{
\left\langle \delta \varphi_1 \delta \varphi_1 \delta \varphi_2^{\rm extra} \right\rangle_{\rm II,1PI, equil}
}{ 
 \left\langle \delta \varphi_1 \delta \varphi_1 \delta \varphi_2 \right\rangle_{\rm I,1PI, equil}
} 
\sim    \ln \frac{H \, \Delta}{\vert \dot{\varphi}_0 \vert} \;.  
\label{ratio-II-Ic}
\end{align} 
as in eq. \eqref{ratio-II-Id}. It then follows that  the  II,1PI contribution can be disregarded in comparison with the  II,1PIR one, in the same way as the   I,1PI  contribution could be disregarded in comparison with the I,1PR one (see eq. \eqref{ratio-c-d}.)

\section{Bispectrum in the EFT computation}
\label{ap:BS-gradient}

The term (\ref{gradient-operator}) produces the additional term in the second order solution  
\begin{eqnarray} 
& &  \!\!\!\!\!\!\!\!  \!\!\!\!\!\!\!\!  \!\!\!\!\!\!\!\!  \!\!\!\!\!\!\!\! 
g_{2s,{\rm grad}} \left( \vec{k} ,\, \tau \right)  =  
   \frac{ 5 \,  H {\tilde \mu}^2}{4 \,  \dot{\varphi}_0  \, k^3} \int \frac{d^3 p}{\left( 2 \pi \right)^{3/2}} \,  p \, \vert \vec{k} - \vec{p} \vert \, \vec{p} \cdot \left( \vec{p} - \vec{k} \right)  \nonumber\\ 
& & \quad\quad  \quad\quad  \quad\quad 
  \int_{-\infty}^\tau d \tau' G \left( - k \, \tau ,\, - k \, \tau'  \right)  \, g_{1s} \left( \vec{p} ,\, \tau' \right)  g_{1s} \left( \vec{k} - \vec{p} ,\, \tau' \right) \,.  
\label{grad-sol}
\end{eqnarray} 
This term gives the bispectrum contribution
\begin{equation}
 \left\langle g_{1s} g_{1s} g_{2s} \right\rangle_{\rm grad} \equiv  \left\langle g_{1s} g_{1s} g_{2s,{\rm grad}} \right\rangle_{\rm grad,1PR}   +   \left\langle g_{1s} g_{1s} g_{2s,{\rm grad}} \right\rangle_{\rm grad,1PI}   \,.   
\label{grad-corr}
\end{equation} 

Inserting (\ref{grad-sol}) into (\ref{grad-corr}), and proceeding exactly as in the previous appendix, we obtain  
\begin{align}
& \left\langle g_{1s} \left( \vec{k}_1 ,\, \tau \right)  g_{1s} \left( \vec{k}_2 ,\, \tau \right)  g_{2s} \left( \vec{k}_3 ,\, \tau \right)  \right\rangle_{\rm grad,1PR}   \simeq  {\rm sign } \left( \dot{\varphi}_0 \right)   \frac{80 \, {\cal C}_2^2  \, g^7 \, {\tilde \mu}^2 \, \vert \dot{\phi}_0 \vert^4}{H^3 \, \Delta^2 } \,  \frac{\delta^{(3)} \left( \vec{k}_1 + \vec{k}_2 + \vec{k}_3 \right)}{\left( 2 \pi \right)^{15/2}} \, \frac{ k_3^2 - k_1^2 - k_2^2}{k_1^5 \, k_2^5 \, k_3^3}  \nonumber\\ 
& \qquad  \int_{-\infty}^\tau d \tau' G \left( - k_3 \, \tau ,\, - k_3 \, \tau'  \right)  \left[   \int_{-\infty}^{\tau'}  \frac{  d \tau_{0}  }{ \left( - \tau_{0} \right)^4} \;    G \left( - k_1 \, \tau ,\, - k_1 \, \tau_{0}  \right)  \;   G \left( - k_1 \, \tau' ,\, - k_1 \, \tau_{0}  \right)  \right] \nonumber \\
& \qquad  \qquad \qquad \left[   \int_{-\infty}^{\tau'}  \frac{  d \tau_{0}  }{ \left( - \tau_{0} \right)^4} \;    G \left( - k_2 \, \tau ,\, - k_2 \, \tau_{0}  \right)  \;   G \left( - k_2 \, \tau' ,\, - k_2 \, \tau_{0}  \right)  \right] \;. 
\end{align} 
and
\begin{align}
& \left\langle g_{1s} \left( \vec{k}_1 ,\, \tau \right)  g_{1s} \left( \vec{k}_2 ,\, \tau \right)  g_{2s} \left( \vec{k}_3 ,\, \tau \right)  \right\rangle_{\rm grad,1PI}   \simeq   {\rm sign } \left( \dot{\varphi}_0 \right)      \frac{80 \,  H {\tilde \mu}^2}{ \Delta } \; \frac{ {\cal C}_4 \, g^{11/2} \vert \dot{\phi}_0 \vert^{3/2}}{k_1^3 \, k_2^3 \, k_3^3} \, \frac{\delta^{(3)} \left( \vec{k}_1 + \vec{k}_2 + \vec{k}_3 \right) }{\left( 2 \pi \right)^{15/2}}  
 \nonumber\\ 
& \qquad 
\int    \frac{d^3 p }{   p^2 \vert \vec{k}_3 - \vec{p} \vert^2   }  \,   \vec{p} \cdot \left( \vec{k}_3 - \vec{p} \right)    \; 
 \int_{-\infty}^\tau d \tau' G \left( - k_3 \, \tau ,\, - k_3 \, \tau'  \right)  \nonumber\\ 
&  \int_{-\infty}^{\tau'}  \frac{ d \tau_{0}  }{ \left( - \tau_{0} \right)^4} 
\;   G \left( - k_1 \, \tau ,\, - k_1 \, \tau_{0}  \right)  G \left( - k_2 \, \tau ,\, - k_2 \, \tau_{0}  \right)     G \left( - p \, \tau' ,\, - p \, \tau_{0}  \right)  G \left( - \vert \vec{k}_3 - \vec{p} \vert \, \tau' ,\, -  \vert \vec{k}_3 - \vec{p} \vert  \, \tau_{0}  \right) \;, 
\end{align}

The 1PR term is evaluated exactly as in appendix \ref{ap:second_order-Id}. In the equilateral limit, and at super horizon scales,  we obtain 
\begin{align}
& \left\langle \delta \varphi_1 \, \delta \varphi_1 \, \delta \varphi_2 \right\rangle_{\rm grad,1PR,equil} + 2 {\rm perm.} \simeq 
-   {\rm sign } \left( \dot{\varphi}_0 \right)   \frac{240 \, \sqrt{2} \, {\cal C}_2^2 \, g^2 \, {\tilde \mu}^6 \, H \,  \vert \dot{\varphi}_0 \vert}{49 \, \pi^{3/2}} 
 \,  \frac{\delta^{(3)} \left( \vec{k}_1 + \vec{k}_2 + \vec{k}_3 \right)}{k^6}   \nonumber\\ 
& \times
  \int^{\infty}_0 d x' c_1 \left(   x'  \right)  \left[  \int^{\infty}_{x'}  \frac{  d x_{0}  }{ \left(  x_{0} \right)^4} 
\;    c_1 \left(  x_{0}  \right)  \;   G \left(  x' ,\,  x_{0}  \right)  \right]^4 
 \simeq - 0.00067 \, g^2 \, {\tilde \mu}^4 \, H \,  \dot{\varphi}_0 \,  \frac{\delta^{(3)} \left( \vec{k}_1 + \vec{k}_2 + \vec{k}_3 \right)}{k^6}    \;. 
\end{align} 

The 1PI term is evaluated exactly as in appendix  \ref{ap:second_order-Ic}. We find that the ratio between the 1PI and the 1PR contributions scales exactly as eq. (\ref{ratio-c-d}), and therefore the 1PI term can be disregarded.

\section{Comparison with previous results}
\label{ap:comparison}

In this appendix we explain the origin of the differences between our results and those of \cite{Green:2009ds}. 
We divide the discussion in two parts, one for the two point function, and one for the three point function. In this discussion, quantities with the suffix ``previous'' refers to results of calculations of \cite{Green:2009ds}. 
The suffix ``used  in  \cite{Green:2009ds}'' refers to quantities defined in \cite{Green:2009ds}.

\subsection{Difference in the two-point function} 

Eq. (\ref{PS-res}) of the main text presents our result for the two point function of trapped inflation 
\begin{equation}
P_{\zeta} \left( k \right)  \simeq 
5.7 \cdot 10^{-4} \, \frac{g^{9/4} \, H}{\Delta^{1/2} \, \vert \dot{\varphi}_0 \vert^{1/4}} 
= 6.9 \cdot 10^{-5} \,   \frac{g^{7/2} \, \vert \dot{\varphi}_0 \vert^{1/2}}{\Delta \, {\tilde \mu}}  \;\;. 
\end{equation}

This disagrees by four orders of magnitude with the result obtained in the original work \cite{Green:2009ds} (their eq. 3.55, rewritten in our notation) 
\begin{equation}
P_{\zeta,{\rm previous}} \simeq \frac{g^{7/2} \, \vert \dot{\varphi}_0 \vert^{1/2}}{\Delta \, {\tilde \mu}} \simeq 15,000 \times P_{\zeta} \;. 
\end{equation} 
The result of the two-point function obtained in Ref. \cite{Green:2009ds} has been quoted and used in Ref. \cite{LopezNacir:2011kk}, so this aspect of the discussion also applies to \cite{LopezNacir:2011kk}. 

We verified that, once rewritten in our notation,  eq. (3.42) of  \cite{Green:2009ds} for the $\delta \varphi_1$ perturbation agrees with the first equation of our system (\ref{eq:system-3rd}). Also their formal solution (3.52) is equivalent to our formal solution (\ref{eq:g1_soln}). Useful intermediate steps in verifying this equivalence are 
\begin{eqnarray} 
&& {\tilde G} \left( x ,\, x' \right) \vert_{\text{used  in   \cite{Green:2009ds}}} = \frac{x}{x'} G \left( x ,\, x' \right) \;, \nonumber\\ 
&& \frac{  - g \left( g \dot{\phi} \right)^{3/4} \sqrt{N_{\rm hits} H^{-1} } }{k^2} \, \Delta {\hat n}  \vert_{\text{used  in   \cite{Green:2009ds}}} = - \frac{g}{k^2 \tau^2 H^2} \Delta n  \vert_{\text{used  in   \cite{Green:2009ds}}} 
= x \, s_1 \;, 
\label{convert-GS}
\end{eqnarray} 
where $G \left( x ,\, x' \right)$ and $s_1$ are, respectively, our Green function (\ref{eq:green}) and source (\ref{eq:sources}). (The difference in the Green functions stems from the fact that our Green function is obtained from the third order differential equation for $g_1$, while theirs is obtained from the third order differential equation for $\delta \varphi_1$.)  

We then have 
\begin{eqnarray}
\!\!\!\!\!\!\!\! \!\!\!\!\!\!\!\! 
\left\langle \zeta \left( \vec{k} \right) \zeta \left( \vec{k}' \right) \right\rangle  &=&  \frac{H^2}{ \dot{\varphi}_0^2 } g^{7/2}  \vert \dot{\varphi}_0 \vert^{3/2} \, N_{\rm hits} H^{-1} 
 \int  \frac{d {\tilde \tau}'}{{\tilde \tau}'} \partial_{{\tilde \tau}'} \left[ {\tilde \tau}' \frac{{\tilde G}_{\text{used  in   \cite{Green:2009ds}}}  \left( 0 ,\, {\tilde \tau}' \right)}{k^2} \right] 
 \int  {\tilde \tau}'' \frac{d {\tilde \tau}}{\tilde \tau'} \partial_{{\tilde \tau}''} \left[ {\tilde \tau}'' \frac{{\tilde G}_{\text{used  in   \cite{Green:2009ds}}}  \left( 0 ,\, {\tilde \tau}' \right)}{k^2} \right] \nonumber\\ 
&&  \times   \left\langle \Delta {\hat n} \left( {\tilde \tau}' ,\, \vec{k} \right)   \Delta {\hat n} \left( {\tilde \tau}'' ,\, \vec{k}' \right) \right\rangle \;. 
\label{zz-agree}
\end{eqnarray}
This expression immediately follows from  (3.52) of   \cite{Green:2009ds}, and uses their notation. Since we have just verified (3.52), this expression can be used as the starting point of our and of their computation of the power spectrum. 

To proceed, one needs to evaluate the two point correlator of the source. This is where our computation starts to differ from that of  \cite{Green:2009ds}. Using our expressions (\ref{Sc-Sd}) and (\ref{Sn-result}) for the  source correlation function, and converting into their notation through (\ref{convert-GS}), we find 
\begin{equation}
 \left\langle \Delta {\hat n} \left( \tau' ,\, \vec{k} \right)  \Delta {\hat n} \left( \tau'' ,\, \vec{k}' \right) \right\rangle =  
 \delta^{(3)} \left( \vec{k} + \vec{k}' \right) \frac{{\cal C}_2}{\pi^3} \, \tau' \, \tau'' \,  
 \int d \tau_0 \frac{\theta \left( \tau' - \tau_{0} \right) \theta \left( \tau'' - \tau_0 \right) }{\left( - \tau_0 \right)^4}   \;, 
\label{SS-us-converted}
\end{equation} 
which, inserted into (\ref{zz-agree}), gives 
\begin{eqnarray}
\left\langle \zeta \left( \vec{k} \right) \zeta \left( \vec{k}' \right) \right\rangle  &\simeq&  
 \frac{g^{7/2}  \vert \dot{\varphi}_0 \vert^{1/2} }{\Delta} \delta^{(3)} \left( \vec{k} + \vec{k}' \right) \frac{{\cal C}_2}{\pi^3} \, 
  \int   \frac{ d \tau_0 }{\left( - \tau_0 \right)^4}   
  \left\{ \int_{\tau_0}  d {\tilde  \tau}'  \frac{1}{k}  \partial_{{\tilde \tau}'} \left[ {\tilde \tau}' \frac{{\tilde G}_{\text{used  in   \cite{Green:2009ds}}}  \left( 0 ,\, {\tilde \tau}' \right)}{k^2} \right] 
\right\}^2 \nonumber\\ 
  &=& \frac{\delta^{(3)} \left( \vec{k} + \vec{k}' \right)}{k^3} \, \frac{{\cal C}_2 \, g^{7/2} \,   \vert \dot{\varphi}_0 \vert^{1/2}}{\pi^3 \, \Delta} \, \int \frac{d x_0}{x_0^4} \, c_1^2 \left( x_0 \right)   \,. 
\label{zz-us-converted}
\end{eqnarray}
We stress that eqs. (\ref{SS-us-converted}) and (\ref{zz-us-converted}) are our results, expressed in the notation of  \cite{Green:2009ds}.

The source correlator used in  \cite{Green:2009ds}  (their eq. (3.43)) is 
\begin{equation}
 \left\langle \Delta {\hat n} \left( \tau' ,\, \vec{k} \right)  \Delta {\hat n} \left( \tau'' ,\, \vec{k}' \right) \right\rangle_{\rm previous} =  \left( 2 \pi \right)^3 \, \delta^{(3)} \left( \vec{k} + \vec{k'} \right) \delta \left( \tau - \tau' \right) \;. 
\label{SS-them}
\end{equation} 
which approximates the two sources as being correlated only at equal time. It is true that the source is the sum of many contributions $\Delta {\hat n}_i$, one from each $\chi_i$ species, and that different $\chi_i$ species are uncorrelated; however, this simply implies that 
\begin{equation}
\left\langle  \Delta {\hat n} \left( \tau'  \right)   \Delta {\hat n} \left( \tau''  \right)  \right\rangle = 
\sum_{ij}  \left\langle \Delta {\hat n}_i \left( \tau'  \right)   \Delta {\hat n}_j \left( \tau''  \right)  \right\rangle = 
\sum_i \left\langle  \Delta {\hat n}_i \left( \tau'  \right)   \Delta {\hat n}_i \left( \tau''  \right)  \right\rangle \;, 
\label{SS-us}
\end{equation} 
and not that the two times need to be equal to have a nonvanishing correlator. 

Inserting (\ref{SS-them}) into (\ref{zz-agree}) one obtains 
\begin{eqnarray}
\left\langle \zeta \left( \vec{k} \right) \zeta \left( \vec{k}' \right) \right\rangle_{\rm previous}   
  &=&   \left( 2 \pi \right)^3 \frac{ \delta^{(3)} \left( \vec{k} + \vec{k}' \right) }{ k^3 } 
 \frac{g^{7/2}    \vert \dot{\varphi}_0 \vert^{1/2}}{\Delta} \,   \int d {\tilde \tau}'   \left\{ \frac{1}{{\tilde \tau}'}  \partial_{{\tilde \tau}'} \left[ {\tilde \tau}' {\tilde G} \left( 0 ,\, {\tilde \tau}' \right) \right] \right\}^2   \nonumber\\ 
    &=&   \left( 2 \pi \right)^3 \frac{ \delta^{(3)} \left( \vec{k} + \vec{k}' \right) }{ k^3 } 
 \frac{g^{7/2}  \, \vert \dot{\varphi}_0 \vert^{1/2}}{\Delta} \,  \int \frac{d x_0}{x_0^2} \left[ c_1' \left( x_0 \right) \right]^2 \,. 
\label{zz-them}
\end{eqnarray}
Eq. (3.53) of   \cite{Green:2009ds} immediately follows from the result in the first line of this expression. In the second  line, we have instead used the relation (\ref{convert-GS}) and the limit (\ref{eq:xG}). 

In comparing (\ref{zz-us-converted}) with (\ref{zz-them}), we disregard the $  \left( 2 \pi \right)^3 $ factor as it simply follows from a different $2 \pi-$convention, and it cancels in the power spectrum. We then note that, according to the scaling (\ref{c1-scaling}), and to the discussion after eq. (\ref{eq:B112-Id-res}), both integrals scale as ${\tilde \mu}^{-1}$, which explains why the power spectrum of   \cite{Green:2009ds} and of the present work have the same parametric scaling. We perform the integral in (\ref{zz-them}) numerically, obtaining the result $\simeq \frac{0.055}{\tilde \mu}$. Ref.  \cite{Green:2009ds} gives the result $\frac{1}{\tilde \mu}$ for the integral. According to our expression (\ref{zz-us-converted}), we then see that the factor $\frac{1}{\tilde \mu}$ should be replaced by 
$ \frac{{\cal C}_2}{\pi^3} \, \int \frac{d x_0}{x_0^4} \, c_1^2 \left( x_0 \right)   \simeq \frac{1}{730 \, \tilde \mu}$. 

A further suppression of our power spectrum results with respect to that of   \cite{Green:2009ds} is the $\frac{1}{2 \pi^2} \simeq \frac{1}{20}$ factor in the relation (\ref{def-PB}) between the power spectrum and the two point function; this factor has been disregarded in  \cite{Green:2009ds}. From these two factors, we obtain $\frac{1}{730} \times 
\frac{1}{20} \simeq \frac{1}{15,000}$ which precisely account for the ratio between our result of the power spectrum and that of   \cite{Green:2009ds,LopezNacir:2011kk}. We note that the inaccurate approximation (\ref{SS-them}) is also quoted and used in ref. \cite{LopezNacir:2011kk}, where it appears as eq. (205).

\subsection{Difference in the three-point function}

Eq. (3.72) of  \cite{Green:2009ds} reads $f_{\rm NL,{\rm equil}} \sim {\tilde \mu}^2$, namely it is a factor 
$ {\tilde \mu}^2$ enhanced with respect our ${\rm O } \left( {\tilde \mu}^0 \right)$ solution (the first relation in  (\ref{fNL-result})) 
 As we now show, we believe that this originates from an overestimate of the $\left\langle \delta \varphi_{1s}  \delta \varphi_{1s}  \delta \varphi_{2s} \right\rangle$ correlator, which in turn originates from the approximation (\ref{SS-them}). Eqs. (3.63) and (3.64) of \cite{Green:2009ds} provide their estimate for the dominant part of the second order perturbation. 
According to  \cite{Green:2009ds}, the two terms give a comparable contribution to the bispectrum (we agree with this statement), so let us concentrate on the (3.63) term, which is the one used in  \cite{Green:2009ds} to estimate the bispectrum. In our notation, with the rescaling $g_i = \delta \varphi_i / x$, their relation (3.63) reads 
\begin{align} 
& \!\!\!\!\!\!\!\! 
g_{2,{\tilde m},{\rm previous}} \sim 
  - \frac{1}{k_3 } \,   \frac{  H {\tilde \mu}^2 }{ \vert \dot{\varphi}_0 \vert  } \int \frac{d^3 p}{\left( 2 \pi \right)^{3/2}} 
\frac{p \vert \vec{k}_3 - \vec{p} \vert}{k_3^2}  \, \int_{-\infty}^\tau d \tau'   \,  G \left( -k_3 \, \tau ,\, - k_3 \, \tau' \right) \nonumber\\
& \quad\quad\quad\quad  \quad\quad\quad\quad 
  \times 
\left[  \partial_{\tau'}   \partial_{\tau''} +  \frac{ \partial_{\tau'} }{\tau'} +  \frac{ \partial_{\tau''} }{\tau'} + \frac{1}{\tau^{'2}}  \right] 
 g_1 \left(  \vec{p} ,\, \tau' \right) \, g_1 \left( \vec{k}_3 - \vec{p} ,\, \tau'' \right) 
\Big\vert_{\tau'' = \tau'} \;. 
\end{align} 

This combination is a subset of the terms included in  our eq. (\ref{g2-I}), and it agrees with this subset  up to an order one coefficient. In our computation, this subset gives a order one contribution to the $\langle \delta \varphi_1  \delta \varphi_1  \delta \varphi_2 \rangle_{I}$ term, which we have shown to be subdominant with respect to the   $\langle \delta \varphi_1  \delta \varphi_1  \delta \varphi_2 \rangle_{II}$ term.  For these (now subdominant) modes, Ref.   \cite{Green:2009ds} obtains $\langle \delta \varphi_1  \delta \varphi_1  \delta \varphi_2 \rangle_{\rm previous} \propto  \langle \delta \varphi_1  \delta \varphi_1  \delta \varphi_2 \rangle_{I} \times {\tilde \mu}^2$. We now discuss the origin of this extra ${\tilde \mu}^2$ factor. 

Evaluating the bispectrum with the term  (3.63) of  \cite{Green:2009ds}, and disregarding the subdominant 1PI contribution, we obtain, in the notation of  \cite{Green:2009ds},   
\begin{eqnarray}
&&  \!\!\!\!\!\!\!\!  \!\!\!\!\!\!\!\!  \!\!\!\!\!\!\!\! 
\left\langle \delta \varphi_1 \left( \tau ,\, \vec{k}_1 \right) 
 \delta \varphi_1 \left( \tau ,\, \vec{k}_2 \right) 
 \delta \varphi_1 \left( \tau ,\, \vec{k}_3 \right) \right\rangle_{\rm prev., equil}  \sim  2 
\left[  g \left( g \vert \dot{\varphi}_0 \vert \right)^{3/4} \sqrt{N_{\rm hits} H^{-1}} \right]^4 
\int d {\tilde \tau}' \, \frac{{\tilde G}_{\text{used  in   \cite{Green:2009ds}}} \left(  {\tilde \tau} ,\, {\tilde \tau}' \right)}{k^2} \, \frac{{\tilde m}^2}{H \, {\tilde \tau}' \vert \dot{\varphi}_0 \vert } \nonumber\\
& & \quad\quad\quad\quad 
\times \int^{\tilde \tau} \frac{d {\tilde \tau}_A}{k^2} {\tilde g} \left( {\tilde \tau} ,\, {\tilde \tau}_A \right) 
\int^{{\tilde \tau}'} \frac{d {\tilde \tau}_B}{p} {\tilde g}^{(1,0)} \left( {\tilde \tau}' ,\, {\tilde \tau}_B \right) 
\int^{\tilde \tau} \frac{d {\tilde \tau}_C}{k^2} {\tilde g} \left( {\tilde \tau} ,\, {\tilde \tau}_C \right) 
\int^{{\tilde \tau}'} \frac{d {\tilde \tau}_D}{\vert \vec{k}_3 - \vec{p} \vert} {\tilde g}^{(1,0)} \left( {\tilde \tau}' ,\, {\tilde \tau}_D \right) \nonumber\\ 
& &  \quad\quad\quad\quad 
\times \left\langle \Delta {\hat n} \left( {\tilde \tau}_A ,\, \vec{k}_1 \right)  \Delta {\hat n} \left( {\tilde \tau}_B ,\, \vec{p} \right) \right\rangle \star  \left\langle \Delta {\hat n} \left( {\tilde \tau}_C ,\, \vec{k}_2 \right)  \Delta {\hat n} \left( {\tilde \tau}_D ,\, \vec{k}_3 - \vec{p} \right) \right\rangle  \;. 
\label{df1df1df2-them}
\end{eqnarray}

Using the exact source correlator (\ref{SS-us}) in this expression reproduces our result \eqref{eq:intT0T1}, without the 
$c_1' \left( x' \right)$ term, and with different numerical coefficients for the terms (this only changes our result by an order one factor). Using instead the approximated relation (\ref{SS-them}) for the source correlators, one instead obtains  eq. (3.69) of  \cite{Green:2009ds}, up to a factor $2$ that was disregarded in their estimate. In our notation, eq. (3.69) of  \cite{Green:2009ds} gives 
\begin{align} 
\langle \delta \varphi_1  \delta \varphi_1  \delta \varphi_2  \rangle'_{\rm previous , equil} & \sim  - 
  \frac{2^{17/3} \, \pi^2  }{7^{11/3} }   \, \frac{ {\tilde \mu}^{22/3}  \, g^{1/3} \, H^{7/3} \, \Delta^{2/3} }{k^6}  \nonumber\\ 
& \quad\quad\quad\quad 
 \int d x' \,  c_1 \left( x' \right) \left[  \tilde{\cal T}_1 \left( x' \right) +  \frac{ \tilde{\cal T}_0 \left( x' \right) }{x'} \right]_2  \left[  \tilde{\cal T}_1 \left( x' \right) + \frac{  \tilde{\cal T}_0 \left( x' \right) }{x'} \right]_3 \;, 
\label{eq:intT0T1-them} 
\end{align}
where the suffixes $2$ and $3$ refer to the $i=2,3$ contributions to the Green function \eqref{eq:green} (these are the terms considered in their estimate), and where we have defined 
\begin{equation}
\tilde{\cal T}_n \left( x' \right) \equiv \int_{x'} \frac{d x_0}{x_0^2}   \, c_1' \left( x_{0a} \right)   \, G^{(n,1)} \left( x' ,\, x_0 \right) \;. 
\end{equation}

In comparing the expression (\ref{eq:intT0T1-them}) with our result (\ref{eq:intT0T1}), we see that the only relevant difference is the replacement of the function ${\cal T}_n$ entering in our expression with the functions 
$\tilde{\cal T}_n$. As compared with  ${\cal T}_n$, the functions $\tilde{\cal T}_n$ have two fewer powers of $x_0$ in the denominator, but two extra derivatives. This mirrors the difference obtained in the two point function (see the discussion after eq.~(\ref{zz-them})), and it is simply due to the different expression used for the source correlator. 

The ${\cal T}_n$ functions entering in our expression scale with ${\tilde \mu}$ according to the relation (\ref{T0T1-scaling}). From this one obtains that all terms in  (\ref{eq:intT0T1}) provide a comparable contribution to the $d x'$ integral, that scales as ${\tilde \mu}^{-4}$. From a numerical investigation, we observe that the  $\tilde{\cal T}_n$ 
functions instead satisfy the scaling 
\begin{align}
\left\vert {\tilde {\cal T}_0 \left( x' \right)} \right\vert \simeq \frac{1}{{\tilde \mu}^2}  f_{\tilde {\cal T}_0}  \left( \frac{x'}{{\tilde \mu}} \right) \;\;\;\; {\rm with \; peak \;
value } \simeq \frac{0.2}{{\tilde \mu}^2} \;\;, \nonumber\\ 
\left\vert {\tilde {\cal T}_1 \left( x' \right)} \right\vert \simeq \frac{1}{{\tilde \mu}^2}   f_{\tilde {\cal T}_0}  \left( \frac{x'}{{\tilde \mu}} \right)  \;\;\;\; {\rm with \; peak \;
value } \simeq \frac{0.4}{{\tilde \mu}^2} \;\;. 
\end{align} 
We note that $ {\tilde {\cal T}_1 \left( x' \right)} = \partial_{x'} \,  {\tilde {\cal T}_0 \left( x' \right)} $ and from the scaling it appears that the derivative mostly acts on the phase of ${\cal T}_0$. This agrees with the discussion present after eq. (3.70) of   \cite{Green:2009ds}. As a consequence, $\tilde{\cal T}_1$ dominates the integral in (\ref{eq:intT0T1-them}), and provides an overall  ${\tilde \mu}^{-2}$ scaling. This is precisely a factor   ${\tilde \mu}^{2}$ greater than our result. 

To summarize, the use of the approximated expression  (\ref{SS-them}) for the source correlator (as opposed to the exact relation  (\ref{SS-us})) results in different expressions for the function appearing in the final time integration, both in two-point and in the three-point case. In the three-point function, this affects the scaling with ${\tilde \mu}$ of this contribution (which, moreover, we have shown to be subdominant). Therefore, the approximation   (\ref{SS-them}) does not appear to be adequate also in this case.

\end{document}